\documentclass[12pt,a4paper]{article}
\pdfoutput=1
\usepackage{amsmath}
\usepackage{color}
\usepackage{amsfonts}
\usepackage{amssymb}
\usepackage{graphicx}
\usepackage{geometry}
\usepackage{amssymb,epsfig,subfigure}
\usepackage{amssymb}
\usepackage[utf8]{inputenc}
\usepackage{hyperref}
\usepackage{wasysym}
\usepackage{comment}
\usepackage[font=footnotesize]{caption}
\usepackage[T1]{fontenc}
\usepackage{latexsym}
\usepackage[english]{babel}
\usepackage{cite}

\usepackage{enumerate}
\usepackage[shortlabels]{enumitem}

\makeatletter
\renewcommand\section{\@startsection {section}{1}{\z@}%
                                 {-3.5ex \@plus -1ex \@minus -.2ex}
                                   {2.3ex \@plus.2ex}%
                                   {\normalfont\large\bfseries}}
\renewcommand\subsection{\@startsection{subsection}{2}{\z@}%
                                   {-3.25ex\@plus -1ex \@minus -.2ex}%
                                     {1.5ex \@plus .2ex}%
                                     {\normalfont\bfseries}}
\renewcommand\subsubsection{\@startsection{subsubsection}{3}{\z@}%
                                   {-3.25ex\@plus -1ex \@minus -.2ex}%
                                     {1.5ex \@plus .2ex}%
                                     {\normalfont\itshape}}
\makeatother

\def\pplogo{\vbox{\kern-\headheight\kern -29pt
\halign{##&##\hfil\cr&{\ppnumber}\cr\rule{0pt}{2.5ex}&\ppdate\cr}}}
\makeatletter
\def\ps@firstpage{\ps@empty \def\@oddhead{\hss\pplogo}%
  \let\@evenhead\@oddhead 
}
\def\maketitle{\par
 \begingroup
 \def\thefootnote{\fnsymbol{footnote}}
 \def\@makefnmark{\hbox{$^{\@thefnmark}$\hss}}
 \if@twocolumn
 \twocolumn[\@maketitle]
 \else \newpage
 \global\@topnum\z@ \@maketitle \fi\thispagestyle{firstpage}\@thanks
 \endgroup
 \setcounter{footnote}{0}
 \let\maketitle\relax
 \let\@maketitle\relax
 \gdef\@thanks{}\gdef\@author{}\gdef\@title{}\let\thanks\relax}
\makeatother

\numberwithin{equation}{section}

\newcommand\eea{\end{eqnarray}}
\newcommand\bea{\begin{eqnarray}}

\def\beq{\begin{equation}}
\def\eeq{\end{equation}}

\newcommand{\be}{\begin{equation}}
\newcommand{\ee}{\end{equation}}
\newcommand{\ba}{\begin{align}}
\newcommand{\ea}{\end{align}}
\newcommand{\bg}{\begin{gather}}
\newcommand{\eg}{\end{gather}}
\newcommand{\bseq}{\begin{subequations}}
\newcommand{\eseq}{\end{subequations}}

\oddsidemargin=-.in

\begin{document}
\setcounter{page}0
\def\ppnumber{\vbox{\baselineskip14pt
}}
\def\ppdate{
} \date{}

\author{Horacio Casini\footnote{e-mail: casini@cab.cnea.gov.ar}, Marina Huerta\footnote{e-mail: marina.huerta@cab.cnea.gov.ar}, Javier M. Mag\'an\footnote{e-mail: javier.magan@cab.cnea.gov.ar}, Diego Pontello\footnote{e-mail: diego.pontello@ib.edu.ar} \\
[7mm] \\
{\normalsize \it Centro At\'omico Bariloche and CONICET}\\
{\normalsize \it S.C. de Bariloche, R\'io Negro, R8402AGP, Argentina}
}

\bigskip
\title{\bf  Entanglement entropy and superselection sectors I. Global symmetries 
\vskip 0.5cm}
\maketitle

\begin{abstract}
Some quantum field theories show, in a fundamental or an effective manner, an alternative between a loss of duality for algebras of operators corresponding to complementary regions, or a loss of additivity. In this latter case, the algebra contains some operator that is not generated locally, in the former, the entropies of complementary regions do not coincide. Typically, these features are related to the incompleteness of the operator content of the theory, or, in other words, to the existence of superselection sectors. We review some aspects of the mathematical literature on superselection sectors aiming attention to the physical picture and focusing on the consequences for entanglement entropy (EE). 
For purposes of clarity, the whole discussion is divided into two parts according to the superselection sectors classification: The present part I is devoted to superselection sectors arising from global symmetries, and the forthcoming part II will consider those arising from local symmetries.
Under this perspective, here restricted to global symmetries, we study in detail different cases such as models with finite and Lie group symmetry as well as with spontaneous symmetry breaking or excited states. We illustrate the general results with simple examples.
As an important application, we argue the features of holographic entanglement entropy correspond to a picture of a sub-theory with a large number of superselection sectors and suggest some ways in which this identification could be made more precise. 
\end{abstract}
\bigskip

\newpage

\tableofcontents

\vskip 1cm


\section{Introduction}

When the set of operators available in a model is not enough to create any given finite energy state from the vacuum, it is said that the theory contains superselection sectors (SS). Typically, this is the case when the algebra ${\cal O}$ does not contain charged operators. Then, charged states cannot be produced or destroyed by acting with  ${\cal O}$ and the full Hilbert space of the theory splits as a sum of different superselection sectors labeled by the charges.  In general, charged operators can be introduced such that they are able to create and destroy charges. As a consequence, this enlarged algebra of ``fields'' ${\cal F}$ can be thought as a more complete theory that does not have superselection sectors.\footnote{Traditionally the algebra ${\cal O}$ is thought to be the algebra of local physical observables, while the charged operators in ${\cal F}$ retain some locality properties but are not physically realizable in local laboratories, e. g. an operator that can change the baryonic number. In the theoretical setting of this paper we do not make this epistemological distinction.} 

If we are interested in studying the model ${\cal O}$ restricting our attention to the Hilbert space of neutral states that are created by acting with these operators on the vacuum (the vacuum sector of the theory), we may naively think we can dispense of with the structure of charged states that the model admits. However, it is the result of a large body of research into the superselection structure of quantum field theory (QFT) that the SS leave a definite imprint in the relations between the different local subalgebras of operators assigned to regions in the theory ${\cal O}$ itself. Indeed, the superselection structure, and the field algebra ${\cal F}$, can be fully reconstructed from the vacuum sector \cite{Doplicher:1990pn,Longo:1994xe}.  
 The physical reason is quite simple to understand. A state of non zero global charge is not locally distinguishable from a state of zero global charge since the charge can be placed very far away. For local algebras, we can place approximately localized charges in the region that are compensated by opposite charges far away. These local charges mimic the SS locally.  Hence the information of the SS must be accessible from the sector of zero total charge itself.  
 
The superselection structure affects the relations between algebras and regions in ${\cal O}$ either violating the property of duality (the algebra of the complement of a region $W$ consists of all operators that commute with the algebra of operators in $W$), or additivity (operators in $W$ are generated by operators in smaller balls inside $W$) for some topologically non trivial regions.   
The superselection structure also affects the vacuum fluctuations through charge-anticharge virtual pairs and hence it is visible in the entanglement entropy (EE). The main focus of this paper is the analysis of the consequences of superselection sectors for EE. 

Charges notoriously come in two types, corresponding to global or gauge symmetry charges. These correspond to two abstract types of superselection sectors, called DHR (because of Haag, Doplicher, Roberts \cite{Doplicher:1969tk,Doplicher:1969kp,Doplicher:1973at}) and BF sectors (because of Buchholz, Fredehagen \cite{Buchholz:1981fj}), respectively. The main difference between these two cases is geometric, global charges creating operators can be localized in a ball, while gauge charges creating operators can be localized in cones to allow the Wilson line to extend to infinity. 

For clarity purposes and taking into account the vast material we have collected and produced on the relevance of the SS in the EE, we have organized the complete analysis in two parts: this paper, Part I, (EE and SS I: Global symmetries) covers the DHR SS analysis and the BF type SS will be described in a future article, Part II (EE and SS II: Local symmetries).

 \begin{figure}[t]
\begin{center}  
\includegraphics[width=0.4\textwidth]{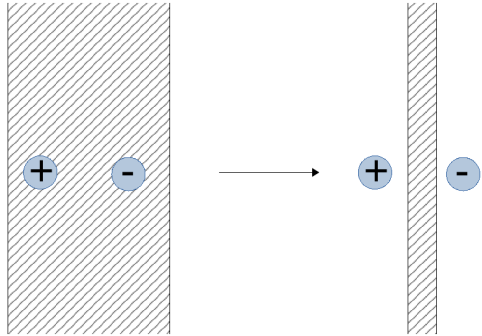}
\captionsetup{width=0.9\textwidth}
\caption{Mutual information between two regions separated by a strip of width $\epsilon$ (shaded region in the figure). For $\epsilon$ wide enough the typical charge anticharge fluctuations are not sensed by mutual informations of both models (left panel). When the width $\epsilon$ becomes small enough to allow for charge anticharge fluctuations to occur on each side of the wall with enough probability (right panel), the mutual information of ${\cal F}$ will take into account these correlations while the neutral model ${\cal O}$ will not. }
\label{fig6}
\end{center}  
\end{figure}

An essential feature of EE for general QFT is that it cannot be defined without the introduction of an ultraviolet regulator, making this quantity inherently ambiguous through the regularization scheme choice. This can be cured by computing (half) the mutual information between nearly complementary regions that are separated by a regulating distance $\epsilon$. This is a natural quantity taking the place of EE that is well defined in the continuum model and has been used in the literature as a regularized entanglement entropy \cite{Casini:2015woa,Casini:2008wt,Casini:2007dk}. 
On the other hand, there is also another source of ambiguities that has been discussed in the literature concerning the assignation of local algebras to regions \cite{Casini:2013rba}. In this sense, local algebras may contain a center related to ambiguities on the choice of algebra at the boundary of the region in a lattice model. This type of ambiguities has attracted attention especially in relation with gauge models (for the  discussion around this topic see for example \cite{Ghosh:2015iwa,Soni:2016ogt,Soni:2015yga,VanAcoleyen:2015ccp,Donnelly:2014fua,Donnelly:2011hn,Donnelly:2015hxa,Donnelly:2014gva,Camps:2018wjf}). However, this kind of local ambiguities does not survive the continuum limit, leaving the mutual information as a well-defined quantity \cite{Casini:2013rba}. 
In the new scenario we are presenting here, where models with SS sectors are considered, the analysis is enriched giving place to more interesting consequences. In models with SS, there is more than one choice for the {\sl macroscopic} algebra of regions that are topologically non-trivial. The possible choices affect mutual information. These mutual informations  however can be reinterpreted as corresponding to different models, with and without SS.   
 
 More concretely, mutual information crucially depends on the physical regulating distance $\epsilon$ that allows us to sense or not the presence of virtual charge pairs according to the comparison of the size of $\epsilon$ with the typical scale $\Lambda$ of these fluctuations. See figure \ref{fig6}. Hence, two possible results may come out in the limits $\epsilon/ \Lambda\gg 1$ or $\epsilon /\Lambda\ll 1$, {\sl independently} of the size $R$ of the region when $R$ is much larger than both $\epsilon$ and $\Lambda$. Then, in terms of the mutual information, it may seem we still have an apparent ambiguity. One of the main results that come from the analysis on SS is the clarification of this issue. Each result for the mutual information corresponds to a particular algebra choice where the SS have been included or not respectively. This means we should not interpret this as an ambiguity but as a consequence of alternative model choices.
 
With this perspective, partially following previous works in the mathematical literature \cite{Longo:2017mbg,Xu:2018fsv}, we develop entropic order parameters capable to sense these differences. More specifically, we study models with finite or Lie group symmetry, spontaneous symmetry breaking and also the consequences of considering charge excited states.

Another important application results in the use of these ideas for holographic theories. These theories have well-known oddities in the assignation of algebras and regions for the low energy sector of the theory. 
 There is also the curious fact that the EE that is non-local on the boundary theory is given by the localized contribution of an area in the bulk at leading order in the central charge \cite{Ryu:2006bv,Lewkowycz:2013nqa}. This localization has been explained either using the picture of bit threads \cite{Freedman:2016zud,Cui:2018dyq} or the idea of quantum error correction \cite{Almheiri:2014lwa,Pastawski:2015qua,Harlow:2016vwg}.     
We interpret both of these features of holographic theories as due to the presence of an effective large number of superselection sectors for the low lying modes. We think this interpretation may open the way to actual computations on how entanglement gets localized in the minimal surface area. We only briefly elaborate on this proposal in this paper and hope to come back to this important problem in the future. 

While throughout the paper we try to keep the discussion as simple and physical as possible, with a mixed degree of mathematical rigor, we are forced to use some specific mathematical tools to avoid making ambiguous statements.
We do not treat explicit examples where the superselection sectors do not come from a symmetry group. This includes models with DHR sectors in $d=2$ (and BF sectors in $d=3$ in part II). This would require more formal developments but would not add to the general physical picture. In the same spirit, the paper does not include a discussion of order parameters in terms of the algebraic index of inclusion of algebras instead of the entropy. The interested reader can consult the important papers \cite{Longo:1994xe,Longo:1989tt,Kawahigashi:1999jz,carpi2008structure} in this subject, and \cite{Longo:2017mbg,Xu:2018uxc} for a connection between the index and the relative entropy.       

The paper is structured as follows. In section \ref{algebra-regions} we describe the problems in the relations between algebras and regions in theories with superselection sectors and briefly introduce, mainly by concrete examples, some elements of the theory of superselection sectors. In section \ref{entropyDHR} we investigate the EE in the case of DHR sectors, describe the relevant order parameters and their mutual relationships, that take the form of entropic certainty and uncertainty relations. We explicitly compute the relevant quantities in several cases of interest. This includes the cases of finite and Lie symmetry groups, compactified scalars, regions with different topologies, charge excitations and thermal states. In section \ref{bounds} we study in concrete examples the behaviour of expectation values of some operators (intertwiners and twists) that are the main witnesses of the superselection sectors and play a major role in the evaluation of entropic quantities. In section \ref{holography} we describe how the holographic case matches the expectations for a theory with a large number of sectors and suggest some ways in which this understanding could be made more concrete. We end in section \ref{conclusion} with a summary and the conclusions.   

As we mentioned before, theories with BF  sectors (pure gauge fields) will be included in Part II,  where we will treat topological models and the case of the Maxwell field. We also leave for the next part the discussion about the relevance and implications of SS in the entropic RG flow analysis. In fact, theories with BF sectors are particularly interesting in this regard.
\section{Algebras, regions, and superselection sectors}
\label{algebra-regions}

We are interested in some particularities of the relation between algebras and regions that affect entanglement entropy. In the algebraic approach to quantum field theory (see for example \cite{Haag:1992hx,Witten:2018lha}), that is well suited for the analysis of these questions, the description is centered around the algebras of (bounded) operators ${\cal A}(W)$ corresponding to spacetime regions $W$.  A preliminary step is to understand some features of the relations between algebras and regions in generic QFT that in some cases depart from the naive expectations. In this section we are reviewing ideas that have been worked out in the mathematical literature about the relation between algebras and regions, that are tightly related to the theory of superselections sectors.

Let us spell in more detail what are these naive expectations. We follow the heuristic presentation in section III.4.2 of \cite{Haag:1992hx}. We define  $W'$ to the set of spacetime points spatially separated from the spacetime subset $W$. A causally complete region $W$ is one such that $W=(W')'$. For bounded regions, these have the form of a domain of dependence of bounded regions on a Cauchy surface. Because of causal evolution, it is expected that an operator in the causal domain of a region belongs to the algebra of that region, and hence different regions with the same causal domain of dependence share the same algebra. Therefore, causally complete regions select a unique region in this class and are the ones that are naturally expected to be associated with algebras in a one to one fashion. Then we will restrict attention to these causally complete regions and assume there is an algebra assigned to any such region.    

 These algebras are subject to some basic relations, for example, the operators inside $W_1$ should be also inside any other region including $W_1$,  
\be
{\cal A}(W_1)\subseteq {\cal A}(W_2)\,,\hspace{1cm}W_1\subseteq W_2\,. \label{isotonia}
\ee
In addition, since local operators commute at spacial distance because of causality, we have
\be
{\cal A}(W')\subseteq {\cal A}(W)'\,,\label{causality}
\ee
where the commutant ${\cal A}(W)'$ is the set of all operators that commute with those of ${\cal A}(W)$. 

An algebra is closed under products and linear combinations. In a specific Hilbert space representation minimal  requirement in QFT is that the operator algebras are von Neumann algebras, which satisfy ${\cal A}''={\cal A}$.\footnote{This is automatically true for finite dimensional algebras.}  The question naturally arises as to whether complementary causal regions $W$ and $W'$ can be consistently assigned commutant algebras in the vacuum representation
\be
 {\cal A}(W')= {\cal A}(W)'\,,   \label{duality}
\ee
that is an enhancement of relation (\ref{causality}). This property is called {\sl duality}. For a free scalar field, it has been shown duality holds for a large class of regions \cite{araki1964neumann}. For the case of topologically trivial regions such as a double cone (the domain of dependence of a ball) and in the vacuum state, this property is expected to hold under general conditions and is called {\sl Haag's duality} \cite{horuzhy2012introduction}  (see also \cite{Bisognano:1975ih,gabbiani1993operator}).    Looking at the relation (\ref{causality}) it may seem rather simple to complete a given net of algebras to have duality, just by enlarging the algebras taking ${\cal A}(W)\rightarrow {\cal A}(W')'$. However, in this process of completion, some problems might appear for topologically non-trivial regions. In particular, some other interesting properties that we usually assume about the algebras may be lost, impeding such an enlargement. 
     
One such expected property is that the operator algebra of a region $W$ could be generated by the algebras of operators of smaller regions included in $W$. This expresses that the algebra is locally generated, and it is the way in which one would form the algebra of a $W$ by taking arbitrary polynomials of smeared fields with support in arbitrary small regions inside $W$. If we have two algebras ${\cal A}_1$ and ${\cal A}_2$ the smallest von Neumann algebra containing the two is the generated algebra ${\cal A}_1\vee {\cal A}_2=({\cal A}_1\cup {\cal A}_2)''$. In the same fashion, given two causally complete regions $W_1$ and $W_2$, the smallest causally complete region containing the two is $W_1\vee W_2=(W_1 \cup W_2)''$.  Hence, this expected {\sl additivity}  property can be written  
\be
{\cal A}(W_1 \vee W_2)={\cal A}(W_1 )\vee{\cal A}( W_2)\,. \label{additivity}
\ee 
Heuristically, this is especially expected for $W_1$ and $W_2$ based on a common Cauchy surface. 

There is a dual way to define the additivity property. This is the idea that if two regions based on the same Cauchy surface have non-trivial intersection, the common elements of their algebras should be elements of the algebra of the intersection of the regions. To put this in a more formal way, recall that the largest algebra contained in two algebras  ${\cal A}_1$ and ${\cal A}_2$ is their set-theory intersection that we call ${\cal A}_1 \wedge {\cal A}_2\equiv {\cal A}_1 \cap {\cal A}_2$. This is again a von Neumann algebra. For regions, the intersection of two causally complete regions is also a causally complete region we can call $W_1\wedge W_2\equiv W_1 \cap W_2$.
Then another intuitive idea about the relation of algebras and regions is the following {\sl intersection property}
\be
{\cal A}(W_1 \wedge W_2)={\cal A}(W_1 )\wedge {\cal A}( W_2)\,. \label{intersection} 
\ee
 It is not difficult to see that we always have the so-called Morgan laws\footnote{This is the terminology of lattice theory. The relation of inclusion $\subseteq$ makes the set of von Neumann algebras and the set of causal regions two partially ordered sets. The existence of a supremum for any two elements, given by the operation $\vee$, and an infimum, given by the operation $\wedge$, make these ordered sets a lattice. The complement ${\cal A}'$ or $W'$ changes the sign of the order relation and satisfies the Morgan laws. If no local algebras have center (what can be expected for the vacuum representation) then ${\cal A}(W)\wedge {\cal A}(W)'=1$, and the complements make both lattices to be orthocomplemented lattices. Then, in its strongest form, the assignation of algebras to regions would be a homomorphism of orthocomplemented lattices. See \cite{Haag:1992hx,Casini:2002ah}. }
\bea
({\cal A}_1\vee {\cal A}_2)'={\cal A}_1'\wedge {\cal A}_2'\,,\\
(W_1\vee W_2)'=W_1' \wedge W_2' \,.
\eea
Then, using these properties,  the intersection property follows if we have unrestricted validity of duality (\ref{duality}) and additivity (\ref{additivity}), and conversely, additivity follows from unrestricted validity of duality and the intersection property. 

Summarizing, properties (\ref{isotonia}) and (\ref{causality}) are elementary axioms forming part of the basic idea of a QFT and will always hold. However, the properties (\ref{duality}), (\ref{additivity}) and (\ref{intersection}), are assumptions that will hold only for ``sufficiently complete'' models, and  fail for some physically significant models. We will see some simple examples below. 

It is important to realize that the failure of these properties does not (necessarily) have to do with the fact that QFT is a continuum theory with infinitely many degrees of freedom or the intricate nature of the specific von Neumann algebras (type III algebras). The problems we want to address are not related to UV problems, but already appear in lattice models.  

\subsection{DHR superselection sectors}
\label{DHR}

A representation of the algebra of operators arises by acting with the operators on a state. A theory can have different disjoint representations, for example, because the local operators are uncharged, and we have charged states. Then we cannot transform between states with different charges by acting with the local operators. The full Hilbert space is then decomposed into superselection sectors. It turns out that some of the expected properties in the relation between algebras and regions fail when there are superselection sectors in the theory.

Some particular superselection sectors are called DHR sectors because of the work of Doplicher, Haag and Roberts \cite{Doplicher:1969tk,Doplicher:1969kp,Doplicher:1973at} (for a review see 
\cite{Haag:1992hx,Halvorson:2006wj}). These are sectors where the charge can be localized inside a sphere, or more precisely, where the representation induced by a charged state on the algebra of operators outside the sphere cannot be distinguished from the vacuum representation. In this case, the charge has to be a global charge, not associated to a gauge symmetry, since gauge charges can be measured by the electric flux through a shell of large radius around the charge, and hence the representation cannot be the same as the vacuum representation outside a sphere.

In order to introduce the main ideas, we find useful to start with a specific simple model where all the involved objects can be displayed explicitly. Once the main ideas are understood we will discuss the general case. Let us then focus on the following model. We take a free Dirac field and consider the subalgebra consisting of all operators with even fermionic number, the bosonic part of the fermion algebra. This is the algebra generated by an even number of fermion fields, $1, \psi(x)\psi(y), \psi^\dagger(x)\psi^\dagger(y), \psi(x) \psi^\dagger (y)$, etc., where all fields have to be smeared with test functions, and we can take arbitrary polynomials with even fermionic number. Call ${\cal O}$ to this algebra and for a bounded region $W$ let us call ${\cal O}_{W}$ to the additive subalgebra generated by the fields contained in balls in $W$. Analogously, we have the global fermion algebra ${\cal F}$ and the one restricted to a region ${\cal F}_{W}$. The fermion net is not local in the sense that spatially separated operators do not commute, but a graded locality can be defined to accommodate it in the algebraic version of QFT \cite{carpi2008structure}. The bosonic net is a local subnet of the fermion one. The vacuum representation of ${\cal O}$ is generated by acting with even operators on the vacuum $|0\rangle$. 

Let us consider the fermionic operator in ${\cal F}_{W}$
\be
V_W=\int d^{d-1}x\, \alpha(x)\, (\psi(x)+\psi^\dagger(x))\,,\label{siste}
\ee   
where $\alpha(x)$ is a real spinor  supported in a ball $W$. In addition we choose $\int d^{d-1}x\, \alpha(x)^2=1$. We are smearing in space only.\footnote{This can be done for a free field. We could have also selected space-time smearing functions as well, at the expense of replacing $\int d^{d-1}x\, \alpha(x)^2=1$ by an integral in $d$ dimensions weighted with the anticommutator distribution.} 
Using the anticommutation relations $\{\psi(x),\psi^\dagger(y)\}=\delta(x-y)$, we have
\be
V_W^\dagger=V_W\,,\hspace{1cm}V_{W}^2=1\,.
\ee
Hence this is a unitary operator in the fermion theory. It also has fermion number $1$. The state $|\psi\rangle=V_{W}^\dagger|0\rangle$ has also fermion number $1$. We can borrow the Hilbert space of the fermionic theory to produce a representation of ${\cal O}$ by acting with operators of ${\cal O}$ on the state $|\psi\rangle$. This will be a different representation than the vacuum one, in particular, there is no normalizable vacuum state in the representation, and if the fermion is massive, the minimum energy in the representation is the mass of the fermion \footnote{Notice that in this case, the representations induced by  $|0\rangle$ and $|\psi\rangle=V_{W}^\dagger|0\rangle$ respectively exahust all the inequivalent representations.}.

We are not really interested in the new representation, but on the vacuum one. However, there are consequences of this superselection structure that are visible already in the vacuum representation. To convince oneself, first notice that the expectation values of operators $b \in {\cal O}$ in the new representation 
are just vacuum expectation values of a transformed operator $\rho(b)\in{\cal O}$
\bea
\langle \psi|b|\psi\rangle &=& \langle 0|\rho(b)|0\rangle\,, \hspace{1cm} b\in {\cal O}\,,\\
\rho(b) &=& V_{W} \, b \,V_{W}^\dagger\,, \label{endo}\\
\rho({\cal O}) &\subseteq & {\cal O}\,.
\eea
In this way the new representation is written as the composition of the vacuum representation with an endomorphism $\rho$ of the algebra ${\cal O}$ in itself (in this particular case it is an automorphism since $V_W$ is invertible).

The beauty and power of the DHR analysis reside in translating the problem of computing different representations to the one of finding endomorphisms of ${\cal O}$. Endomorphisms can then be composed, which is equivalent to the addition of charges, or the tensor product of symmetry group representations.

 In the present example, the fact that the fermionic charge is localized in a ball means that the endomorphism leaves the elements of the complement of the ball unaffected,  
 \be
 \rho(b)=b\,, \hspace{1cm} b\in {\cal O}_{W_1}\,, \,\, W_1\subset W' \,,\label{defi}
\ee
 where $W_1$ is any ball included in the complement of $W$.
 By Haag duality for the sphere, it carries the algebra of the ball in itself
\be
\rho({\cal O}_W)\subseteq {\cal O}_W\,.
\ee

\begin{figure}[t]
\begin{center}  
\includegraphics[width=0.55\textwidth]{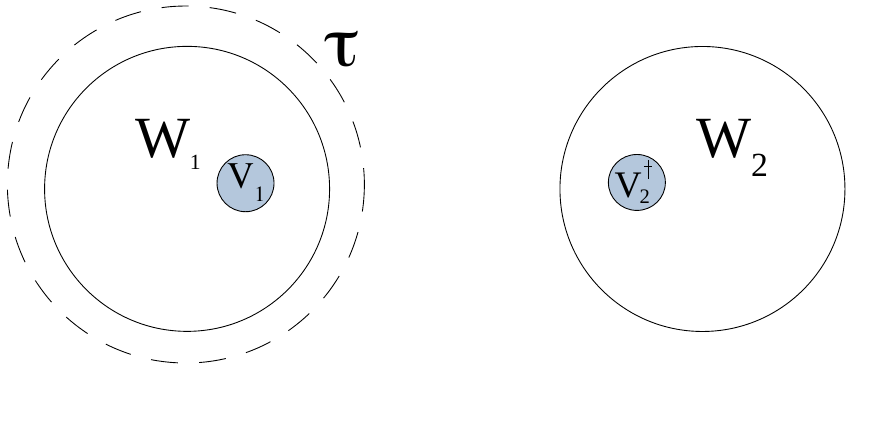}
\captionsetup{width=0.9\textwidth}
\caption{The intertwiner $V_1 V_2^\dagger$ commutes with the additive algebra of the exterior of the two balls $W_1$ and $W_2$. It cannot be formed additively with the algebras of these balls. The twist $\tau$ cannot be formed additively on the exterior of the two balls but commmutes with the algebras of the balls. The intertwiner and the twist do not commute with each other. }
\label{f3}
\end{center}   
\end{figure}

The endomorphism $\rho$ maps ${\cal O}$ in itself and can be analyzed by studying  ${\cal O}$ without the knowledge of the bigger algebra ${\cal F}$.  
 The endomorphisms are not given by operators in ${\cal O}$, but we would like to understand these endomorphisms in terms of certain limit of operators in ${\cal O}$. With this aim consider now two disjoint balls $W_1$, $W_2$, and endomorphisms $\rho_1$ and $\rho_2$ localized in  $W_1$, $W_2$, as in (\ref{siste}). Since these endomorphisms correspond to representations one can ask about operators that intertwine these representations, that is
\be
{\cal I}_{W_1, W_2}\,  \rho_1(b)=\rho_2(b)\, {\cal I}_{W_1, W_2}\,.  
\ee 
 For unitary equivalent representations ${\cal I}_{W_1, W_2}$ is just a unitary operator.   
 This operator translates the charge from one position to the other, but it is not a translation since it leaves intact all operators that are localized outside $W_1$ and $W_2$. In the present example (see figure \ref{f3})
\be
{\cal I}_{W_1, W_2}=V_{W_2} V_{W_1}^\dagger\,. \label{isit}
\ee 
Note that this is a bosonic operator; it belongs to ${\cal O}$. When there is a unitary intertwiner in $\cal O$ for an endomorphism localized in a ball to an endomorphism localized in any other ball, the endomorphism is called transportable, and the DHR analysis deals with transportable endomorphisms.  
In fact, the intertwiner (\ref{isit}) belongs to the algebra of a ball $W_{3}$ containing both $W_1$ and $W_2$, and this is always the case by Haag duality for balls, since ${\cal I}_{W_1W_2}$ commutes with operators localized in balls in the complement of  $W_{3}$.  

One importance of the intertwiner is that it can be used to produce the endomorphism creating the charge as a limit of elements in ${\cal O}$. Essentially, the intertwiner changes the position of the charge, and then one can bring a charge from infinity as
\be
\rho_1(b)= V_{W_1} \, b \,V_{W_1}^\dagger=\lim_{a\rightarrow \infty}{\cal I}_{W_1, W_2+a}^\dagger\, b\,  {\cal I}_{W_1, W_2+a}\,.
\ee

From our point of view, the existence of the intertwiner is important because it shows that duality does not hold for the topologically non-trivial region formed by the union of two disjoint spheres. This is because ${\cal I}_{W_1, W_2}$ is an operator of the algebra ${\cal O}$, it belongs to the commutant $({\cal O}_{(W_1 \cup W_2)'})'$ of the additive algebra ${\cal O}_{(W_1 \cup W_2)'}$ of the complement of the union of the two balls because it leaves intact operators on balls localized in the complement of the two spheres. See figure \ref{f3}. However, it does not belong to the additive algebra of the two spheres. Otherwise, the endomorphisms would have been trivially produced by unitary operators in ${\cal O}(W_1)$, ${\cal O}(W_2)$, and would not be related to non-trivial superselection sectors. 

Hence, the corollary is: for any theory with a non-trivial DHR superselection sector (a charge that can be localized in a ball) duality fails for two balls, due to the intertwiners. This is a consequence of global superselection sectors in the algebraic structure of the net already in the vacuum sector.  

Let us see with a bit more detail how this happens in the present example. The intertwiner belongs to  $({\cal O}_{(W_1 \cup W_2)'})'$, or  $\int dx\,dy\, \alpha_1(x) \alpha_2(y)\,(\psi(x)+\psi^\dagger(x))  \,(\psi(y)+\psi^\dagger(y)) $ belongs to $({\cal O}_{(W_1 \cup W_2)'})'$ where $\alpha_1$ and $\alpha_2$ are localized in $W_1$ and $W_2$. But what impedes this operator to be in the algebra of the two balls? That is, how do we know that this operator will not appear in the double commutant of the algebra ${\cal O}_{W_1}\cup {\cal O}_{W_2}$? The reason is that associated to the intertwiner     
 linking $W_1$ and $W_2$ there is ``dual'' operator that does not commute with it, and belongs to the commutant of  ${\cal O}_{W_1}\cup {\cal O}_{W_2}$. This is a {\sl twist} operator that essentially senses the total fermionic charge on $W_1$ (see figure \ref{f3}). Placing the spheres $W_1,W_2$ at time $t=0$, a choice for the twist can be written as
 \be
 \tau=e^{i\pi \int dt\, d^{d-1}x\, \gamma(t)\,\beta(\vec{x}) J^0(x)}\,, \label{eje}
 \ee 
 with a smearing function   $\gamma(t)\,\beta(\vec{x})$ with support on  $W_2'$ (in particular it vanishes on $W_2$), has $\int dt\, \gamma(t)=1$, and $\beta(\vec{x})=1$ for all $\vec{x}$ on $W_1$. $J^0(x)$ is the charge density operator $:\psi^\dagger(x)\psi(x):$.
 This unitary operator $\tau$ belongs to the global algebra ${\cal O}$,   trivially commutes with ${\cal O}_{W_2}$, and anti-commutes with fermion operators in $W_1$. Indeed the anticommutation relation is really due to the fact that $\tau$ implements the group operation in $W_1$, namely $\tau \psi \tau^{-1}=-\psi$. Therefore, it is clear it commutes with ${\cal O}_{W_1}$ too. However, it does not belong to the additive algebra ${\cal O}_{(W_1 \cup W_2)'}$ of the complement of the two balls. The twist operator does not commute with the intertwiner
 \be
 [\tau, {\cal I}_{W_1 W_2}]\neq 0\,.\label{conn}
\ee 
 Therefore if we start with the operators in the two spheres ${\cal O}_{W_1}\cup {\cal O}_{W_2}$ the commutant will have the twist operator, and the double commutant will not have the intertwiner. Conversely, if we start with the algebra of the two spheres plus the intertwiner, the commutant will not contain the twist. In other words, writing as ${\cal O}_W$ the additive algebra for any region $W$ we should have \footnote{Notice that these relations are not tensor products.}
\bea
{\cal O}_{W_1\cup W_2}' &=& {\cal O}_{(W_1\cup W_2)'}\vee \tau\,,\\
{\cal O}_{(W_1\cup W_2)'}' &=& {\cal O}_{W_1\cup W_2}\vee {\cal I}_{W_1 W_2}\,.
\eea 
Therefore, duality for the two spheres or its complement would require that  
 we enlarge the additive algebras of the complementary regions with the twist operator or the intertwiner respectively losing additivity. We cannot have both properties together when there are DHR superselection sectors for regions with these non-trivial topologies.
   
Notice that we could change the definition of the twist by choosing different smearing functions satisfying the stated requisites. But these different twist operators differ by elements of the additive algebra of the complement of the two balls. Therefore they give place to the same algebra ${\cal O}_{(W_1\cup W_2)'}\vee \tau$.  In the same line, another twist operator can be defined that crosses through $W_2$ instead of $W_1$. However, this combined with $\tau$ and operators in the algebra of $(W_1\cup W_2)'$ is the operator $e^{i \pi \int d^{d-1}x \, J^0}$, where the exponent is integrated on the full space. This just commutes with all operators of ${\cal O}$ and is the identity in the vacuum sector. Analogously, the precise smearing function in the definition of the intertwiner is irrelevant for the purpose of generating the algebra.  
 
The intersection property also conflicts with additivity since we can take two regions $A_1$, $A_2$, which are topologically spheres but whose intersection is the union of two spheres, $A_1\cap A_2 =W_1\cup W_2$, where $W_1$ and $W_2$ are two disjoint spheres. The intertwiner then belongs to the algebras of each of the two regions $W_1$, $W_2$ because it is an even element locally generated inside these regions. Therefore it will belong to the intersection of the algebras. If we accept the intersection property the intertwiner will belong to the algebra of the two spheres, and this algebra will not be additive. 

The existence of the twist explains why in (\ref{defi}) we cannot just put $\rho(b)=b$ for all $b\in {\cal O}_{W'}$. The algebra ${\cal O}_{W'}$ is defined to be ${\cal O}_{W'}=({\cal O}_W)'$ in the vacuum sector giving  Haag's duality for spheres. It is the completion (double commutant) of $\cup_{W_1\subseteq W'} {\cal O}_{W_1}$ over all balls $W_1$, and this includes the twist operator ``crossing'' through $W$. This is the same operator that results integrating the current up to spatial infinity and does not belong to the union of any finite number of balls. It appears only as a result of the double commutant operation.  

\subsection{The general DHR case}
\label{sec:esa}

For a general interacting theory with DHR (ball localized) superselection sectors it would be difficult to write in explicit way the endomorphisms and intertwiners.\footnote{Another example is a unitary charge creating operator for a $U(1)$ symmetry of a free charged scalar field given by $\phi_f/(\phi_f \phi_f^\dagger)^{1/2}$ where $\phi_f$ is a smeared scalar mode \cite{bogolubov1975introduction}. This, in contrast to the example of the previous section, is more difficult to handle to evaluate expectation values.}   However, general arguments show that the story is analogous to the example described above. See for example \cite{Haag:1992hx,Halvorson:2006wj,araki1999mathematical}. The theory proceeds from the endomorphisms generated by the superselection sectors, to the spin-statistics theorem, and the construction of a bigger field algebra containing the operators that create charged states.   

The final result is that a theory with DHR superselection sectors can always be thought as the charge neutral sector of a theory without superselection sectors. This later is called the field algebra ${\cal F}$. This has a global compact symmetry group $G$ (for $d>2$), and the observable algebra    
 ${\cal O}\subset {\cal F}$ is the orbifold under the action of $G$, that is, it is formed by the operators invariant under the actions of group elements $g\in G$.  The vacuum Hilbert space ${\cal H}_{\cal F}$ of the theory ${\cal F}$ reduces under the action of ${\cal O}$ into a direct sum over the Hilbert spaces of the superselection sectors,
\be
 {\cal H}_{\cal F}= \oplus_{r,i}  {\cal H}_{r,i}\,,\label{dec}
\ee  
where $r$ are different irreducible representations under $G$, and $i=1,\cdots,d_i$ is an index that span the dimension of the representation. One of these sectors is the vacuum sector of ${\cal O}$ that we call ${\cal H}_0$. 
The label $r$ associated to the group representation $r$ appears $d_r$ times in the decomposition (\ref{dec}) since the elements of ${\cal O}$ cannot move between the different base elements of the representation, say, isospin projection. 

A state in a charged sector associated to a generic (not necessarily irreducible) representation $r$  localized in a sphere $W$ can be written by using operators of ${\cal F}$ in $W$ acting on the vacuum. The operators will be labeled by an index corresponding to the representation dimension
\be 
V_{r}^{i\, \dagger} |0\rangle\,.
\ee 
It is possible to choose these operators such that they transform in a (unitary) representation of the symmetry group 
\be
U(g)^{\dagger} \,V_i^r\, U(g)=D_r(g)_{ij} V_j^r\,, \label{beca}
\ee
where $D_r(g)_{ij}$ is the representation matrix. These operators do not only help to construct a complete basis for the Hilbert space of $\mathcal{F}$ but indeed any element $a$ of the  algebra ${\cal F}$ can be written as
\be
a= \sum_{r,i} b_{r,i}\, V_r^i \,, \label{esis}
\ee
with $b_{r,i}\in {\cal O}$. It is important to remember that even if it is useful to think in terms of the field algebra and the algebra ${\cal O}$ as its subalgebra, the field algebra does not contain new information that is not present in ${\cal O}$ itself. This will be clearer as we develop the necessary mathematical and physical tools.

To obtain an endomorphism of  ${\cal O}$ associated with the representation $r$  define 
\be
\rho_r (b)= \sum_i V_{r}^i\, b\,  (V_{r}^i)^\dagger\,,
\ee
such that $\rho_r(b)\in {\cal O}$ because of (\ref{beca}). In order that this is an endomorphism and respect the product of operators, and that it maps the identity in itself, we need additionally\footnote{In particular $\rho$  is a completely positive mapping to the image.}
\bea
(V_{r}^i)^\dagger V_{r}^j &=& \delta_{ij}\,,\label{s}\\
\sum_i V_r^i (V_r^i)^\dagger &=& 1\,,\label{t}
\eea
such that the $V_r^i$ are partial isometries. For the case of one dimensional representations $V_r$ is unitary, as in the example of the previous section. 

This endomorphism arises when considering the state 
\be
|\psi_i\rangle=\sqrt{d_r} V_r^{i\, \dagger}| 0 \rangle\,, \label{dietrich}
\ee
corresponding to the irreducible representation $r$. The factor $\sqrt{d_r}$ 
 is necessary to have a properly normalized state since
 \bea
 \langle 0 | V_r^{i\,} V_r ^{i\, \dagger}| 0 \rangle &=& \frac{1}{|G|} \sum_g \langle 0 | U(g) V_r^{i\,}U(g)^\dagger U(g) V_r ^{i\, \dagger} U(g)^\dagger| 0 \rangle \nonumber \\
&=& \frac{1}{|G|} \sum_g D_r(g)_{ij}D_r(g)_{il}^* \langle 0 | V_r^{j\,} V_r ^{l\, \dagger} | 0 \rangle= \frac{1}{d_r} \langle 0 | \sum_j V_r^{j} V_r ^{j\, \dagger} | 0 \rangle=  \frac{1}{d_r} \,,
\eea
where we have used  the orthogonality relation for irreducible representations,    
\be
\sum_{g\in G} D^{r_1}_{l_1l_2}(g)D^{r_2\,*}_{l_3l_4}(g)= \frac{|G|}{d_{r_1}} \delta_{r_1 r_2}\delta_{l_1 l_3}\delta_{l_2 l_4}\,.
\ee
The same type of algebraic manipulations show that for any element $b$ of ${\cal O}$
\be
\langle \psi_i | b|\psi_i \rangle= \langle 0| \rho(b)|0\rangle=\omega\circ\rho(b)\,,
\ee
where $\omega$ is the vacuum state and  
\be
\rho(b)= \sum_j V_r^{j} b V_r^{j\,\dagger}
\ee
the corresponding endomorphism. Notice $|\psi_i\rangle$ is pure in ${\cal F}$ but not in ${\cal O}$.

The operators $V_r^i$ with the relations (\ref{s}) and (\ref{t}), generate what is called a Cuntz algebra. This cannot be represented in finite dimensions. However, the algebra of operators of the form \cite{Longo:1989tt}
\be 
(a)=\sum_{ij} a_{ij} V_r^i (V_r^j)^\dagger
\ee
closes with a matrix multiplication for the coefficients,  
\be
(a) (b)=(a\cdot b)\,.
\ee
Hence it is a finite subalgebra of the Cuntz algebra of matrices of $d_r\times d_r$. 

 If $V_1^i$ and $V_2^i$ are the charge generating operators localized in two disjoint balls $W_1$ and $W_2$ associated to the same representation (up to unitary transformations in ${\cal O}$), an  intertwiner between the two is
\be
{\cal I}_{W_1W_2}=\sum_i V_1^i  (V_2^i)^\dagger\,.
\ee
It follows from (\ref{s}) and (\ref{t}) that ${\cal I}_{W_1W_2}$ is unitary. 
This belongs to ${\cal O}$ because it is invariant under the symmetry group, commutes will all operators localized outside the two spheres but is not generated by ${\cal O}_{W_1}$ and ${\cal O}_{W_2}$. Therefore duality for the two spheres does not hold in ${\cal O}$.  

The twist operators appear in the commutant of the algebra of the two spheres and are labeled by elements of the group, $\tau_g$.  These commute with the algebra ${\cal O}_{W_1}$ and ${\cal O}_{W_2}$ but do not commute with ${\cal F}_{W_1}$ (or, alternatively, with ${\cal F}_{W_2}$). We can think of them as the implementation of the group symmetry by unitary operators localized in a region of the space bigger than $W_1$ and non intersecting with $W_2$. Their action on each operator localized in $W_1$ coincides with the action of the global symmetry. In particular, on the charge generating operators $V_i$ in $W_1$, we have 
\be
\tau_g^{\dagger} \,V^r_i\, \tau_g=R_r(g)_{ij} V^r_j\,.\label{cinco}
\ee

Twist operators can be chosen such that they satisfy the group operation \cite{Doplicher:1984zz}\footnote{This is not the case of the simple twist operator (\ref{eje}). To construct twists with these special properties one needs to use the split property that allows to include the two algebras  $W_1$ and $W_2$ in each of the two type I factors of a tensor product decomposition of the full operator algebra. See \cite{Doplicher:1984zz,araki1999mathematical}. }
\be
\tau_g \tau_{h}=\tau_{g h}\,, 
\ee
and are transformed covariantly under the group action
\be
U(g) \tau_{h} U(g)^\dagger= \tau_{g h\,g^{-1}}\,. \label{seis}
\ee

The twist operators are not elements of ${\cal O}$ in general, i.e. for non Abelian groups. We can form (generally non-unitary)     
 elements of ${\cal O}$ by taking linear combinations $\tau_c=\sum_g c_g \tau_g$ and demanding   
 \be
 U(g) \tau_c U(g)^\dagger= \tau_c\,.
\ee
Then the invariant twist elements are naturally associated to the center of the group algebra $C_G$ where the coefficients $c_g=c_{h g h^{-1}}$ are invariant under conjugation. The dimension of this center $n_C$ is equal to the number of irreducible representations or the number of different conjugacy classes.
The group algebra is equivalent to a direct sum of matrix algebras where the group is represented with matrices $\oplus_r D^r(g)$. The center of the group algebra is then clearly spanned by all diagonal matrices which are linear combinations of the projectors on each irreducible representation. These projectors are precisely\footnote{That these operators are projectors follows from the convolution property of the characters
\be
\sum_g \chi_{r_1}(g) \chi_{r_2}(h g^{-1})=\frac{|G|}{d_{r_1}}\delta_{r_1 r_2}\,\chi_{r_1}(h)\nonumber\,.
\ee}  
\be
P_r=\frac{d_r}{|G|} \sum_g \chi_r^*(g) \tau_g \,,\label{projec}
\ee
where $\chi_r(g)$ is the character of the irreducible representation $r$.
Then the invariant twists are written 
\be
\tau_c=\sum_r c_r P_r\,. \label{center}
\ee

For $d=2$ there is a difference with respect to higher dimensions. We can divide the compactified line into four intervals. Let $I_1$, $I_2$ be two disjoint intervals and $I_3$, $I_4$ the two disjoint intervals forming the complement of $I_1\cup I_2$. The intertwiner between $I_1$ and $I_2$ belongs to 
\be
{\cal O}_{(I_3\cup I_4)}'\,, 
\ee
but the twist operator crossing $I_3$ (or $I_4$) also belongs to this algebra. Therefore, the number of additional elements in the algebra of ${\cal O}_{(I_3\cup I_4)}'$ with respect to the additive one ${\cal O}_{I_1\cup I_2}$ is larger. The difference with respect to higher dimensions is because the topology of the two intervals and its complement is the same in $d=2$. In higher dimensions the intertwiner between two spheres $W_3$, $W_4$, placed inside $(W_1\cup W_2)'$ is trivially included in the additive algebra of $(W_1\cup W_2)'$ because $W_3,W_4$ can be deformed to coinciding position without crossing $W_1,W_2$.  In addition, in $d=2$ the DHR sectors do not necessarily come from a group symmetry, and we have a more general theory of sectors determined by their fusion rules under composition that replace the decomposition of tensor product of group representations as a sum of irreducible representations. The reasons for this difference with higher dimensions are related to the complications that appear when analyzing the spin statistic theorem since charged operators cannot smoothly interchange its positions without crossing each other in $d=2$. See for example \cite{Halvorson:2006wj}.  

As a final remark, notice that lattice models with global symmetries are easily constructed. We have a Hilbert space ${\cal H}_a$ for each vertex $a$ of the lattice on which there is a faithful representation of the group $G$. The global Hilbert space is the tensor product ${\cal H}=\otimes_a {\cal H}_a$ and the group acts with the tensor product representation. The algebra ${\cal F}$ is the full algebra of operators in ${\cal H}$ and ${\cal O}$ the subalgebra of invariant operators.  
\section{Entropy and DHR sectors}
\label{entropyDHR}

We are interested in the mutual information between the two topologically trivial regions $W_1$ and $W_2$. It can be written (with a cutoff in place) as 
\be
I(W_1,W_2)=S(W_1)+S(W_2)-S(W_1 W_2)\,. \label{asr}
\ee
 We have seen that in the model ${\cal O}$ we can have two different algebras for $W_1 W_2$, one with and one without the intertwiners.
 The algebra ${\cal O}(W_1 W_2)$ without the intertwiners is additive and hence is the appropriate one to produce the mutual information $I_{\cal O}(W_1,W_2)$ in ${\cal O}$. However, it is of obvious interest to look for an information theoretic quantity that senses the contributions of the intertwiners in the algebra of the union. To start with, the simplest thing to do is to focus on the mutual information corresponding to the field algebra $I_{\cal F}(W_1,W_2)$. The algebra of ${\cal F}(W_1W_2)$ naturally contains the intertwiners while retaining additivity in ${\cal F}$. 
 
 Hence, as an order parameter indicative of the presence of DHR SS in ${\cal O}$ we can compute
\be
I_{{\cal F}}(1,2)- I_{{\cal O}}(1,2)\,.\label{sity}
\ee 
This is always positive by monotonicity. We emphasize that even if $I_{{\cal F}}(1,2)$ and $I_{{\cal F}}(1,2)- I_{{\cal O}}(1,2)$ look like quantities that depend on $\mathcal{F}$, they are in fact properties of ${\cal O}$ itself (for invariant states), from which ${\cal F}$ can be reconstructed. The ultimate physical reason is that the only non zero vacuum expectation values in ${\cal F}$ are equal to expectation values in ${\cal O}$. In fact, we will later show in detail how both quantities are directly written in terms of the model  ${\cal O}$.

To put this in a firmer ground we will follow some ideas presented in \cite{Longo:2017mbg}. We first need to review some quantum information tools that will also be useful in the rest of the paper. This is done in the next section \ref{tools}. Next, in section \ref{lower}, we describe the order parameter in terms of a relative entropy that is determined by the intertwiner expectation values. This allows us to put useful lower bounds. The description of the order parameter in terms of twist expectation values is done in section \ref{upper}. 
 This gives us a tool for computing upper bounds on $\Delta I$. We show that the difference of mutual informations saturate to $\log |G|$, where $|G|$ is the number of elements in the symmetry group, for any finite group, in the limit when the two regions touch each other. Twist and intertwiners do not commute and satisfy entropic certainty and uncertainty relations. This is described in section \ref{certainty}. After that we make different computations using the main ideas developed in sections \ref{lower} and \ref{upper}, such as treat the case of Lie group symmetries in section \ref{U1}, the case of regions with different topologies in section \ref{OT}, states with excitations of non-Abelian sectors in section \ref{EX}. We treat the case of spontaneous symmetry breaking in section \ref{SSB}, the thermofield double state in section \ref{thermofield}. The procedure for computing the entropies of the symmetric model ${\cal O}$ with the replica trick is reviewed in section \ref{RE}, and, finally, in section \ref{dosd} we make some remarks on the special case of $d=2$.

\subsection{Some quantum information tools}
\label{tools}

The material of this section can be found for example in \cite{petz2007quantum,ohya2004quantum}.

Given two algebras ${\cal O}\subset {\cal F}$ a conditional expectation $E$ from ${\cal F}$ to ${\cal O}$ is a linear map that carries positive elements to positive elements (self-adjoint operators with positive spectrum or elements of the form $a a^\dagger$) and
\bea
E(1) &=& 1\,\\
E(b_1\, a \,b_2) &=& b_1\, E(a)\, b_2\,,\hspace{1cm}b_1,b_2 \in {\cal O}\,, \, a\in {\cal F}\,. 
\eea
In particular $E$ leaves ${\cal O}$ invariant.\footnote{Any  positive unital map to a subalgebra which is the identity on the image is automatically a conditional expectation and completely positive.} 

A simple example is the partial trace when  ${\cal O}$ is a tensor factor in ${\cal F}$. For the present case, the natural conditional expectation from the field algebra ${\cal F}$ to the invariant one ${\cal O}$ is 
\be
E(a)=\int_G dg\, g \, a\, g^\dagger\,, \label{cee}
\ee
using the normalized measure for a compact group $G$ acting unitarily in ${\cal F}$. This is replaced by $|G|^{-1}\sum_g$ for a finite group. 
All the properties stated above about conditional expectations are easily seen to hold for $E$. Essentially $E$ takes the part of an element that is invariant under the group. For the case of the even part of the fermion algebra, a general element is of the form $a= a_0 + a_1\, \psi$ with $a_0, a_1$ even operators and $\psi$ any smeared fermion field. Then $E(a)=a_0$. 

 Now suppose we have a state $\omega$ in the algebra ${\cal F}$. It generates a state $\omega_{\cal O}$ in ${\cal O}$ just by evaluating expectation values in ${\cal O}$. If we have a state $\phi$ in ${\cal O}$ we can form a state in ${\cal F}$ by using the conditional expectation as $\phi\circ E$. Given a state $\omega$ in ${\cal F}$ we can form an invariant state in ${\cal F}$ by $\tilde{\omega}=\omega\circ E$.  
 Thus, we have the following important property of relative entropy (conditional expectation property)
\be
S_{\cal F}(\omega|\phi\circ E)-S_{\cal O}(\omega_{\cal O}|\phi)=S_{\cal F}(\omega|\omega_{\cal O}\circ E)\,. \label{dfh}
\ee
The difference in the relative entropies on the left-hand side is clearly positive because of monotonicity since the two states $\omega_{\cal O}, \phi$ in ${\cal O}$ are the restrictions from the two states $\omega,\phi\circ E$ in ${\cal F}$. What is interesting is that this positive difference can itself be expressed in terms of a relative entropy which in addition does not depend on $\phi$.  
A usefull property that follows from this relation is that given two states invariant under the conditional expectation, $\omega_1=\omega_1\circ E$, $\omega_2=\omega_2\circ E$,  we have $S_{\cal F}(\omega_1|\omega_2)=S_{\cal O}(\omega_1|\omega_2)$, or, more simply, for any two states 
\be
S_{\cal F}(\omega_1\circ E|\omega_2\circ E)=S_{\cal O}(\omega_1|\omega_2)\,.\label{igualitos}
\ee
 That is, the  distinguishability of two invariant states under $E$ is not improved in the the bigger algebra.

Another useful property that we use for an algebra ${\cal A}={\cal A}_1\otimes {\cal A}_2$ is 
\be
S(\omega|\phi_1\otimes \phi_2)=S(\omega_1|\phi_1)+S(\omega_2|\phi_2)+S(\omega|\omega_1\otimes \omega_2)\,.\label{another}
\ee

For matrix algebras, we can use expressions in terms of density matrices.  One interest in looking at finite dimensional algebras is the following. One may entertain the idea that even if the entropies have ambiguities in continuum limit in QFT, the particular difference $S_{\cal F}(W)-S_{\cal O}(W)$ of the complete and the neutral models in the same ball and for the vacuum could be well defined in the continuum limit, independently of the chosen lattice and exact definition of the algebras. In such a case we could use just this difference as an order parameter. With a focus in investigating this question, we collect some formulas for matrix algebras that will be useful along the paper.

First we have a general property valid for any state $\phi$ and conditional expectation $E:{\cal A}\rightarrow {\cal B}$ that preserves the trace $\textrm{tr}_{\cal A}(x)=\textrm{tr}_{\cal A} (E(x))$ in matrix algebras.\footnote{Given a subalgebra of a given algebra there is a unique conditional expectation mapping the two that preserves the trace in this sense.} This is not the general case for conditional expectations but the conditional expectation (\ref{cee}) preserves the trace in any subalgebra since it is an average over automorphisms. In this case we have
\be
S_{\cal A}(\phi|\phi\circ E)=S_{\cal A}(\phi\circ E)-S_{\cal A}(\phi)\,.\label{estasi}
\ee 
 To show this we write the relative entropy as $\Delta \langle H\rangle-\Delta S$, where $\Delta S$ is the difference between the entropies of the two states and $\Delta \langle H\rangle$  is the difference in expectation values of $H=-\log(\rho_{\phi\circ E})$, where $\rho_{\phi\circ E}$ is the density matrix corresponding to the state $\phi\circ E$. Then $\textrm{tr}_{\cal A}(\rho_{\phi\circ E} \,a)=\textrm{tr}_{\cal A}(\rho_\phi E(a))=\textrm{tr}_{\cal A}(E(\rho_\phi) E(a))= \textrm{tr}_{\cal A}(E(\rho_{\phi}) a)$  
 and we get $\rho_{\phi\circ E}=E(\rho_\phi)$ (as an element in ${\cal A}$). We then have $E(H)=H$ and   
\be
\textrm{tr}_{\cal A} (\rho_\phi \,H)= \textrm{tr}_{\cal A} (E(\rho_\phi \, H))=\textrm{tr}_{\cal A} (E(\rho_\phi) H)=\textrm{tr}_{\cal A} (\rho_{\phi\circ E} \,H)\,.\label{xilo1}
\ee 
 Then it follows $\Delta \langle H\rangle=0$ and eq. (\ref{estasi}). 

Another property we are using is that for a subalgebra ${\cal A}$ belonging to a full matrix algebra and a global pure global state $\varphi$, the entropy for commutant algebras coincide $S_{\cal A}(\varphi)=S_{{\cal A}'}(\varphi)$. 

Eq. (\ref{estasi}) is the difference in entropies in the same algebra (the bigger one ${\cal A}$) between a state and the corresponding invariant one. In contrast, the quantity  $S_{\cal F}(W)-S_{\cal O}(W)$ refers to an entropy difference between an invariant state in two different algebras. 
Let $\varphi_1$ and $\varphi_2$ be two states invariant under some conditional expectation $E:{\cal A}\rightarrow {\cal B}$, that is, $\varphi_i^{\cal A}=\varphi_i^{\cal B}\circ E$. We have 
\bea
S_{{\cal A}}(\varphi_1) &=& -S_{{\cal A}}(\varphi_1|\varphi_2)- \textrm{tr} \rho^{\cal A}_{\varphi_1} \log \rho^{\cal A}_{\varphi_2} \,, \\
S_{{\cal B}}(\varphi_1) &=& -S_{{\cal B}}(\varphi_1|\varphi_2)-\textrm{tr} \rho^{\cal B}_{\varphi_1} \log \rho^{\cal B}_{\varphi_2}  \,.
\eea
For invariant states we always have  $S_{{\cal A}}(\varphi_1|\varphi_2)=S_{{\cal B}}(\varphi_1|\varphi_2)$. Hence subtracting these equations we get
\be
S_{{\cal A}}(\varphi_1)-S_{{\cal B}}(\varphi_1)=\textrm{tr} \rho^{\cal B}_{\varphi_1} \log \rho^{\cal B}_{\varphi_2} -\textrm{tr} \rho^{\cal A}_{\varphi_1} \log \rho^{\cal A}_{\varphi_2}=\langle \log \rho^{\cal B}_{\varphi_2}\rangle_{\varphi_1}- \langle \log \rho^{\cal A}_{\varphi_2}\rangle_{\varphi_1}\,. \label{tirania}
\ee
This holds independently of the invariant state $\varphi_2$ we choose and is linear in $\varphi_1$.

Eq. (\ref{tirania}) shows that expecting $S_{\cal F}(W)-S_{\cal O}(W)$ to be well defined is incorrect. In particular, it is not ordered by inclusion as is the case of (\ref{estasi}). A simple example shows the problems that may occur. Consider the fermion algebra at a site with basis given by the operators $1, c, c^\dagger, c^\dagger c$, with $c, c^\dagger$ creation and annihilation operators, and the fermion symmetry as a symmetry group. For an even state such as the vacuum, the entropy in this algebra ${\cal F}$ is equal to the one of the neutral algebra ${\cal O}=\{1,c^\dagger c\}$, that is $S_{{\cal F}}-S_{{\cal O}}=0$ for any even state. However, if we choose the algebra $\{1, c+c^\dagger \}$, the entropy will be $\log(2)$ for any even state, while the entropy of the even part of the algebra, which is the trivial one $\{1\}$,  is zero. Hence $S_{{\cal F}}-S_{{\cal O}}=\log 2$ for any even state.  Hence we expect that in a lattice, as we enlarge the algebras to arrive to the continuum limit,  $S_{{\cal F}}-S_{{\cal O}}$ can be fluctuating depending on the precise detail on which the algebras are chosen. 
This highlights the necessity of using the mutual information difference to get unambiguous results.

\subsection{Intertwiner version. Lower bound}
\label{lower}
 
If we apply (\ref{dfh}) to the vacuum states $\phi\rightarrow \omega_{{\cal O}_1}$ in ${\cal O}_{W_1}$ and the vacuum $\omega\rightarrow\omega_{1}$ in ${\cal F}_{W_1}$, with ${\cal O}_1\subset {\cal F}_1$, then  eq. (\ref{dfh}) is trivial. In fact $\omega$ and $\phi\circ E$ are the same state in ${\cal F}_{W_1}$, since the vacuum is an invariant state under $G$. 

Now we apply (\ref{dfh}) to the case of the algebra ${\cal F}_1\otimes {\cal F}_2$ and its subalgebra ${\cal O}_1\otimes {\cal O}_2$, corresponding to two disjoint regions $W_1$, $W_2$, in order to gain information about the differences of mutual information.\footnote{We are using the tensor product of algebras. Technically this can be done because of the split property. See \cite{Haag:1992hx}.} Note that ${\cal F}_1\otimes {\cal F}_2$ contains the intertwiners that belong to ${\cal O}$ on top of the elements of ${\cal O}_1\otimes {\cal O}_2$. This last algebra does not contain the intertwiners, that belong to the global neutral algebra but not to the one formed additively in $W_1 \cup W_2$. To exploit this fact we use the conditional expectation $E_{12}=E_{1}\otimes E_2$ that maps these two algebras. Notice that in $E_1\otimes E_2$ the group average is done on each factor independently. To do so we can use the twist operators rather than the global group transformations, as in (\ref{cee}).   

 Let us call  $\omega_{12}$ to the vacuum in the algebra ${\cal F}_1\otimes {\cal F}_2$, and $\phi_{12}$ to the vacuum in ${\cal O}_1\otimes {\cal O}_2$.
  The states we choose for using in (\ref{dfh}) will be $\omega_{12}$ in ${\cal F}_1\otimes {\cal F}_2$ and the state $\phi_1\otimes \phi_2$ in the algebra ${\cal O}_1\otimes {\cal O}_2$, where $\phi_1$ and $\phi_2$ are the vacuum in the algebras ${\cal O}_1$ and ${\cal O}_2$ respectively. We have $(\phi_1\otimes \phi_2)\circ E_{12}=\omega_1\otimes \omega_2$ because both states are invariant under the group transformations on each region separately. They give the same expectation value for any operator. We also have trivially $\omega_{12}|_{{\cal O}_1\otimes {\cal O}_2}=\phi_{12}$. Hence from (\ref{dfh})
\be
I_{{\cal F}}(1,2)- I_{{\cal O}}(1,2)= S_{\cal F}(\omega_{12}|\phi_{12}\circ E_{12})=S_{\cal F}(\omega_{12}|\omega_{12}\circ E_{12})\,.\label{labe}
\ee

The mutual information $I_{{\cal F}}(1,2)$ measures vacuum correlations between regions $W_1$ and $W_2$ including the intertwiners,  while $I_{{\cal O}}(1,2)$ does not include correlations coming from the intertwiners. When regions $W_1$ and $W_2$ are near to each other, and the set of intertwiners is finite, there will be plenty of correlations but these will be essentially the same in ${\cal F}$ and ${\cal O}$, and the leading divergent terms of the mutual informations will cancel, only the effect of the intertwiners will make a change. On the other hand $S_{\cal F}(\omega_{12}|\omega_{12}\circ E_{12})$ is not a mutual information. This will measure the difference between two states on the algebra ${\cal F}_1\otimes {\cal F}_2$, one is the vacuum $\omega_{12}$ and the other is essentially the same state but where the intertwiners have been projected to the neutral algebras on each region. Intuitively, the conditional expectation $E_{12}$ kills the intertwiners by destroying their vacuum correlations.

Now we have the necessary tools to show how both the order parameter $S_{\cal F}(\omega_{12}|\omega_{12}\circ E_{12})$ and even the full mutual information $I_{{\cal F}}(1,2)$ are indeed objects that pertain directly to the theory $\mathcal{O}$. In relation to the order parameter, notice first that the two states appearing in the relative entropy, namely $\omega_{12}$ and $\omega_{12}\circ E_{12}$, are invariant under the action of the global symmetry group. Second, we have that $E({\cal F}_1\otimes {\cal F}_2)=({\cal O}_{(12)'})'$, where the commutants are taken in ${\cal O}$.  Therefore, because of (\ref{igualitos}), we conclude that the order parameter can be computed equivalently as:
\be
S_{\cal F}(\omega_{12}|\omega_{12}\circ E_{12})=S_{({\cal O}_{(12)'})'}(\omega_{12}|\omega_{12}\circ E_{12})\,.
\ee
This formula shows us transparently that the order parameter is an intrinsic quantity of the model ${\cal O}$ itself.

Even more surprisingly, the same is true for the mutual information $I_{{\cal F}}(1,2)$. This can be written in the theory ${\cal O}$ more succinctly as 
\be\label{IO}
I_{{\cal F}}(1,2)=S_{({\cal O}_{(12)'})'}(\omega_{12}|(\omega_1\otimes \omega_2)\circ E_{12})\;.
\ee
Such relation follows by applying again formula (\ref{dfh}) to the present scenario, where it leads to
\begin{eqnarray}
S_{({\cal O}_{(12)'})'}(\omega_{12}|(\omega_1\otimes \omega_2)\circ E_{12})&=&S_{{\cal O}_{(12)}}(\omega_{12}|\omega_1\otimes \omega_2)+S_{({\cal O}_{(12)'})'}(\omega_{12}|\omega_{12}\circ E_{12})=\nonumber\\
=I_{{\cal O}}(1,2)+S_{\cal F}(\omega_{12}|\omega_{12}\circ E_{12})&=&I_{{\cal F}}(1,2)\;.
\end{eqnarray}
Although we have defined the conditional expectation by means of the field algebra $\mathcal{F}$, the conditional expectation $E_{12}$ can be defined directly in the algebra $\mathcal{O}$ as well. It is basically the same conditional expectation, just acting on the smaller algebra $({\cal O}_{(12)'})'$. More quantitatively, the action of $E_{12}$ in ${\cal O}$ can be expressed in the following way. A generic element of ${\cal O}$ can be written $b=\sum_r b_r \, {\cal I}^r_{12}$ as an expansion in intertwiners of different irreducible representations and where the $b_r$  commute with the twists (or the invariant twists in ${\cal O}$). Then $E_{12}(b)=b_1$.  

Having shown that at the end of the day, even if one computes relative entropies in the field algebra $\cal F$, one actually ends up with relative entropies of the invariant algebra $\mathcal{O}$, it turns out to be technically and conceptually simpler to work with the field algebra $\cal F$, and we will do so in what follows.

The consequence of expressing this difference of mutual informations as a relative entropy is that we can use monotonicity of relative entropy to put lower bounds.  In particular, to produce a lower bound we can restrict the states to a subalgebra  ${\cal C}_{12}$ of ${\cal F}_1\otimes {\cal F}_2$, 
\be
I_{{\cal F}}(1,2)- I_{{\cal O}}(1,2)\ge S(\omega_{12}|\omega_{12}\circ E_{12})|_{{\cal C}_{12}}\,.\label{left}
\ee
Moreover, since the expectation value of the intertwiners is the main difference between states,
 we have to find a useful ${\cal C}_{12}$ that contains the relevant information about the intertwiners. 
 
 If we have a finite dimensional ${\cal C}_{12}$ (or more generally a type I subalgebra) the left hand side of (\ref{left}) can be written in terms of the entropies if we further require that the conditional expectation maps the algebra in itself, $E_{12}({\cal C}_{12})\subseteq {\cal C}_{12}$. Using (\ref{estasi}) we get a lower bound given by a difference of entropies,
\be
I_{{\cal F}}(1,2)- I_{{\cal O}}(1,2)\ge S(\omega_{12}|\omega_{12}\circ E_{12})|_{{\cal C}_{12}}=S(\omega_{12}\circ E_{12})|_{{\cal C}_{12}}-S(\omega_{12})|_{{\cal C}_{12}}\,.\label{left1}
\ee
The fact that this difference is positive is because the charged operators on $W_1$ and $W_2$ can have entanglement in vacuum , what is reflected in  the expectation values of the intertwiners. This entanglement will count for the entropy of the first state on the right hand side of (\ref{left1}) but not for the second.

To improve the lower bound we can try to maximize the entropy difference over all choices of intertwiner operators (or the algebra ${\cal C}_{12}$). To see what we can do, suppose we have an intertwiner ${\cal I}_{12}=V_1 V_2^\dagger$ for an Abelian sector, with $V_1$, $V_2$ unitaries in each region. An obvious idea is to try to maximize the expectation value, that is, 
\be
\langle  V_{1} V_2^\dagger\rangle \rightarrow 1\,.\label{11}
\ee
This is to say that both $V_1$ and $V_2$, acting on complementary regions, create essentially the same state acting on the vacuum.
If $V_1$ and $V_2$ where inverse to each other we would get $1$ but this is not possible since they have disjoint supports. 

Of special interest is the case where the region $W_2\rightarrow W_1'$, and both regions cover the full space. In this case, we will be able to get the maximum value (\ref{11}). Then, let us think directly in this case. 
By using the modular reflection operator $J$ of the region $W_1$ (and the theory ${\cal F}$) we can convert\footnote{See \cite{borchers2000revolutionizing} for a review of modular theory. $J$ is an antiunitary operator mapping the algebra ${\cal F}_1$ to its commutant ${\cal F}_2$. For the case of a Rindler wegde $J$ is the CRT operator \cite{Bisognano:1975ih,Witten:2018lha}.}   
\be
\langle  V_{1} V_2^\dagger \rangle= \langle  V_{1} J \tilde{V}_2 J\rangle\,,
\ee
with $\tilde{V}_2=J V_2^\dagger J$ now belongs to the algebra of $W_1$. By Tomita-Takesaki modular theory this is the same as 
\be
 \langle V_{1}V_2^\dagger\rangle=\langle  V_{1} \Delta^{\frac{1}{2}} \tilde{V}_2^\dagger \rangle\,,
\ee
with $\Delta$ the modular operator, that is positive definite.\footnote{Heuristically $\Delta =\rho_1\otimes \rho_2^{-1}$, with $\rho_1$, $\rho_2$ the reduced density matrices. } 
Using Schwarz inequality 
\be
|\langle V_{1}V_2^\dagger\rangle|^2 =|\langle  V_{1} \Delta^{\frac{1}{2}} \tilde{V}_2^\dagger \rangle|^2\le \langle  V_{1} \Delta^{\frac{1}{2}} V_1^\dagger  \rangle\, \langle \tilde{V}_2 \Delta^{\frac{1}{2}}\tilde{V}_2^\dagger \rangle\,.
\ee
Therefore to maximize the expectation value we can choose either $V_1 J V_1 J$ or $\tilde{V}_2 J \tilde{V}_2 J$ as intertwiners. Without loss of generality we write
\be
V_2^\dagger=J V_1 J\,.
\ee
 Note that $V_1$ and $V_2^\dagger$ will be formed by representations of opposite charge because of the action of $J$, and this is exactly what we need to produce an intertwiner.

Therefore we need to maximize 
\be
\langle  V_{1} \Delta^{\frac{1}{2}} V_1^\dagger \rangle\,.
\ee 
If we could choose $V_1$ commuting with $\Delta^{\frac{1}{2}}$, because $\Delta|0\rangle=|0\rangle$, we would get the desired $\langle V_{1}V_2^\dagger\rangle= \langle  V_{1} V_1^\dagger \rangle=1$.
Intuitively, this commutation can be achieved by writing $V_1$ in the base that diagonalizes the modular Hamiltonian or the density matrix.
We can always write a unitary operator that commutes with the density matrix by choosing phases in the basis that diagonalizes the density matrix. However, this unitary will have zero charge because the density matrix commutes with the charge operator. Hence, we can solve the problem only in an approximate sense, choosing charge creating operators corresponding to modes of the modular Hamiltonian with modular energy tending to zero, or as much invariant under the modular flow as possible.  
 In QFT we can always approach $\langle  V_{1} V_2^\dagger\rangle \rightarrow 1$ as much as we want for complementary regions (in many different ways) since zero is included in the spectrum of the modular Hamiltonian which is continuous in $(-\infty,\infty)$. In the next section we explore the physical content of this requirement with some explicit examples.    

Note this cannot be done if $W_1$ and  $W_2$ are at a finite distance since in that case $J$ would take us from $W_1$ to $W_1'$ that is bigger than $W_2$. Then the maximal correlator cannot be achieved exactly in general for non zero distance. However, if the regions touch along some part of the boundary, no matter how small, we can think in putting highly localized excitations very near this region of the boundary where the modular energy is small. In a sense, in this region we can think the states are similar to the case where the full space $W_1'$ is covered by $W_2$. Then we expect for any such case the maximal correlation can be achieved for a convenient choice of excitations approaching the boundary. 

Now, coming back to the bound on the mutual information difference, we can have a universal bound for a finite group $G$ when the two regions $A$ and $B$ are complementary to each other or touch in a $d-2$ dimensional piece of the boundary. In this case we expect we can maximize the value of the intertwiner expectation values. We will see this bound depends only on the number of elements $|G|$ of the group.

To see this, let us think we have a finite subalgebra of operators on each region which is isomorphic to the algebra of matrices of $N\times N$ and we further require this algebra is kept in itself by group transformations. Let us call $P^1_{ij}$ and $P^2_{ij}$ to the operators forming the matrix basis of these algebras in $W_1, W_2$. That is
\be
P^1_{ij}P^1_{kl}=\delta_{jk}\, P^1_{il}\,,\hspace{1cm} (P^1_{ij})^\dagger=P^1_{ji}\,,\hspace{1cm} \sum_i P^1_{ii}=1\,,\label{416}
\ee    
and analogously for $W_2$. 
One way to generate these finite algebras is to use the charge generating operators $V_r^i$ for some representation (not necessarily irreducible). These close an infinite dimensional algebra in general. However, the finite dimensional algebra 
 (discussed in section (\ref{sec:esa})) formed by the operators 
\be 
(a)=\sum_{ij} a_{ij} V_r^i (V_r^j)^\dagger \label{esa1}
\ee
form a matrix algebra. However, one can produce a subalgebra without worrying about the partial isometries $V_r^i$. We will give examples in the next section. 

We want to maximize the entanglement between these two algebras, and then we choose $P^1_{ij}=J P^2_{ij} J$ and think these operators approximately commute with the modular operator. Under this choice, we notice that if $D^{\left(1\right)}\left(g\right)$ is the unitary
matrix representation of the global group transformations $U\left(g\right)$
in the algebra $\{ P_{ij}^{\left(1\right)}\} $, then $D^{\left(2\right)}\left(g\right)=(D^{\left(1\right)}\left(g\right))^*$
is the representation of $G$ in the algebra $\{ P_{ij}^{\left(2\right)}\} $.

The density matrix of the vacuum state $\omega$ on this algebra writes 
\be
\rho^{\omega}_{jl,ik}=\langle P^1_{ij}P^2_{kl}\rangle=\langle P^1_{ij} J P^1_{kl} J\rangle=\langle P^1_{ij} \Delta^{\frac{1}{2}} P^1_{lk} \rangle\simeq \langle P^1_{ij} P^1_{lk} \rangle= \delta_{jl} \langle P^1_{ik}\rangle\,.
\ee
 Hermiticity of $\rho^\omega_{jl,ik}$ implies that, under these assumptions for the state,  
\be 
\langle P^1_{ik}\rangle=N^{-1} \, \delta_{ik}\,,
\ee
and
\be
\rho^{\omega}_{jl,ik}=N^{-1} \, \delta_{ik}\delta_{jl}\,.\label{md}
\ee
This state is invariant under conjugation with any  unitary transformation matrix of the form 
\be
D\otimes D^*\,,
\ee
and in particular it is invariant under global group transformations that have this form given our choice of algebras. 
This is a pure state 
\be
S(\omega)=0\,,\label{twisted11}
\ee
and $\omega$ is maximally entangled between $W_1$ and $W_2$, as expected.

In order to compute the state $\phi$ we need to know how the group acts on each of the algebras. Let us decompose the action of the group on each algebra (\ref{416}) in irreducible representations. We have representations $r$ of dimension $d_r$ and multiplicity $n_r$. Hence
\be
\sum_r n_r d_r=N\,.
\ee
Without loss of generality we take the basis vectors that decompose the group representation into irreducible ones, and rename the indices of the basis as $i\rightarrow (r,s,l)$, where $s=1\,\cdots, n_r$, $l=1,\cdots,d_r$. The state $\phi=\omega\circ E_{12}$ has density matrix
\bea
&&\rho^{\phi}_{(r_1s_1l_1)(r_2s_2l_2),(r_3s_3l_3)(r_4s_4l_4)}\\&&=\frac{1}{|G|^2}\sum_{g_1, g_1\in G} D^{r_1}_{l_1l_1'}(g_1)D^{r_2\,*}_{l_2l_2'}(g_2) \rho^{\omega}_{(r_1s_1l_1')(r_2s_2l_2'),(r_3s_3l_3')(r_4s_4l_4')}D^{r_3\,*}_{l_3l_3'}(g_1)D^{r_4}_{l_4l_4'}(g_2)\nonumber\\
&& = \frac{1}{d_{r_1}\,N} \delta_{r_1 r_2}\delta_{r_2 r_3}\delta_{r_3 r_4}\delta_{s_1 s_2}\delta_{s_3 s_4} \delta_{l_1 l_3}\delta_{l_2 l_4}\,.\nonumber 
\eea
In the last equation we have used the orthogonality relation for irreducible representations,    
\be
\sum_{g\in G} D^{r_1}_{l_1l_2}(g)D^{r_2\,*}_{l_3l_4}(g)= \frac{|G|}{d_{r_1}} \delta_{r_1,r_2}\delta_{l_1 l_3}\delta_{l_2 l_4}\,, \label{orthogonality}
\ee
and the formula (\ref{md}). Therefore the non zero part of the density matrix has the structure of a direct sum of blocks labelled by the irreducible representations. The density matrix is 
\be
\rho^{\phi}=\bigoplus_r \frac{n_r d_r}{N} \,\, \left[\frac{1}{n_r}(1)_{n_r\times n_r}\oplus (0)_{n_r^2-n_r\times n_r^2-n_r}\right]\otimes \left[ \frac{1}{d_r^2} \textrm{I}_{d_r^2\times d_r^2}\right] \,.\label{irredi}
\ee 
The first factor is proportional to a matrix with all entries equal to $1$ (a one dimensional projector), except for zero blocks,  and the second factor is proportional to an identity matrix. Both of these factors are normalized to have unit trace. Hence, 
writing the fraction of basis vectors with representation $r$ as
\be
q_r=\frac{n_r d_r}{N}\,, \hspace{1cm} \sum_r q_r=1\,,\label{con}
\ee
 the entropy is
\be
S(\phi)=-\sum_r q_r \log q_r + \sum_r q_r \log d_r^2\,. \label{twisted}
\ee
We can vary the frequency $q_r$ of the representation $r$ in order to achieve maximal entropy difference $S(\phi)-S(\omega)\equiv S(\phi)$, taking into account the constraint (\ref{con}). We get the maximum is achieved for
\be
q_r=\frac{d_r^2}{|G|}\,,\label{proji}
\ee
where we used the relation $|G|=\sum_r d_r^2$ valid for finite groups. This implies
\be
n_r=d_r\,\frac{N}{|G|}\,,
\ee
and from (\ref{twisted})
\be
S(\phi)-S(\omega)=\log |G|\,.
\ee

Therefore, the optimal multiplicity of a representation is proportional to the dimension of the representation. This is exactly the case of the regular representation of the group. The optimal representation then consists of any number of copies of the regular one. Other representations will give weaker constraints. Notice that there is no increase in the entropy by arbitrarily multiplying the representations and enlarging the Hilbert space. The conditional expectation will take into account that redundant copies are not measuring any new difference between models since they are produced by the neutral algebra.\footnote{It is interesting to consider the Renyi entropies of the state (\ref{irredi}) of the intertwiner algebra. These Renyi entropies are all equal to the same constant $\log |G|$  when taking the regular representation and in this limit of maximal entanglement. This feature of a state is named ``flat spectrum'' in the literature. Pressumably this leads to a flat spectrum of the difference of Renyi mutual informations between the two models in the limit of touching regions. }   

With the regular representation we have the best lower bound (for complementary regions)
\be
I_{{\cal F}}(1,2)- I_{{\cal O}}(1,2)\ge \log |G|\,.
\ee
As we will see below, $\log|G|$ is also an upper bound for the difference of mutual informations.

In the appendix \ref{regular} we show formally that the regular representation can always be achieved using the charge generators $V^i_r$ of all irreducible representations. But from a physical standpoint, in general, we remark that the regular representation is naturally constructed with high frequency by fusion. We will use this idea in the example in section \ref{freexamples}. 
 The reason is that the character of the regular representation is $\chi_R(g)=|G| \delta_{g,1}$ and then the regular representation is stable under fusion. The tensor product of a regular representation with another representation of dimension $K$ has character $\chi(g)=K |G| \delta_{g,1}$, and then decomposes into exactly $K$ copies of the regular representation. This is not the case for other representations.  For any representation of dimension $d>1$ the character satisfies
 \be
 \frac{\chi(1)}{d}=1\,, \hspace{1cm} \left|\frac{\chi(g)}{d}\right|< 1\,,
\ee  
and then  for the product $r_{12}=r_1 \otimes r_2$ of two representations
\be
\frac{\chi_{12}(1)}{d_1 d_2}=1\,, \hspace{1cm} \left|\frac{\chi_{12}(g)}{d_1 d_2}\right| =\left|\frac{\chi_1(g)}{d_1}\right|\left|\frac{\chi_2(g)}{d_2}\right|\,,
\ee
the normalized character always approaches the one of the regular representation. 

Another way to see this is to realize that the tensor product of arbitrary representations $R$ with some fix representation $R_0$ can be thought of as a stochastic process in the space of the probabilities $q_r$. In fact, the new representation $R'=R_0\otimes R$ will have
\be
q^{R'}_{i}=\sum_j M^0_{ij} \, q_j^{R}\,,  
\ee
where
\be
M^0_{ij}=\sum_k \frac{N^i_{kj}}{d_k d_j} d_i\, q_k^{R_0}\,,
\ee
and $N^i_{kj}$ is the fusion matrix giving the number of irreducible representations of type $i$ that appear in the tensor product of representations $k$ and $j$. The matrix $M^0$ is stochastic, and represents a stochastic process since it has positive entries and $\sum_i M^0_{ij}=1$. Since for any fixed $k$ we have\footnote{This follows from the fact that the tensor product of the regular representation with any other one is proportional to the regular representation.} $\sum_j N^i_{kj}d_j\sim d_i$ it follows that the probability vector $q_i=\frac{d_i^2}{|G|}$ is the fixed point of the stochastic process, an eigenvector of $M^0$ of eigenvalue $1$. As for any stochastic process, applying it repeatedly will approach the fixed point rapidly. 

Roughly speaking, the infinite algebra of QFT in a region is formed by infinitely many products of subalgebras and the group representation is closed under fusion. Hence the frequency of each irreducible representation must be that of the regular representation. 
In the regular representation the basis elements $|g\rangle$ are treated on equal footing by the group transformations, and the subspace of the irreducible representation $r$ has dimension $d_r^2$.   
Then the probability of each irreducible sector in vacuum must be given by (\ref{proji}).    

\subsection{Twist version. Upper bound}
\label{upper}

The simplest upper bound  for $\Delta I$ uses the following convexity property of relative entropy \cite{ohya2004quantum}. Let $\sigma_i$ and $\varphi$ be states on a given algebra and $0\le\lambda_i\le 1$, $\sum_i \lambda_i=1$. We have
\be
\sum_i \lambda_i S(\sigma_i|\varphi)- S(\sum_i \lambda_i \sigma_i| \varphi)\le -\sum_i \lambda_i \log \lambda_i\,.\label{tirsos}
\ee

To use this property in the present context, note that 
\be
\omega_{12}\circ E_{12}= \frac{1}{|G|^2}\sum_{g_1\in G_1,g'_2\in G_2} \omega_{g_1g'_2}=\frac{1}{|G|}\sum_{g_1\in G_1} \omega_{g_1} \label{accor}\,,
\ee
where we are writing $\omega_{g}=\omega \circ g$, and the labels $1$ and $2$ in $g$ mean that the group transformations act on the two regions independently (we can use the twists) and in the second equality we have used the invariance of $\omega$ under the group transformations, which implies that $\omega_{g_{1}g_{2}}=\omega$. We apply (\ref{tirsos}) with $\lambda_i=1/|G|$, the different $\sigma_i$ given by the states $\omega_{g_i}$ for different $g_i$, and $\varphi=\omega_{12}\circ E_{12}$. The second relative entropy in (\ref{tirsos}) vanishes while the relative entropies $S(\sigma_i|\varphi)=S(\omega_{g_i}|\frac{1}{|G|}\sum_{g_1\in G_1} \omega_{g_1})$ for different $g_i$ are all equal, because we can transform any one into any other by a group automorphism, which is just a unitary tranformation into each of the states appearing in the relative entropy.

Therefore we get the upper bound \footnote{This upper bound might be considered an intertwiner or twist upper bound, depending on the focus one is taking. But this bound is not tight in general. The tightest upper bound, which we are deriving below, comes from analyzing the problem from a twist perspective.}
\be
I_{{\cal F}}(1,2)- I_{{\cal O}}(1,2)=S(\omega_{12}|\omega_{12}\circ E_{12}) \le \log|G|\,, \label{supp}
\ee
which together with the lower bound of the previous section allows us to conclude that as the two boundaries touch each other the bound becomes saturated for finite $|G|$,
\be
I_{{\cal F}}(1,2)- I_{{\cal O}}(1,2)= \log|G|\,.\label{fini}
\ee
Defining the quantum dimension ${\cal D}$ by ${\cal D}^2=\sum_r d_r^2$,  which in the present case is equal to $|G|=\sum_r d_r^2$, we can also write this same result in the form
\be
\Delta I= \log({\cal D}^2)=\log(\sum_r d_r^2)
\ee
and for the regularized entropy
\be
\Delta S=\frac{\Delta I}{2}=\log ({\cal D})\,.
\ee
Written in this way the contribution coincides with the formula for the topological entanglement entropy \cite{Kitaev:2005dm,Levin:2006zz}. We will come back to this identification in Part II.  

It is interesting to note that (\ref{fini}) is a purely topological contribution and does not depend on the interactions or whether the models are massive or massless. Of course, the size of $\epsilon$ where saturation is achieved depends on the typical size where the intertwiners have appreciable expectation values. For a conformal theory and two spheres, $\Delta I$ will be a function of the cross ratio determining the geometry, while for a massive theory we need to cross the scale of the gap to see some difference between the mutual informations to arise, independently of the size of the regions $W_1,W_2$.

In the context of RG flows, in general $\Delta I$ should be attributed to the mutual information of ${\cal O}$ as a negative contribution $-\log|G|$ (a lack of entanglement that ${\cal F}$ posses). For example, for a massive complete model in the IR we expect there is no constant term in the entropy in odd dimensions (the $F$ term in EE of a sphere). However, for the orbifold, we get $-\log|G|$ as a constant topological term. We will come back to these issues in Part II (a companion paper), where we discuss implications for the renormalization group. 

According to the derivation of (\ref{supp}) saturation is only possible if the supports for the states $\omega_g$ become disjoint for different $g$. This requires the vacuum expectation values for the squeezed twists that implement group operations in $W_1$ and not in $W_2$ to go to zero in this limit. We will see later this is also implied by uncertainty relations between twist and intertwiners that do not commute with each other.

An improved upper bound can be obtained by considering the dual version of (\ref{labe}) where the relative entropy is based on the complementary algebra of the two regions, namely the shell.  This requires a more specific property that we could not find in the mathematical literature. We are proving this property in the lattice and taking the continuum limit afterward. 

We again consider the algebra  ${\cal F}_{W_1W_2}={\cal F}_{W_1}\otimes {\cal F}_{W_2}$, and call ${\cal F}_S=({\cal F}_{W_1W_2})'$, where for notational convenience we have called $S=(W_1W_2)'$ to the ``shell'' complementary to the two balls. For simplicity we take $W_1$ and $W_2$ to be two disjoint sets of vertices on the lattice and take as algebras ${\cal F}_{W_1}$, ${\cal F}_{W_2}$ the full set of operators at these vertices. These algebras are in tensor product with the rest of the lattice operators. 
We take a group of twists $G_\tau$ acting on $W_1$. The invariant part of ${\cal F}_{W_1W_2}$ under $G_\tau$ is ${\cal O}_{W_1}\otimes {\cal F}_{W_2}$. The commutant of this algebra is $({\cal O}_{W_1}\otimes {\cal F}_{W_2})'={\cal F}_S\vee G_\tau$. We have two conditional expectations. The first one is 
\be
E_1: {\cal F}_{W_1}\otimes {\cal F}_{W_2}\rightarrow {\cal O}_{W_1}\otimes {\cal F}_{W_2}\,,
\ee
 which follows by acting with the twists in region $W_1$. The ``dual'' conditional expectation maps
 \be
 E_\tau: {\cal F}_S\vee G_\tau\rightarrow {\cal F}_S\,.
\ee
To describe the action of $E_\tau$ note that any element $a\in {\cal F}_S\vee G_\tau$ can be written
\be
a=\sum_g a_g \, \tau_g\,,
\ee
where the $a_g\in {\cal F}_S$. The decomposition of the element $a$ is unique. Then we take
\be
E_\tau (a)=a_1\,.
\ee
 $E_\tau$ defines a conditional expectation. Further, the definition of ${\cal F}_S\vee G_\tau$ and $E_{\tau}$ does not depend on the precise form of the twists chosen. In this lattice setting we can just choose  $G_{\tau}$ as the elements of the group acting on the vertices of $W_1$, such that $G_{\tau}$ commutes with ${\cal F}_{S}$. Without loss of generality we then make this choice of $G_{\tau}$. Then ${\cal F}_S\vee G_\tau={\cal F}_S\otimes \hat{G}_\tau$, where $\hat{G}_\tau $ is the group algebra. 

Because of the invariance of the global vacuum, we have as in (\ref{accor}),  
\be
S_{{\cal F}_{W_1}\otimes {\cal F}_{W_2}}(\omega|\omega\circ E_{12})=S_{{\cal F}_{W_1}\otimes {\cal F}_{W_2}}(\omega|\omega\circ E_{1})\,.
\ee
Using (\ref{estasi}) this is
\be
S_{{\cal F}_{W_1}\otimes {\cal F}_{W_2}}(\omega|\omega\circ E_{1})=S_{{\cal F}_{W_1}\otimes {\cal F}_{W_2}}(\omega\circ E_{1})-S_{{\cal F}_{W_1}\otimes {\cal F}_{W_2}}(\omega)\,.
\ee

Using the purity of the global state $\omega$ twice, we transform this successively as 
\bea
&& S_{{\cal F}_{W_1}\otimes {\cal F}_{W_2}}(\omega\circ E_{1})-S_{{\cal F}_{W_1}\otimes {\cal F}_{W_2}}(\omega)= S_{{\cal F}_{W_1}\otimes {\cal F}_{W_2}}(\omega\circ E_{1})-S_{{\cal F}_{S}}(\omega)\nonumber
\\
&& =S_{{\cal O}_{W_1}\otimes {\cal F}_{W_2}}(\omega)-S_{{\cal F}_{S}}(\omega)+(S_{{\cal F}_{W_1}\otimes {\cal F}_{W_2}}(\omega\circ E_{1})-S_{{\cal O}_{W_1}\otimes {\cal F}_{W_2}}(\omega))\nonumber\\
&&=S_{{\cal F}_S\vee G_\tau}(\omega)-S_{{\cal F}_{S}}(\omega) +(S_{{\cal F}_{W_1}\otimes {\cal F}_{W_2}}(\omega\circ E_{1})-S_{{\cal O}_{W_1}\otimes {\cal F}_{W_2}}(\omega))\nonumber \\
&& =S_{{\cal F}_S\vee G_\tau}(\omega)-S_{{\cal F}_S\vee G_\tau}(\omega\circ E_\tau)+\\
&&\hspace{2cm}+(S_{{\cal F}_{W_1}\otimes {\cal F}_{W_2}}(\omega\circ E_{1})-S_{{\cal O}_{W_1}\otimes {\cal F}_{W_2}}(\omega))+(S_{{\cal F}_S\vee G_\tau}(\omega\circ E_\tau)-S_{{\cal F}_{S}}(\omega))\,.\nonumber
\eea

Since the conditional expectation $E_\tau$ does not preserve the trace unless the group is Abelian, we cannot convert the first two terms into a relative entropy using (\ref{estasi}). However, here we can use the fact that $\omega\circ E_\tau$ is a product state in ${\cal F}_S\otimes \hat{G}_\tau$. In fact this state is equal to $\omega_{{\cal F}_S}\otimes \sigma$, where $\sigma$ is the state in $\hat {G}_\tau$ defined by $\sigma(\tau_g)=\delta_{g,1}$. Then we  write
\be
S_{{\cal F}_S\vee G_\tau}(\omega)-S_{{\cal F}_S\vee G_\tau}(\omega\circ E_\tau)=-S_{{\cal F}_S\vee G_\tau}(\omega|\omega\circ E_{\tau})+\Delta \langle H_\sigma\rangle\,,
\ee
where $\Delta \langle H_\sigma\rangle=-\omega(\log \rho_\sigma^{\hat{G}_\tau})+ \sigma(\log \rho_\sigma^{\hat{G}_\tau})$. 
We get
\bea
&& S_{{\cal F}_{W_1}\otimes {\cal F}_{W_2}}(\omega|\omega\circ E_{12})+S_{{\cal F}_S\vee G_\tau}(\omega\circ E_\tau)\label{45}\\&&\hspace{1cm}=\Delta \langle H_\sigma\rangle+(S_{{\cal F}_{W_1}\otimes {\cal F}_{W_2}}(\omega\circ E_{1})-S_{{\cal O}_{W_1}\otimes {\cal F}_{W_2}}(\omega))+(S_{{\cal F}_S\vee G_\tau}(\omega\circ E_\tau)-S_{{\cal F}_{S}}(\omega))\,.\nonumber
\eea

The two last terms within brackets in the right-hand side are formed by differences in entropies between states that are invariant under the conditional expectations but computed in the algebra and its fixpoint subalgebra under the conditional expectations. The last term in brackets, since the state $\omega\circ E_\tau$ is a tensor product of states in ${\cal F}_S\otimes \hat{G}_\tau $, gives the entropy of the state $\sigma$ in the algebra of the group. To compute it we note the algebra of the group is a sum of full matrix algebras of dimensions $d_r$, $\oplus_r M_{d_r\times d_r}$. The projectors on the different blocks are the $P_r$ in (\ref{projec}), which have expectation values $\langle P_r \rangle_\sigma=d_r^2/G$. Then the density matrix is block diagonal with elements $d_r/|G|$ on the diagonal in each block. The entropy is
\be
S_{{\cal F}_S\vee G_\tau}(\omega\circ E_\tau)-S_{{\cal F}_{S}}(\omega)=S(\sigma)=\log |G|-\sum_r \frac{d_r^2}{|G|} \log d_r\,.  \label{360}
\ee

To evaluate the first bracket in the right hand side of (\ref{45}) we note that inside the full matrix algebra ${\cal F}_{W_1}\otimes {\cal F}_{W_2}$ the common center of ${\cal O}_{W_1}\otimes {\cal F}$ and $\hat{G}_\tau$ is again formed by the algebra of projectors $P_r$ of the center of the group algebra. Then, diagonalizing these projectors, we have a representation $\oplus_r M_{d_r\times d_r}\otimes N_r$, where the group acts with $D_r(g)$ in each block in the first factor, and $N_r$ represents matrix algebras of invariant elements. An invariant state like $\omega\circ E_1$ has density matrix 
\be
\oplus_r q_r \, \frac{1_{d_r\times d_r}}{d_r}\otimes \rho_r\,, 
\ee
where $q_r=\omega(P_r)$ are the frequencies with which each sector appears in the algebra ${\cal F}_{W_1}$ and $\rho_r$ are density matrices in $N_r$. We get
\be
S_{{\cal F}_{W_1}\otimes {\cal F}_{W_2}}(\omega\circ E_{1})-S_{{\cal O}_{W_1}\otimes {\cal F}_{W_2}}(\omega)=\sum_r q_r \log d_r\,.
\ee
Moreover, taking into account that the vacuum is invariant under global group symmetries $\omega_{\hat{G}_\tau}=\oplus_r q_r \frac{1_{d_r\times d_r}}{d_r}$, and
\be
\Delta \langle H_\sigma \rangle=-\sum_r q_r \log d_r+\sum_r \frac{d_r^2}{|G|} \log d_r \,.
\ee
Therefore, adding all together we get 
\be
S_{{\cal F}_{W_1W_2}}(\omega|\omega\circ E_1\otimes E_2)=\log|G|-S_{{\cal F}_S\vee G_\tau}(\omega|\omega\circ E_\tau)\,.\label{sedien}
\ee
Since this relation holds in any lattice discretization, it should hold also in the continuum limit. This is because the terms in the equation are all well defined in such a limit. Indeed, we notice that the conditional expectation $E_{\tau}$ can be obtained directly in the continuum with the help of the charged operators $V^g$ in $W_1$, corresponding to the regular representation of the group, in the following way
\be
\frac{1}{|G|} \sum_{h\in G} V^{h\, \dagger} a V^{h}=  \frac{1}{|G|} \sum_{h\in G} \sum_{g\in G} a_g \, V^{h\, \dagger} \tau_g V^{h} = \frac{1}{|G|} \sum_{h\in G} \sum_{g\in G} a_g \, V^{h\, \dagger} V^{gh} \tau_g=a_1=E_\tau(a)\,. 
\ee
Finally, collecting all results together we arrive to
\be
I_{{\cal F}}(1,2)- I_{{\cal O}}(1,2)=\log|G|-S_{{\cal F}_S\vee G_\tau}(\omega|\omega\circ E_\tau)\,.\label{sedaa}
\ee
 When $W_1$ and $W_2$ increase (\ref{sedaa}) increases and the relative entropy on the shell decreases as it must be. This is why the relative entropy on the shell and twists appears with a minus sign.
 
The last equation again expresses an upper bound $\log |G|$ to $\Delta I$ but improves it by the relative entropy in the right hand side. As in the case of the intertwiners, we can take any subalgebra $\tilde{{\cal S}}_\tau$ of ${\cal F}_S\vee G_\tau$ of the shell and the twists to get a convenient upper bound
\be
I_{{\cal F}}(1,2)- I_{{\cal O}}(1,2)\le \log|G|-S_{\tilde{{\cal S}}_\tau}(\omega|\omega\circ E_\tau)\,.\label{sedien1}
\ee
Any set of twists $\tau_g$ that close a representation of the group form the linear basis of an algebra $\hat{G}_\tau$ and we can restrict to this algebra.\footnote{Note these are smeared twists, as spread as possible to increase expectation values, in contrast to the sharp twists we used above.} In this case, recalling that the twist algebra $\omega_{\hat{G}_\tau}=\oplus_r q_r \frac{1_{d_r\times d_r}}{d_r}$ and  $\sigma=\oplus_r \frac{d_r}{|G|} \, 1_{d_r\times d_r} $, we get
\be
I_{{\cal F}}(1,2)- I_{{\cal O}}(1,2)\le -\sum_r q_r \log q_r+\sum_r q_r \log(d_r^2) \,.\label{sed}
\ee
Eq. (\ref{sed}) is the same expression (\ref{twisted}) which is bounded below by $0$ and above by $\log |G|$. 
It is a function of the twist expectation values through  (see (\ref{projec}))
\be
q_r=\langle P_r \rangle=\frac{d_r}{|G|}\sum_g \chi^*_r(g) \langle \tau_g \rangle\,. \label{370}
\ee
We get $\log |G|$ for sharp twists satisfying $\langle \tau_g\rangle=\delta_{g,1}$. These expectation values imply the regular representation probabilities through the previous relation. In a realistic scenario, the smallest upper bound will be for the most spread out twists, where the expectation values of the twists are bigger and the relative entropy in the twist algebra is larger. On the other side of the story, the upper bound goes to zero when $\langle \tau_g\rangle =1$ for all  $\tau_g$. This is the case for the vacuum and the global group transformations, which satisfy $q_r=\delta_{r,1}$. Finally, notice that for abelian groups (\ref{sed}) is just the entropy in the twist algebra since the second term vanishes. This is not the case of a non Abelian group where the  entropy in the twist algebra is $-\sum_r q_r \log q_r+\sum_r q_r \log(d_r)$ rather than (\ref{sed}). Hence there is an additional contribution in (\ref{sed}). This is necessary to match the intertwiner relative entropy in special cases where upper and lower bounds coincide.  

We want to remark that the expression (\ref{sed}) for an upper bound should remain valid for continuous groups as far as the group is compact and the statistics of the sectors give a finite result. In fact, it will turn out this expression is generally finite for Lie group symmetries in QFT.

\subsection{Entropic certainty and uncertainty relation}
\label{certainty}

Recall eq. (\ref{sedien}), 
\be
S_{{\cal F}_{W_1W_2}}(\omega|\omega\circ E_1\otimes E_2)+S_{{\cal F}_S\vee G_\tau}(\omega|\omega\circ E_\tau)=\log|G|\,.
\ee
For large intertwiner expectation values (small $\epsilon$) the first relative entropy will approach $\log|G|$ implying the twist one goes to zero, while the opposite is true for large $\epsilon$ where there are some twists with large expectation values. Since we are using the full algebra  and a global pure state this is a ``certainty relation'', but reducing to subalgebras $\tilde{{\cal C}}_{12}$, $\tilde{{\cal S}}^\tau$ that contain at least some closed algebra of intertwiners and some closed algebra of twists respectively, we have the entropic uncertainty relation
\be
S_{\tilde{{\cal C}}_{12}}(\omega|\omega\circ E_{12})+S_{\tilde{{\cal S}}^\tau}(\omega|\omega\circ E_\tau) \le \log|G|\,.
\ee
Similar entropic uncertainty relations occur for generalized measurements \cite{coles2017entropic,berta2016entropic}.
Notice that the maximal relative entropy for each term needs minimal uncertainty: expectation values for the twist operators or for intertwiners equal to maximal ones. In the case of minimal uncertainty, each relative entropy can achieve $\log |G|$.

Therefore, minimal uncertainty cannot be achieved at the same time for intertwiners and twists. 
The non-trivial commutation relations between twists and intertwiners is what prevents the left-hand side of this inequality to reach $2 \log |G|$,  while $\log |G|$ would be the maximum that can be achieved for each of the two terms.  

In the same way, if we have an impure global state that is invariant under the group (i. e. a thermal state), we can purify it in a larger Hilbert space and upon reduction we get  
\be
S_{{\cal F}_{W_1W_2}}(\omega|\omega\circ E_1\otimes E_2)+S_{{\cal F}_S\vee G_\tau}(\omega|\omega\circ E_\tau)\le \log|G|\,.
\ee

Uncertainty relations may be derived for operator expectation values rather than entropies using the commutations relations between twists and intertwiners. For example, in the case of the even fermionic subalgebra described above we have just one twist and one intertwiner satisfying 
\be
\tau \,{\cal I}_{W_1 W_2}=-{\cal I}_{W_1 W_2} \, \tau\,.
\ee
 The usual uncertainty relation for non-commuting operators gives  
\be
1-|\langle \tau\rangle|^2 -|\langle {\cal I}_{W_1 W_2}\rangle|^2 \ge 0\,.
\ee 
Then, when the twist has maximal expectation value  $|\langle \tau\rangle|=1$ the expectation value of the intertwiner is zero, and vice-versa. More generic scenarios include commutators that are controlled by the group representations and will be considered in Part II.

\subsection{Lie Group}
\label{U1}

When the group is not finite  $\Delta I(1,2)$ will be divergent in the limit when the regions touch each other. The interest lies in understanding how this quantity depends on $\epsilon$. 

Let us first analyze the case of a group $U(1)$. We have a continuum of twist operators  
\be
\tau_k=e^{i k Q_1}\,,
\ee
 where $Q_1$ is the generator of the twist algebra crossing $W_1$ and $k\in (-\pi,\pi)$.

In general, computing the exact operators $Q_1$ and the expectation values of the twists on a specific theory will be a problem depending on the dynamics. However, we are interested in the $\epsilon\rightarrow 0$ limit, and we will argue the leading divergent term of the result is universal. We know that inside the ball $W_1$
\be
Q_1\sim \int d^dx\, J^0(x) \alpha(x)\,,
\ee
where $J^0$ is the charge density and $\alpha(x)$ is a convenient smearing function. This integrates to $1$ in time, and it is spatially constant inside the ball.  On the shell the operator content and the smearing changes, such as to give $\tau$ the desired group properties. 

In the small $\epsilon$ limit the leading term of the total charge fluctuation inside the ball will come from short distance charge fluctuations distributed all along the surface, with a particle-antiparticle on each side the wall separating the two regions, see figure \ref{fig6}. We can then picture the fluctuations of the total charge contributing to $Q_1$ as given by a large sum of independent random variables since short distance fluctuations that are separated by a macroscopic distance along the surface of the sphere will not see each other. We will come back to this point in section \ref{twistcontinuo} where we elaborate a bit more on general properties of twists expectation values. 

Then, because of the central limit theorem we can use a formula for Gaussian distributions in the space of charges where the probabilities for different $Q_1$ is
\be
p_q=\frac{1}{\sqrt{2\pi\langle Q_1^2\rangle}}e^{-\frac{q^2}{2\langle Q_1^2\rangle}} \label{prubi}
\ee
where $\langle Q_1^2\rangle\gg 1$ for small $\epsilon$.  
Therefore, using these probabilities, since the Abelian algebra of the twists $\tau_k$ is represented by $e^{i k q}$ in the space of (integer) charges we have 
\be
\langle \tau_k \rangle=\sum_q p_q\, e^{i q k}\simeq e^{-\frac{1}{2}k^2\langle Q_1^2\rangle}\,, \hspace{1cm} k\in (-\pi,\pi)\,.\label{proyi}
\ee
We have used an approximation for  $\langle Q_1^2\rangle\gg 1$ applying a continuous Fourier transform. This is why the result is not periodic in $k$, but it will hold very approximately in the limit that we are studying. 

An upper bound for $\Delta I$ is then easily computed from (\ref{sed}) to be the entropy of this distribution 
\be
-\sum_q  p_q \log(p_q)= 1/2  \log  \langle Q_1^2\rangle + \textrm{cons.}  \,.\label{thesame}
\ee

Notice that even if the twist algebra has a continuum of operators the upper bound is well defined because it is the entropy of a classical discrete set of charges or, equivalently, because the group is compact. 
We expect the difference of mutual informations is divergent in the non-compact case. But there should be no problem with the mutual information of ${\cal O}$ but rather the one of ${\cal F}$ is the one not well defined in this case. The problem, we think,  is that ${\cal F}$ contains too many sectors that would make fail the splitting property that guarantees we can take the algebra of two regions as a tensor product. This splitting property is related to the finiteness of a nuclearity index \cite{Haag:1992hx}, which in turn is related to the partition function. Similar observations have been made recently using other arguments \cite{Harlow:2018tng}.   

The best upper bound corresponds to the lowest $\langle Q_1^2\rangle$. This corresponds to the most spread out twist. As the smearing function on the shell becomes wider, the probability of charge fluctuations on each side of the shell decreases and the charge fluctuations inside the smearing region are averaged to zero.  We give a more direct calculation of $\langle Q_1^2\rangle$ in section \ref{twistcontinuo} below. Here we just notice that the result must be proportional to the area since bulk virtual fluctuations of the charge are suppressed because they will appear with both signs and the total charge average zero. For a  current that is conformal in the UV the area $A$ must be compensated by powers of the cutoff, $\langle Q_1^2 \rangle\sim A/(\epsilon)^{d-2}$. We then have 
\be 
\Delta I \le \frac{1}{2}  \log  \frac{A}{\epsilon^{d-2}} + \textrm{cons}\sim\frac{(d-2)}{2}  \log  \frac{R}{\epsilon} + \textrm{cons}
\,.\ee

A lower bound can be given by thinking in the intertwiners. There is one for each integer number $Q$ representing the charge which labels the irreducible representations of the group. This Abelian algebra is represented as the  multiplicative algebra of periodic functions on the elements of the group labeled by $k\in (-\pi,\pi)$ ($e^{i k Q}$ is the representative of ${\cal I}_Q$).\footnote{For any Abelian group $G$ the intertwiners are labeled by the representations, and we can represent the Abelian algebra of the intertwiners with the algebra of functions on the group. This coincides with the algebra of the characters, $\chi_{r_1}(g)\chi_{r_2}(g)=\chi_{r_1 \otimes r_2}(g)$.} Then this algebra is the Abelian continuous algebra of functions on $(-\pi,\pi)$. The classical entropy is not well defined on this algebra but the relative entropy is not ambiguous. The state $\omega\circ E_{12}$ gives zero expectation to any $Q\neq 0$, and then is represented by the state $1/(2 \pi)$, constant on $k\in (-\pi,\pi)$. 
 We have to select the intertwiners such as to maximize the relative entropy. This can be achieved by concentrating the probability around $k=0$ as much as possible. 
 This means the probability of the different charges is as much flat as possible. In particular, to sense the probability distribution (\ref{prubi}) of charge fluctuations in vacuum our intertwiners will have to be spread out on the surface of the sphere. 
   Any smaller localization will lead to a less flat distribution of probabilities of charges. Heuristically, the intertwiner of charge $Q$ will then carry a state $\sqrt{p_q} \,|q\rangle_1 \otimes |-q\rangle_2$ to $\sqrt{p_q}\, |q+Q\rangle_1 \otimes |-q-Q\rangle_2$. The expectation value will be
 \be
 \langle {\cal I}_Q\rangle \sim \sum_q \sqrt{p_q p_{q+Q}}=e^{-\frac{Q^2}{8 \langle Q_1^2\rangle}}\,.\label{distic}
 \ee  
 The probability for each $k$ such that $\langle {\cal I}_Q\rangle=\int dk\, e^{i k Q} \, p_k$ is therefore 
 \be
 p_k = \sqrt{\frac{2\langle Q_1^2\rangle}{\pi}} e^{-2 \langle Q_1^2\rangle k^2}\,.
 \ee
   The relative entropy with the constant state $1/(2\pi)$ then gives the same leading order calculation (\ref{thesame}). 
  We then get that the asymptotic behaviour is in fact 
\be 
\Delta I \simeq  \frac{1}{2}\log\frac{A}{\epsilon^{d-2}}\sim \frac{(d-2)}{2}  \log  \frac{R}{\epsilon} 
\,.
\label{epifa}
\ee
This term should be attributed to the orbifold model as a contribution $-\frac{(d-2)}{2}  \log  \frac{R}{\epsilon}$ to the mutual information.  This logarithmic term is ``topological'' in the sense that it appears in odd dimensions as well as in even dimensions, and it does not depend on the curvature of the boundary as the usual logarithmic anomaly terms.  

In $d=2$ we have to replace $(R/\epsilon)^{(d-2)}\rightarrow \log(R/\epsilon)$ and the leading term is 
 \be
 \Delta I=\frac{1}{2}\log (\log(R/\epsilon))\,.\label{cft}
\ee 
 However, in $d=2$ this is correct for two intervals that touch each other, while in the case of nearly complementary regions the shell consists of two intervals and the coefficient gets duplicated for massive fields while it is still (\ref{cft}) for CFT. See section \ref{dosd}. 

For a non-Abelian compact Lie group, we have different twist generators $L_i$, $i=1\,\cdots, {\cal G}$, where ${\cal G}$ is the dimension of the Lie algebra. For each of these charges we expect to have a Gaussian probability of charges as in (\ref{prubi}) for the same reasons as above. 
The group is non-commutative though. However, the typical expectation values of the charges are very large in the limit of small $\epsilon$, and therefore we are in the regime of ``large numbers'' where the non-commutativity is not relevant. Then the intertwiner version gives us a picture of ${\cal G}$ independent charges with 
\be
\Delta I \simeq \frac{ 1 }{2}\,(d-2)\, {\cal G} \,\log  \frac{R}{\epsilon}\,.\label{nona}
\ee
The twist version matches this expectation but there is a subtlety. A twist $e^{i k_i L_i}$ has appreciable expectation value only for small parameters $k_i$ as in (\ref{proyi}). This means only the neighborhood of the identity is probed in the group. Therefore we might expect to have effectively the case of ${\cal G}$ Abelian generators. This is correct, but the conditional expectation knows that these different directions in the Lie algebra can be connected by group transformations and cannot be considered independent. Hence, the entropy in the group algebra is, in fact, smaller than what is expected for the case of ${\cal G}$ Abelian generators. However, the formula (\ref{sed}) contains an additional piece on top of the twist entropy in the non-Abelian case and taking into account this contribution the calculation with the twists matches the expectation (\ref{nona}) from the intertwiners. 

Let us see how this work in a concrete example. Consider the case of $SO\left(3\right)$. According to
the discussion above, for small $\epsilon$, the expectation values
of the twist $\tau_{g}$ are non-zero only for those
corresponding to group elements near the identity element $g\approx1$.
In this situation, as in the Abelian case, it is useful to parametrized
the twist operators with a 3-vector $\bar{k}$ according to
\begin{equation}
\tau_{\bar{k}}=\mathrm{e}^{i\bar{k}\cdot\bar{L}}\,,\;\label{twist_so3}
\end{equation}
where $\bar{L}=\left(L_{1},L_{2},L_{3}\right)$ are like angular momentum operators with commutation relations
$\left[L_{j},L_{k}\right]=i\epsilon_{jkl}L_{l}$ . As argued above,
the vacuum expectation value of such twist operators, in the small
$\epsilon$ limit is Gaussian, and has to be rotationally invariant
\begin{equation}
\left\langle \tau_{\bar{k}}\right\rangle =\mathrm{e}^{-\frac{1}{2}\left|\bar{k}\right|^{2}\left\langle \bar{L}^{2}\right\rangle }\,.\label{exp_so3}
\end{equation}
Then it behaves as if they were the twist operators associated
to 3 independent generators of the Abelian group $U\left(1\right)^{3}$.
The computation using these expectation values is straightforward. First, we have
 that the irreducible representations of $SO\left(3\right)$
are labeled by a non-negative integer $l\in\mathbb{Z}_{\geq0}$.
The $l$-representation has dimension $d_{l}=\left(2l+1\right)$ and
its character $\chi_{l}$ is given by \cite{hamermesh2012group}
\begin{equation}
\chi_{l}\left(\theta\right)=\frac{\sin\left(\left(l+\frac{1}{2}\right)\left|\theta\right|\right)}{\sin\left(\frac{1}{2}\left|\theta\right|\right)}\,,\label{char_so3}
\end{equation}
where $\theta$ is the angle of rotation from the identity. This coincides with $\theta\sim |\vec{k}|$ for small $\theta$. 
To compute the desired upper bound using equation \eqref{sed}, we need
first to calculate the probabilities $q_{l}$ attached to the $l$-representation.
For that we use the Lie group continuum version of \eqref{370}
\begin{equation}
q_{l}=\left(2l+1\right)\frac{1}{\pi}\int_{0}^{\pi}dk\, (1-\cos(\theta))\chi_{l}\left(\theta\right)\,\mathrm{e}^{-\frac{1}{2}\theta^{2}\left\langle \bar{L}^{2}\right\rangle }\,,\label{ql_so3}
\end{equation}
where the finite sum was replaced by the integral over the full group
$SO\left(3\right)$ using the normalized Haar measure (see \cite{hamermesh2012group}) and we are assuming $\langle \bar{L}^{2}\rangle \gg 1$. Replacing \eqref{char_so3}
into \eqref{ql_so3} we can compute analytically the probabilities,
which are given in terms of $\textrm{Erf}$ functions. At the end, replacing
such probabilities into \eqref{sed} we can check 
\begin{equation}
I_{\mathcal{F}}\left(1,2\right)-I_{\mathcal{\mathcal{O}}}\left(1,2\right)\leq-\sum_{l=0}^{\infty}q_{l}\log\left(q_{l}\right)+\sum_{l=0}^{\infty}q_{l}\log\left(d_{l}^{2}\right)\sim\frac{3}{2}\log\left\langle \bar{L}^{2}\right\rangle +\mathrm{const.}\,,\label{bound_so3}
\end{equation}
as we claimed above. 

We notice that the contribution of each term in \eqref{bound_so3} separately for large $\left\langle \bar{L}^{2}\right\rangle $ (small $\epsilon$) is
\begin{eqnarray}
-\sum_{l=0}^{\infty}q_{l}\log\left(q_{l}\right) & \sim & \frac{1}{2}\log\left\langle \bar{L}^{2}\right\rangle \,,\\
2\sum_{l=0}^{\infty}q_{l}\log\left(d_{l}\right) & \sim & \log\left\langle \bar{L}^{2}\right\rangle \,.
\end{eqnarray}
For an invariant state, the
density matrix for the twist algebra decomposes according to the irreducible
representations as
\begin{equation}
\rho=\bigoplus_{l=1}q_{l}\cdot\frac{\mathbf{1}_{d_{l}}}{d_{l}}\,,\label{dm_so3}
\end{equation}
where $\mathbf{1}_{d_{l}}$ is the identity matrix in the full matrix
algebra $\mathbb{C}^{d_{l}\times d_{l}}$. The entropy of this algebra is then
\be
S_\tau=-\sum_{l=0}^{\infty}q_{l}\log\left(q_{l}\right)+\sum_{l=0}^{\infty}q_{l}\log\left(d_{l}\right)\,.\label{asa}
\ee
Then this entropy contributes only with a $\log \langle \bar{L}^{2}\rangle$ and the missing $1/2 \log \langle \bar{L}^{2}\rangle$ comes from the fact that the last term in (\ref{asa}) has a factor of $2$ in the correct formula (\ref{bound_so3}). 
 This is in contrast with the Abelian case where (\ref{bound_so3}) gives the entropy in the twist algebra.

\subsection{Other topologies}
\label{OT}

\begin{figure}[t]
\begin{center}  
\includegraphics[width=0.55\textwidth]{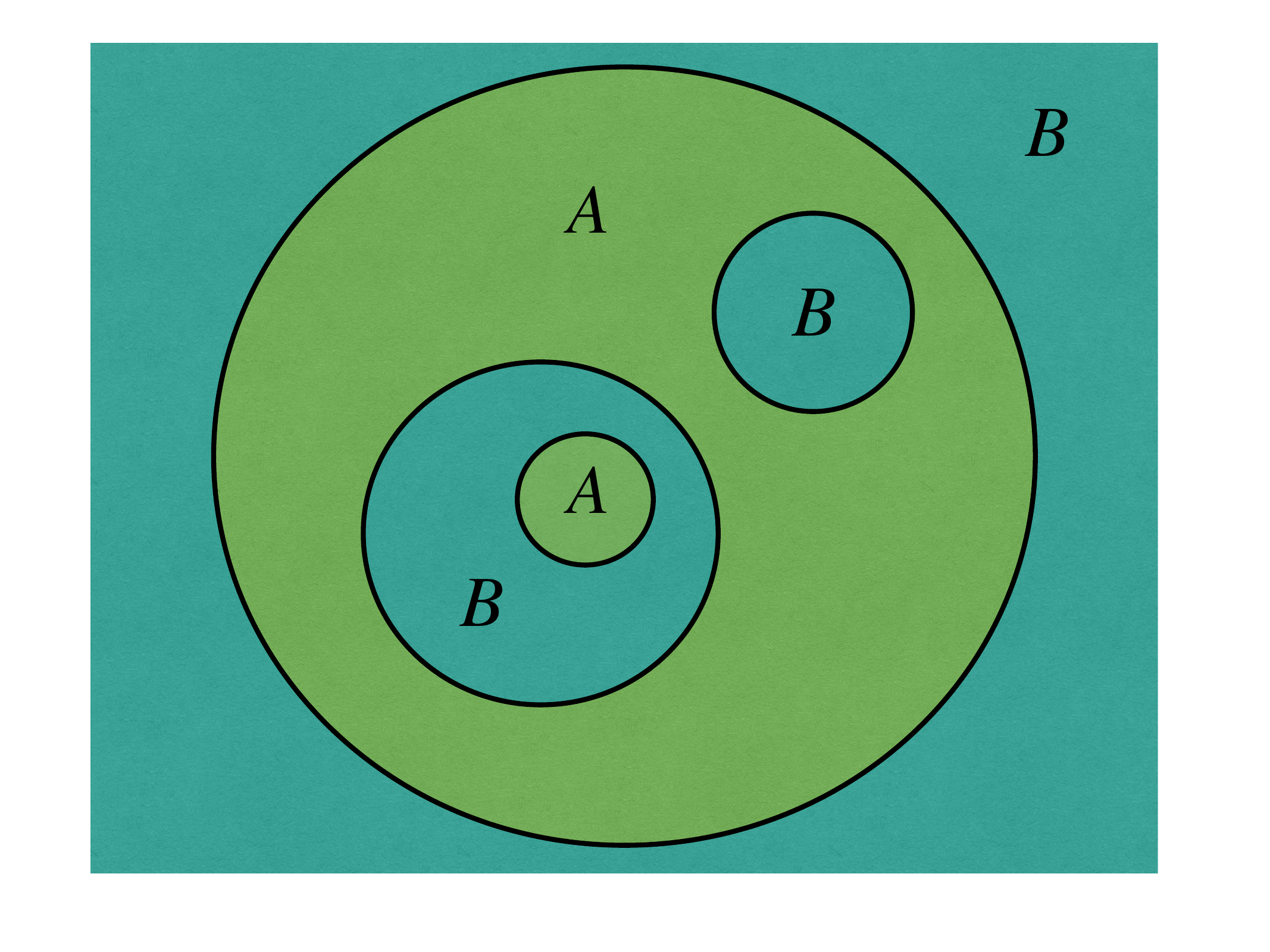}
\captionsetup{width=0.9\textwidth}
\caption{Two complementary regions $A$ and $B$ with non trivial topology. There is an independent set of intertwiners and twists for each connected component of the common boundary between $A$ and $B$. In the figure the number of connected components of the boundary is $n_\partial=4$.}
\label{topofig}
\end{center}  
\end{figure}

The same type of ideas can be used to try to understand the difference in mutual information between the models ${\cal F}$ and ${\cal O}$ for regions with different topologies, such as the one shown in Fig (\ref{topofig}).

Let us first make some general remarks. Suppose we have a region $A$ with connected components $A_1,\cdots,A_n$ and we think in lattice models where these regions rather correspond to mutually commuting finite dimensional algebras. We can use the same type of manipulations used in section \ref{upper} to get 
\bea
&& S_{\cal F}(\omega_A)-S_{\cal O}(\omega_A)\label{fori}\\
&& =S_{\cal F}(\omega_A)-S_{\cal F}(\omega_A\circ E_{A_1}\otimes\cdots\otimes E_{A_n})-(S_{\cal O}(\omega_A)-S_{\cal F}(\omega_A\circ E_{A_1}\otimes\cdots\otimes E_{A_n}))
\nonumber \\
&&=-S_{\cal F}(\omega_A|\omega_A\circ E_{A_1}\otimes\cdots\otimes E_{A_n})-(S_{\cal O}(\omega_A)-S_{\cal F}(\omega_A\circ E_{A_1}\otimes\cdots\otimes E_{A_n}))\,. \nonumber
\eea
Notice that formula (\ref{fori}) is valid even if the global state $\omega$ is not invariant under global group transformations.

This naive expression should give a well defined expression in QFT once we make combinations of entropies of different regions which can be rewritten in terms of relative entropies. For example, for two single component regions we obtain
\be
 I_{\cal F}(1,2)-I_{\cal O}(1,2)=S_{\cal F}(\omega_{12}|\omega_{12}\circ E_1\otimes E_2)-S_{\cal F}(\omega_{1}|\omega_{1}\circ E_1)-S_{\cal F}(\omega_{2}|\omega_{2}\circ E_2)\,,\label{ecui}
\ee
since the remaining term coming from the brackets in (\ref{fori}) is zero,
\be
S_{\cal O}(\omega_{A_1})-S_{\cal F}(\omega_{A_1}\circ E_{A_1})+S_{\cal O}(\omega_{A_2})-S_{\cal F}(\omega_{A_2}\circ E_{A_2})-S_{\cal O}(\omega_{A_1A_2})-S_{\cal F}(\omega_{A_1A_2}\circ E_{A_1}\otimes E_{A_2})=0\,.
\ee
This can be shown with the help of (\ref{tirania}). We choose an invariant state $\phi_2$ in that formula that is the tensor product of its reductions to $A_1$ and $A_2$. Alternatively, we remark that eq. (\ref{ecui}) can be shown directly in the continuum by combining relations (\ref{dfh}) and (\ref{another}), without taking a path that uses the entropies.

Eq. (\ref{ecui}) coincides with (\ref{labe}) once we notice that for the vacuum state the two last terms in the right hand of (\ref{ecui}) side vanish. However, (\ref{ecui}) remains valid for any state $\omega$.  

Then this is an economical way of producing relative entropy identities.  
Now we can think in the case of two regions $A$ and $B$ with $n$ and $m$ connected components respectively. By using (\ref{fori}) we get
\be
 I_{\cal F}(A,B)-I_{\cal O}(A,B)=S_{\cal F}(\omega_{AB}|\omega_{AB}\circ \otimes_i E_{A_i}\otimes_j  E_{B_j})-S_{\cal F}(\omega_{A}|\omega_{A}\circ \otimes_i E_{A_i})-S_{\cal F}(\omega_{B}|\omega_{B}\circ \otimes_i E_{B_i})\,.\label{rhs}
\ee
In contrast to the single component case the two last terms do not vanish for the vacuum when $m,n>1$. 

 Let us try to understand the value of this mutual information difference. We are mainly interested in the limit case of two regions $A$ and $B$ that are nearly complementary to each other. 
 
Let us focus first on the first term of the right-hand side of (\ref{rhs}). In general, there will be a complicated pattern of interference between the intertwiners crossing pairs of the $m+n$ regions, but under the current assumptions this will be dominated by the intertwiners crossing between adjacent boundaries of $A$ and $B$.  
In the particular limit we are focusing here, each connected component of the boundary of $A$ meets with a connected component of the boundary of $B$, see Fig (\ref{topofig}). Therefore the number of connected components of the boundary of $A$ and the ones of $B$ agree. Let us call this number $n_\partial$. Since each connected boundary divides the space in two, the different boundaries form a tree under inclusion. This leads to the fact that the sets of intertwiners crossing each of these boundaries are algebraically independent to each other. They are also statistically independent because they are well localized in different boundaries. The total number of independent set of intertwiners is given by $n+m-1$, because it is given by the total number of independent charge creating operator algebras, minus one to account for the neutrality of the operator. This coincides with the number of boundaries, $n_\partial=n+m-1$,  since in going from the interior to the exterior, each time we cross a boundary we can add a unique connected component of $A$ or $B$. Then the maximization of intertwiner entropy will indeed select the ones crossing each boundary with no free choices left, and we have 
\be
S_{\cal F}(\omega_{AB}|\omega_{AB}\circ \otimes_i E_{A_i}\otimes_j  E_{B_j})=  n_{\partial} \, \log|G|.\label{instance}
\ee
This has the form of a topological contribution. 

There is a parallel twist version of this story. The difference between the entropies of the involved states in this relative entropy must come from $C=(AB)'$. This has exactly $n_\partial$ connected components which are thin regions at the interfaces between $A$ and $B$. Each of these connected surfaces divides the space in two and carries one set of independent twist operators that make the difference between the algebras of the two models in $C$. In the limit of small $\epsilon$, independent sets of twists all have vanishing expectation value and are statistically independent. Hence, again,  each boundary contributes $\log|G|$.  

Thinking in  the mutual information difference in the generic case, the first term in (\ref{rhs}) is bounded above by (\ref{instance}). From (\ref{rhs}) we then have the general bound
\be
0\le I_{\cal F}(A,B)-I_{\cal O}(A,B)\le n_{\partial} \, \log|G|.  \label{instance1}
\ee
 
The second term of the right-hand side of (\ref{rhs})  depends on the intertwiners crossing the different connected components of $A$, which can form elements of ${\cal O}$ which are not in the algebra $A_{\cal O}$.
This term is in fact equal to the difference of generalized mutual informations\footnote{In finite dimensional algebras $S(\omega_A|\omega_{A_1}\otimes \cdots\otimes \omega_{A_n})=S(\omega_{A_1})+\cdots +S(\omega_{A_n})-S(\omega_A)$. } 
\be
S_{\cal F}(\omega_{A}|\omega_{A}\circ \otimes_i E_{A_i})=S_{\cal F}(\omega_A|\omega_{A_1}\otimes \cdots\otimes \omega_{A_n})-S_{\cal O}(\omega_A|\omega_{A_1}\otimes \cdots\otimes \omega_{A_n})\,.
\ee
This is in general difficult to compute but we can simplify this contribution if we focus in the case where the theory is gapped, and we are in the infrared regime, with regions much larger than the gap scale. 
The expectation values of the intertwiners crossing components of $A$ are exponentially small in the regime of large mass because the typical distances between components are large compared to the mass scale. In consequence, this contribution vanishes in this approximation.  The same holds for the third term in (\ref{rhs}), which concerns the multicomponent region $B$. Then, in this limit and for $\epsilon$ smaller than the gap scale, saturation is achieved and we obtain
\be
I_{\cal F}(A,B)-I_{\cal O}(A,B)= n_{\partial} \, \log|G|.  \label{instance2}
\ee
\subsection{Excitations}
\label{EX}

We want to investigate how the entropy changes in a ball when we insert a  well localized charged excitation. Evidently, in the case this charged excitation corresponds to a sector of dimension $d_r=1$ (such as the excitations with Abelian group symmetry) there will not be any change in entropy because this is represented by a unitary endomorphism. 

Then let us think in a state 
\be
|\psi_i\rangle=\sqrt{d_r} V_r^{i\, \dagger}| 0 \rangle\,, 
\ee
corresponding to the irreducible representation $r$, already introduced in (\ref{dietrich}). 
For any element $b$ of ${\cal O}$
\be
\langle \psi_i | b|\psi_i \rangle= \langle 0| \rho(b)|0\rangle=\omega\circ\rho(b)\,,
\ee
with 
\be
\rho(b)= \sum_j V_r^{j} b V_r^{j\,\dagger}
\ee
the corresponding endomorphism. Notice $|\psi_i\rangle$ is pure in ${\cal F}$ but not in ${\cal O}$.

To measure the entropy of this impurity we can compute 
\be
S_{\cal F}(\psi_i|\psi_i\circ E)=S_{\cal F}(\psi_i|\omega\circ\rho\circ E) \,.
\ee

To get an upper bound to this quantity we notice that as in section \ref{upper} changing the first state by a group transformation does not change the relative entropy, 
\be
S_{\cal F}(\psi_i|\psi_i\circ E)=S_{\cal F}(\psi_i\circ g|\psi_i\circ E)
\ee
because this is a unitary transformation in the algebra on both states at the same time (the second state being invariant). Then we can average  over $g$ the first state to obtain the second. Actually we can do it better, since it is possible as well to average over only $d_r$ group elements that just change basis elements in the representation $r$ to get the second state. Using again convexity of relative entropy (\ref{tirsos}) we get the upper bound
\be
S_{\cal F}(\psi_i|\psi_i\circ E)=\frac{1}{d_r}\sum_{g_k} S_{\cal F}(\psi_i\circ g_k|\psi_i\circ E) \le S_{\cal F}(\frac{1}{d_r}\sum_{g_k}\psi_i\circ g_k|\psi_i\circ E)+\log (d_r)=\log (d_r)\,.
\ee
The vectors in the $|\psi_i\rangle$ are orthonormalized because they belong to different 
 superselection sectors 
 \be
 \langle \psi_i |\psi_j \rangle =\delta_{ij}\,. 
 \ee 
As we make the excitation support smaller or the radius of the ball bigger, the reduced states for each $i$ in this sum become disjoint. This is the condition for the bound to become saturated,
\be
S_{\cal F}(\psi_i|\psi_i \circ E)=\log d_r\,. 
\ee

A completely localized excitation is produced by the operators $V_r^i$, 
\be
|\tilde{\psi}_i\rangle= V^i_r |0\rangle\,.\label{vector}
\ee
This time there is no factor $\sqrt{d_r}$ since $\langle \tilde{\psi}_i |\tilde{\psi}_j \rangle =\delta_{ij}$ because of $V_r^{i\, \dagger}V_r^{j}=\delta_{ij}$. The vector (\ref{vector}) corresponds to the conjugate representation $\bar{r}$. This state in ${\cal O}$ is equivalent to the global state
\be
\tilde{\rho}=\frac{1}{d_r} \sum_i |\tilde{\psi}_i\rangle \langle \tilde{\psi}_i |\,. \label{tirosami}
\ee
  In the same way as above we have an upper bound $\log d_r$ for the relative entropy. The lower bound can be obtained by reducing to the subalgebra  (\ref{esa1}) generated by  
$ V_r^i V_r^{j\, \dagger} $. We have for the two states on this subalgebra
\bea
 \langle \tilde{\psi}_i|V_r^k V_r^{l\, \dagger}|\tilde{\psi}_i\rangle &=& \delta_{ik}\delta_{il}\,,\label{tirso}\\
 \frac{1}{d_r} \sum_i\langle \tilde{\psi}_i|V_r^k V_r^{l\, \dagger}|\tilde{\psi}_i\rangle &=& \frac{1}{d_r}\delta_{kl}\,.\label{tarso}
\eea 
The relative entropy in the subalgebra is $\log(d_r)$. Then we have 
\be
S_{\cal F}(\tilde{\psi}_i|\tilde{\psi}_i\circ E)=\log d_r
\ee
exactly, as soon as the operator $V_i$ is inside the region.  

Instead of using this relative entropy we can try to compute the change in mutual information for touching regions. The excitation does not change correlations of operators in ${\cal F}$ outside the support of $V_i$. Then, if this support is small and well inside $W_1$ we expect  
\be
I^{\tilde{\psi}_i}_{\cal F}(1,2)-I^{0}_{\cal F}(1,2)\simeq 0\,.
\ee
To compute the change in the model ${\cal O}$ we use the formula 
 \be
I^{\tilde{\psi}_i}_{\cal F}(1,2)-I^{\tilde{\psi}_i}_{\cal O}(1,2)=S_{12}(\tilde{\psi}_{i}|\tilde{\psi}_{i} \circ E_{12})-S_1(\tilde{\psi}_i|\tilde{\psi}_i\circ E_1)-S_2(\tilde{\psi}_i|\tilde{\psi}_i\circ E_2)\,.
\label{dista}\ee
The last term is zero because the two states are equal in $W_2$. The second term is $\log(d_r)$ as we have seen. The first term is upper bounded by $\log |G|+ \log d_r$ since it is the minimal number of transformed states $\tilde{\psi}_i$ we need to mix to get $\tilde{\psi}_{i} \circ E_{12}$. 
But it is also lower bounded by the same number since we can use an algebra formed by the one in (\ref{tirso}), (\ref{tarso}) plus some intertwiner algebra near the boundary of $W_1,W_2$. Expectation values for these algebras are uncorrelated and the effect of the conditional expectation can also be decoupled. Then we conclude 
\be
I^{\tilde{\psi}_i}_{\cal F}(1,2)-I^{\tilde{\psi}_i}_{\cal O}(1,2)=\log |G|\,.
\ee
In consequence, the mutual information in ${\cal O}$ does not change with respect to the vacuum, as happens with ${\cal F}$.
The excitation is impure in the model ${\cal O}$, but its impurity is due to a transformation of the vacuum well inside the region $W_1$, that does not modify correlations with ${\cal O}_{W_2}$. Therefore it will not change the mutual information. 

If we instead create a particle in $W_1$ and an antiparticle in $W_2$ with $ |\varphi\rangle =V_{r,1}^i V^{i}_{\bar{r},2}|0\rangle$ the mutual information in ${\cal F}$ will not change with respect to the vacuum because of the same reasons as above. For the model ${\cal O}$ we can again use (\ref{dista}). Now the last two relative entropies are $\log d_r$. The first one is again upper bounded and lower bounded by $\log |G|+ \log d_r$. Then we conclude
\be
I^{\varphi}_{\cal F}(1,2)-I^{\varphi}_{\cal O}(1,2)=\log |G|- \log d_r\,,
\ee
and 
\be
I^{\varphi}_{\cal O}(1,2)-I^{0}_{\cal O}(1,2)=\log d_r\,.
\ee
Remarkably this does not depend on how far separated are the excitations.  We would have expected $2 \log d_r$ for the mutual information of a maximally entangled state of a Hilbert space of $d_r$ dimension. But here the effect is rather the classical mutual information of variables with perfect correlation and maximal uncertainty (the effective state is analogous to (\ref{tirosami}) with $|\tilde{\psi}_i\rangle=V_{r,1}^i V^{i}_{\bar{r},2}|0\rangle$), which gives half this number, and it is not produced by entanglement. For a pure state $|\varphi\rangle =1/\sqrt{d_r} \sum V_{r,1}^i V^{i}_{\bar{r},2}|0\rangle$ we get along the same lines 
\be
I^{\varphi}_{\cal F}(1,2)-I^{\varphi}_{\cal O}(1,2)=\log |G|\,.
\ee
The state is invariant under the group and then, as in vacuum, $S_{12}(\tilde{\psi}_{i}|\tilde{\psi}_{i} \circ E_{12})=\log |G|$ rather than $\log|G|+\log d_r$ as above, and the two last terms in (\ref{dista}) vanish. Then we expect
\be
 I^{\varphi}_{\cal O}(1,2)-I^{0}_{\cal O}(1,2)=I^{\varphi}_{\cal F}(1,2)-I^{0}_{\cal F}(1,2)=2 \log d_r\,,
\ee
 as corresponds to a pure state in both models.  

Several results about the entropy of charged states analogous to the ones in this section have previously appeared in the literature for specific models. See for example \cite{Lewkowycz:2013laa,Dong:2008ft,Caputa:2014eta,Nozaki:2014hna,Alcaraz:2011tn,Longo:2018obd,Hollands:2017dov}.   
 
\subsection{Spontaneous symmetry breaking} 
\label{SSB}

  When the symmetry is broken it is not true anymore that the relative entropy for the vacuum in the two models is zero for one component regions. This is because the vacuum expectation values do not generally vanish for the charged operators.\footnote{Both models satisfy clustering. However, the expectation value of the intertwiners in $\cal O$ does not go to zero for large distances between the charge creating fields. The non-vanishing of this expectation value is the indication of SSB in ${\cal O}$ itself. This does not mean a failure of clustering in $\cal O$ since the intertwiner is not the product of operators in $W_1$ and $W_2$ in ${\cal O}$. However, the model ${\cal F}$ with a mixed state in the different possible choices of vacuum does not satisfy clustering and we have to choose only one vacuum.} In fact this relative entropy $S_{\cal F}(\omega|\omega \circ E) $ is an interesting quantity to compute in this case and serves as an order parameter for symmetry breaking.

Let us first discuss the case of a finite group $G$.   As in section \ref{upper}, the relative entropy is upper bounded because of convexity,
\be
S_{\cal F}(\omega|\omega \circ E)\le \log|G|\,.
\ee
The different vacuum states are transformed into each other by the group elements.\footnote{The representation of the group in different vacua cannot transform one vacuum into a linear combination of several vacua because it would transform a state with clustering into another without clustering.} If we have $|G|$ vacua we have a regular representation of the group. If the symmetry is not completely broken the bound is improved to $\log (|G|/|H|)$, where $H$ is the subgroup that still keeps the vacuum invariant. As a curiosity, we note this is a relative entropy for a single region which is invariant under Lorentz transformations of the region. This rare luxury is possible precisely because of the existence of more than one vacuum.    

The relative entropy is increasing with size. We can take the entropy difference in any finite subalgebra stable under $E$ to get a lower bound. We expect that for size $R$ small with respect to the scale of the symmetry breaking the symmetry is effectively restored, and the relative entropy approaches zero. In other words, there are no operators inside the ball that are able to efficiently distinguish the two states. For regions larger than the symmetry breaking scale we expect saturation of the bound. For example, take a theory with broken $Z_2$ symmetry, where the order parameter is a scalar field $\phi$ such that $\langle \phi\rangle=\mu$. As an order parameter we can use a smeared mode $\phi_\alpha=\int d^d x\, \alpha(x) \phi(x)$ such that $\int d^dx\, \alpha(x)=1$. We have $\langle \phi_\alpha \rangle=\mu$ in the state $\omega$ and $\langle \phi_\alpha \rangle=0$ in the state $\omega\circ E$, but the fluctuations of this mode for small support of the test function $\alpha$ will be much bigger than $\mu$ and of order of $R^{-1}$. Hence we cannot efficiently distinguish the states in a small region.\footnote{To get a rough estimate of the behaviour we may assume Gaussian fluctuations. The relative entropy for a classical continuous variable with Gaussian distribution of width $R^{-1}$ centered around the origin and another Gaussian distribution centered in $\mu$ is $\sim (R \mu)^2$.} 

To understand the behavior of the mutual information difference we use formula (\ref{ecui})
\be
I_{\cal F}(1,2)-I_{\cal O}(1,2)=S_{\cal F}(\omega_{12}|\omega_{12}\circ E_1\otimes E_2)-S_{\cal F}(\omega_{1}|\omega_{1}\circ E_1)-S_{\cal F}(\omega_{2}|\omega_{2}\circ E_2)\,,\label{ecui4}
\ee
valid for general states. We are mainly interested in the case of regions $W_1$, $W_2$, which are nearly complementary, and let us think $W_1$ is a ball of radius $R$. In that case, the last term, for an unbounded region $W_2$, saturates to $\log|G|$. The term $S_{\cal F}(\omega_{12}|\omega_{12}\circ E_1\otimes E_2)$ is bounded above by $2 \log |G|$ because, in contrast to the case where $\omega$ is invariant under $G$, here both conditional expectations have to be used to bound the relative entropy by convexity. In the present limit, we can argue this term always saturates the bound and is in fact equal to $2 \log |G|$. This is because we can use as a lower bound the relative entropy of a subalgebra formed by far away charged operators (the same subalgebra that one can use to show $S_{\cal F}(\omega_{2}|\omega_{2}\circ E_2)= \log |G|$) and an intertwiner subalgebra around the common boundary between the regions. Expectation values are independent for these subalgebras because the charged operators in the intertwiner have large fluctuations and we get $2 \log |G|$ for this term, independently of the size of $R$. Therefore we expect
\be
I_{\cal F}(1,2)-I_{\cal O}(1,2)=\log |G|-S_{\cal F}(\omega_{1}|\omega_{1}\circ E_1)\,.
\ee   
Hence, this is controlled by the same order parameter discussed above.  This approaches the result for unbroken symmetry $\Delta I=\log |G|$ for small size with respect to the symmetry breaking scale $\mu$, and $\Delta I\rightarrow 0$ in the infrared where the symmetry is completely broken. The mathematical necessity of this last limit can also be deduced because in the large size limit the two last terms in (\ref{ecui4}) saturate, pushing to saturation the first term of the right-hand side. 

If we have two separated regions $W_1$, $W_2$, with $R\mu \gg 1$ and $\epsilon\mu\gg 1$ we also have $\Delta I=0$. The intertwiner in ${\cal F}$ has non-vanishing expectation value but it does not convey any entanglement since the charged fields are already set to their vacuum expectation value.

Going through the derivation of the twist version of the order parameter in section \ref{upper} we conclude that this still applies\footnote{This is in the continuum limit. In one of the steps in that derivation we computed the entropy of the vacuum in the algebra of the twist and assumed this state was invariant under symmetries. This is still correct here in the continuum limit for the sharp twists used in the derivation.} but for the relative entropy
\be
S_{{\cal F}_{12}}(\omega|\omega \circ E_1)=\log |G|-S_{{\cal F}_S\vee G_\tau}(\omega|\omega\circ E_\tau)\,.
\ee 
The left-hand side is upper bounded by $\log |G|$ in the case of a finite group and will be dominated by the intertwiner entropy in competition with the entropy of the charged algebra inside $W_1$. For small enough $\epsilon$ it is expected that the intertwiner dominates and, even for continuous groups,  we get the same representation as in the symmetric case of the intertwiner relative entropy in terms of twists. 
 In the opposite case, we can take the ball $W_2$ to infinity and because of clustering we get the twist representation of the SSB order parameter
 \be
S_{{\cal F}_{1}}(\omega|\omega \circ E_1)=\log |G|-S_{{\cal F}_S\vee G_\tau}(\omega|\omega\circ E_\tau)\,,
\ee  
where now the twists are allowed to be as wide as we want outside $W_1$. These, however, do not gain by being wider than the symmetry breaking scale. The state $\tau |0\rangle$ represents a domain wall state with one vacuum in $W_1$ and another one in $W_2$. In consequence, there is an optimal width, and for large $R$ or large width, the twist expectation value will be exponentially suppressed.    

In order to understand the case of SSB of a Lie group symmetry we first study the simple model of a compactified free scalar that will play the role of the Goldstone boson in the IR.  

 \subsubsection{Free compactified scalar.}
\label{scalar}

 Let us take the algebra ${\cal O}$ of operators generated by the derivatives $\partial_\mu \phi$ of a free massless scalar field. The model contains a conserved current $J_\mu=\partial _\mu \phi$, where $J^0(x)=\dot{\phi}(x)=\pi(x)$, the conjugate momentum of the scalar.  The charge corresponding to this current is 
\be
Q=\int d^{d-1}x \, \pi(x) \,,
\ee
and we have  
\be
e^{i s Q} \phi(x) e^{- i  s Q }= \phi(x)+ s\,.
\ee
Then ${\cal O}$ is the subalgebra of the full scalar field corresponding to the elements invariant under this symmetry $\phi(x)\rightarrow \phi(x)+ s$. 

The net ${\cal O}$ has superselection sectors. Consider operators of the form
\be
V=e^{i \lambda^{-1} \, \int d^dx \, \alpha(x)\, \phi(x)}\,,\hspace{1cm}\int d^dx\, \alpha(x)=1\,. 
\ee 
 The parameter $\lambda$ has dimension $(d-2)/2$. Taking ${\cal I}=V_1 V_2^\dagger$, with the support of the smearing functions $\alpha_1$ and $\alpha_2$ included in  $W_1$ and $W_2$,  we see this operator belongs to the algebra ${\cal O}$, commutes with all operators outside $W_1$ and $W_2$, but cannot be generated additively in $W_1W_2$ inside ${\cal O}$. $V_1$ generates a superselection sector with automorphism $\rho(x)=V_1 x V_1^\dagger$ and ${\cal I}$ is an intertwiner. The endomorphism can be composed to give charges for all integers, $V_1^n$, $n\in Z$. 

We can form a field algebra ${\cal F}_\lambda$  generated  by ${\cal O}$
 and all operators $V^n$ for different $n$ and smearing functions.   
 ${\cal F}_\lambda$ corresponds to the fix point of the full algebra of the scalar field under the automorphisms $\phi\rightarrow \phi+2 \pi n  \lambda$, $n\in Z$. Hence, ${\cal F}_\lambda$ describes a compactified scalar, with compactification radius $\lambda$.  ${\cal O}$ is obtained from ${\cal F}_\lambda$ by taking the fix point under the rest of the  transformations
\be
\phi\rightarrow \phi+ k \lambda\,, \hspace{.5cm} V\rightarrow e^{i k} V\,,\hspace{1cm} k\in (-\pi,\pi)\,. 
\ee
Therefore there is a $U(1)$ symmetry between ${\cal F}_\lambda$ and  ${\cal O}$. Products of different $V$ are the charged operators. 

Algebras ${\cal F}_\lambda$ and ${\cal F}_{\lambda'}$ are not included in one another if $\lambda$ and $\lambda'$ are not integer multiples of one another.  ${\cal F}_\lambda$ has superselection sectors since it is the fixed point of ${\cal F}_{ m \lambda}$ for integer $m>2$ and the transformations $\phi\rightarrow \phi+ 2 \pi n \lambda$, $n=1,\cdots, m-1$. The full field algebra of the scalar field is reduced to ${\cal O}$ by the action of a non-compact group corresponding to the line $R$. 

The expectation values of charged operators in ${\cal F}_{\lambda}$ with respect to the $U(1)$ symmetry have non zero vacuum expectation values
\be
\langle V_\lambda^n \rangle=e^{-\frac{n^2}{2 \lambda^2} \alpha\cdot G\cdot \alpha}\,,\label{vn}
\ee 
where $G(x)\sim |x|^{-(d-2)}$ is the scalar correlator function. Therefore the $U(1)$ is broken spontaneously.\footnote{The Lagrangian $\frac{1}{2} \partial_\mu \phi \partial^\mu \phi$ is invariant under the symmetry. }

We investigate the difference in mutual informations between ${\cal F}_\lambda$ and ${\cal O}$ which can be investigated using the same tools developed so far. Let us first understand what to expect for $S_{{\cal F}_\lambda}(\omega_{1}|\omega_{1}\circ E_1)$. We can estimate this quantity with the relative entropy in the algebra of a set of operators $V^n$ included in $W_1$, and maximizing over the possible smearing functions. 
 As happens with the intertwiner algebra for a $U(1)$ symmetry in section \ref{U1} this algebra is represented as an Abelian multiplicative algebra of functions on $k=(-\pi,\pi)$, where the state $\omega_{1}\circ E_1$ is just the constant distribution $(2\pi)^{-1}$. The other state depends on the vacuum expectation values of $V^{n}$, eq. (\ref{vn}). We have to take wide smearing functions to get the maximal relative entropy. In analogy with section \ref{U1}, if the coefficient $\alpha\cdot G\cdot \alpha/ \lambda^2$ of $n^2$ in the exponent of (\ref{vn}) is small we get a relative entropy $\sim -1/2 \log (\alpha\cdot G\cdot \alpha/ \lambda^2)$.
 Calling 
 \be
 \mu=\lambda^{\frac{2}{d-2}}
 \ee 
 to the energy scale of $\lambda$,  
  this is the case when $R \mu \gg 1$.  We get
\be
S_{{\cal F}_\lambda}(\omega_{1}|\omega_{1}\circ E_1)\sim  \frac{(d-2)}{2}\log (R \mu)\,.
\ee 
 Then we see this order parameter goes slowly to infinity for large radius.
 For smaller radius the coefficient of the exponent in (\ref{vn}) is large, and the probability is concentrated in $n=0$ as for $\omega_{1}\circ E_1$. The relative entropy has a change of regime at $R\mu\sim 1$ and goes to zero for $R \mu\rightarrow 0$ as happens for the finite groups.

 In order to evaluate the mutual information difference let us investigate the contribution of the intertwiners. This model has the nice feature that we can explicitly compute their expectation values. 
We form a subalgebra of intertwiners using the integer powers of one mode $(V_1V_2^\dagger)^n$. This Abelian algebra ${\cal C}_{12}$ is represented by the functions $e^{i k n}$ with pointwise multiplication and range $k\in (-\pi,\pi)$. The expectation values are
\be
 \langle V_1^n V_2^{-n}\rangle= e^{-\frac{n^2}{2} \sigma^2}\,, \label{ene}
\ee
with 
\be
\sigma^2=\lambda^{-2} (\alpha_1\cdot G\cdot \alpha_1+\alpha_2\cdot G\cdot \alpha_2-2 \, \alpha_1 \cdot G \cdot \alpha_2 )\,.
\label{minsig}
\ee
 This has the general Gaussian form of (\ref{distic}) but here the expression is exact. 
Again, this gives, for small $\sigma^2\ll 1$ and through a Fourier transform, a Gaussian probability distribution in the variable $k$. The relative entropy with the state $\omega\circ E_{12}$, that has uniform probability density $1/(2\pi)$, is given by
\be
S(\omega|\omega\circ E_{12})_{{\cal C}_{12}} \simeq - \log(\sigma)\,.
\ee

In order to minimize $\sigma$ in (\ref{minsig}) $\alpha_1$ and $\alpha_2$ have to be near to each other lying along the boundary. The minimization depends on the geometry, essentially the total area available $A\sim R^{d-2}$ and the separating distance $\epsilon$, but is independent of the compactification radius $\lambda$. Then, we can use symmetric test functions approximately translational invariant along the boundary surface to get
\be
\sigma^2\sim (\lambda^2 R^{d-2}\,f(R/\epsilon))^{-1} \,.
\ee
The area factor within the brackets is dictated by dimensional reasons and the extensivity of the problem along the area. The factor $f(R/\epsilon)$ should be a slowly varying function resulting from the minimization in the shape of the test functions in the direction perpendicular to the boundary. This factor should ensure $\sigma\rightarrow 0$ for $\epsilon\rightarrow 0$, though at a slow pace.   
What we want to emphasize is that this cannot be further improved to be of the order $(\epsilon/R)^{d-2}$ as in the case of a the $U(1)$ symmetry with a conformal current in the UV studied in section \ref{U1}. 
It can also be checked it cannot be improved by taking a larger algebra formed by charge creating operators for different modes along the surface.

The twist version  tells a parallel story but it is easier to compute the dependence on $\epsilon$. The current corresponding to the symmetry is $J_\mu=\partial _\mu \phi$. The twists are then constructed with integrals of $j^0=\dot{\phi}=\pi$, 
\be
\tau_k= e^{i \lambda \,k\, \int d^{d}x \, \alpha(x)\, \pi(x)} \,,\label{df}
\ee
with $\alpha(x)$ integrating to $1$ in the time direction on $W_1$ and vanishing on $W_2$. The expectation value is
\be
\langle \tau_k \rangle= e^{- \frac{1}{2}\lambda^2 \,k^2\, \alpha\cdot G_\pi \cdot \alpha} \,.
\ee
As in  section \ref{U1}, these expectation values (and the twists (\ref{df})) are very good approximations for an exponent which is large around $|k| \gtrsim\pi$, because the result is not really symmetric under $k\rightarrow k+2 \pi$. It is important however that these expectation values are interpreted in terms of discrete probabilities for a conjugate variable $q\in Z$, and the twist represented as $e^{i q k}$. This allow us to set the value of the compactification radius.   
Following the same reasoning as in section \ref{U1} the twist algebra entropy is 
\be
S_\tau\sim \frac{1}{2}\log(\lambda^2 \alpha\cdot G_\pi \cdot \alpha)\,.  \label{484}
\ee
To estimate the argument of the logarithm and avoid integrals over coinciding points we can use the fact that the twist belongs to the neutral algebra and the vacuum state on the neutral algebra is invariant under transformations with the global charge. Then, up to terms depending of the precise smearing functions, $\alpha\cdot G_\pi \cdot \alpha$ is approximated by $\sim - \int_{W_1} dx\,\int_{W_2} dy\,\langle J_0(x) J_0(y)\rangle$, because when acting on the vacuum the integral $\int dx \alpha(x) J_0$ can be converted into the total charge that annihilates the vacuum in ${\cal O}$ by adding some complementary term with a smearing function crossing $W_2$. Then, as $\langle J_0(x) J_0(y)\rangle\propto -|x-y|^{-d}$, doing the two integrals in the direction perpendicular to the boundary first, the integral will be concentrated along the boundary and will be proportional to the area. But it will also have a subleading factor $\sim \int_{\partial W_1} d^{d-2}y\, |x-y|^{-(d-2)}$. Therefore we get
\be
\alpha\cdot G_\pi \cdot \alpha \simeq  R^{d-2} \log(R/\epsilon)\,.
 \ee
Hence we expect
\be
S_\tau\simeq \frac{d-2}{2}\log(\mu R)+ \frac{1}{2} \log(\log(R/\epsilon))\,.\label{tyu}
\ee
This is similar to the case of a general $U(1)$ symmetry with a conformal current in the UV given by (\ref{epifa}) but the cutoff has been replaced by the scale of compactification. The dependence on the cutoff is subleading but still divergent, as it must be since the size of the group is infinite. The reason of the difference with (\ref{epifa}) is clearly that the conserved current is not conformal, it has dimension $d/2$ instead of $d-1$, giving a much smaller charge fluctuation rate for short distances.  

The result (\ref{tyu}) also holds for $R$ smaller than the compactification scale provided the argument in the logarithm in (\ref{484}) is still large or equivalently (\ref{tyu}) is still positive. This curiously seems to require very small $\epsilon$ as we decrease the radius. For smaller $R$ and fix $\epsilon$ the intertwiner expectation value is very concentrated in the identity. We cannot use a continuous charge approximation to get the probability $p_k$, and these probabilities are given by a Fourier series with coefficients proportional to (\ref{ene}). The relative entropy goes to zero fast with $ \mu R \rightarrow 0$.

For the mutual information difference, we can follow the same reasoning as above for finite groups. For $W_2$  bounded the result is finite and then we take the limit of a large region $W_2$ with the rest of the geometry fixed. Then the last term in (\ref{ecui4}) should be canceled by a contribution to the first term given by the same charged fields as the ones contributing to the last term. After this cancellation, there is a competition between the intertwiner and a charged operator in $W_1$ to the first term. For $\epsilon$ small enough, as corresponds to a cutoff, the intertwiner dominates, and we should have      
\be
\Delta I \simeq \frac{d-2}{2}\log(\mu R)+ \frac{1}{2} \log(\log(R/\epsilon)) - S_{\cal F}(\omega_{1}|\omega_{1}\circ E_1)\,.
\ee
Then, according to the preceding discussion, for small $R \mu \ll 1$ we get   
\be
\Delta I \simeq \frac{d-2}{2}\log(\mu R)+ \frac{1}{2} \log(\log(R/\epsilon))\,.\label{123}
\ee
Note the negative sign of the first term for small $R\mu$ must be supported by a compensating sign of the second, and we need an exponentially small cutoff, as already remarked previously. For larger $\epsilon$ we expect $\Delta I$ to vanish in this regime of small $R\mu$.  
For large  $R \mu\gg 1$ we have instead
\be
\Delta I\simeq \frac{1}{2} \log(\log(R/\epsilon))\,.\label{321}
\ee
This does not vanish, in contrast to the case of a finite group. It has the same dependence on $\epsilon$ as in the UV.

We make some comments on previous results in the literature. 
In \cite{Casini:2014aia} there is a numerical study of the EE of a free Maxwell field in $d=3$. This model is equivalent to the algebra of derivatives of a free massless scalar through $\varepsilon_{\mu\nu\delta} F^{\nu\delta}=\partial_\mu \phi$. Then the relation between the scalar and the Maxwell models is the same as between the scalar and the algebra of its derivatives. The symmetry $\phi\rightarrow \phi+k$ is uncompactified and the model does not contain any scales. This is equivalent to the model ${\cal F}_\lambda$ in the decompactification limit $\lambda \rightarrow \infty$. 
Because of that if we evaluate the relative entropy  $S_{\cal F}(\omega_{1}|\omega_{1}\circ E_1)$ it is not finite and we get the divergent quantity $\frac{1}{2}\log (R \mu)$ as $\mu\rightarrow \infty$.  
In presence of a cutoff $\delta$, and with the naive lattice interpretation of the difference in entropies between the two models in place of the relative entropy (see the discussion in section \ref{tools}), the compactification radius should get trade off by the cutoff, and we get up to lattice ambiguities
\be
S_{\textrm{Maxwell}}(R)-S_{\textrm{scalar}}(R)=\frac{1}{2}\log(R/\delta)\,.\label{dife}
\ee
This is what was find in \cite{Casini:2014aia} numerically. 
 For the mutual information difference in the limit of small $\epsilon$ we get
\be
I_{\textrm{scalar}}(1,2)-I_{\textrm{Maxwell}}(1,2)\sim \frac{1}{2}\log(\log (R/\epsilon))\,.\label{mild}
\ee
This does not contain a $\log(R/\epsilon)$ term and because of that it does not reproduce the difference in lattice entropies (\ref{dife}) which would have given the contradictory result that the mutual information of the smaller model would have been bigger than the one of the larger model. We have checked numerically in the lattice following the methods in \cite{Casini:2014aia} this dependence of the difference of mutual information in the short $\epsilon$ limit and found agreement with (\ref{mild}).  This term should be attributed as a term $-1/2 \log (\log(R\epsilon))$ to the Maxwell field mutual information rather than the scalar which has a finite constant term. An analogous result is expected between the scalar and its derivatives (dual to higher form gauge fields) in any dimensions. 
Notice the mutual information of the free scalar is finite (as well as the one of ${\cal O}$)  even if there is a non-compact symmetry relating it to ${\cal O}$. This is different from what we expect for a non-compact symmetry that is not spontaneously broken. 

In \cite{Agon:2013iva} the authors study the change in entropy between a free compact scalar and an uncompactified scalar using the replica trick.  
For the mutual information this is given by the subtraction of (\ref{123}) and (\ref{321}) with (\ref{mild}). We get
\bea
I_{\textrm{compact scalar}}(1,2)-I_{\textrm{scalar}}(1,2) &\sim & \frac{d-2}{2}\log(\mu R)\hspace{1cm} R\mu\ll 1\,,\\
I_{\textrm{compact scalar}}(1,2)-I_{\textrm{scalar}}(1,2) &\sim  & 0\hspace{3.4cm} R\mu\gg 1\,.
\eea
 This coincides with the result of \cite{Agon:2013iva} for the difference in entropies. Then, using $\Delta I/2$ as a proper renormalized entropy our result differs from the one in \cite{Agon:2013iva} by a factor $1/2$. This factor is typical of the difference between $S$ and $\Delta I/2$ for classically correlated variables.  

In $d=3$ eqs. (\ref{123}) and (\ref{321}) are about the difference between a compact free scalar and an uncompactified Maxwell field. The compact scalar is dual to a compact Maxwell field, and hence eqs. (\ref{123}) and (\ref{321}) are about the difference between the mutual informations of a compact and an uncompactified Maxwell field. This change is essentially due to the existence of magnetic charges in the compact Maxwell field (which are DHR sectors in $d=3$). The change in universal terms of the entropy due to charges is a similar phenomenon that explains the difference in logarithmic terms for free Maxwell field and the Maxwell field in presence of charges in $d=4$. We will show in Part II, a companion paper, where we discuss gauge field SS, that this explains the difference in the coefficient of the logarithmic term of a free Maxwell field with respect to the anomaly that has been much discussed in the literature.  The logarithmic term in (\ref{123}) for the compact Maxwell field in $d=3$ is necessary for the validity of the $F$ theorem \cite{Klebanov:2011td,Klebanov:2011gs}.   

\subsubsection{SSB of Lie group symmetry}

Now we consider SSB for the case of a Lie group symmetry with a conformal current in the UV. Let us first discuss the order parameter $S_{\cal F}(\omega_{1}|\omega_{1}\circ E_1)$. For large radius, we can compute this quantity with the relative entropy in the algebra of compactified scalars, that are the Goldstone modes. As in the previous discussion we get
\be
S_{\cal F}(\omega_{1}|\omega_{1}\circ E_1)\sim \frac{{\cal G} (d-2)}{2}\log (R \mu)\,,\label{3121}
\ee
where $\mu$ is the SSB scale that is taken of the same order as the compactification radius, and ${\cal G}$ is the number of Goldstone bosons. 
 For smaller radius, we cannot use the Goldstone boson approximation any more, but we expect that the relative entropy has a change of regime at $R\mu\sim 1$ and goes to zero for $R\mu\rightarrow 0$ as happens for the finite groups and compact scalars. 

For the mutual information difference, we can follow the same reasoning as above for the compact scalar and finite groups. We take the limit of a large region $W_2$ with the rest of the geometry fixed and then $\epsilon$ small, as corresponds to a cutoff. The intertwiners dominate the rest of the contribution to the first term of (\ref{ecui4}) and using the results on section \ref{U1} we have      
\be
\Delta I= \frac{{\cal G} (d-2)}{2}\log (R/\epsilon) - S_{\cal F}(\omega_{1}|\omega_{1}\circ E_1)\,.\label{67}
\ee
Then, according to the preceding discussion, for small $R \mu$ we get   
\be
\Delta I \simeq \frac{{\cal G} (d-2)}{2}\log (R/\epsilon)
\ee
 as in the case with no SSB. For large $R\mu$ we have instead
\be
\Delta I\simeq -\frac{{\cal G} (d-2)}{2}\log (\epsilon \mu)\,.
\ee
This does not vanish, in contrast to the case of a finite group. It has the same dependence on $\epsilon$ as in the UV but the dependence on the radius has been replaced by the SSB scale. Note that both models contain the massless scalar contribution of the Goldstone modes in the IR on top of this difference. 

The EE in models with SSB of continuous symmetries was discussed in \cite{Metlitski:2011pr}. The authors compute the entanglement entropy in the case of SSB for a finite volume space. At finite size, symmetry is restored and the physics is the one of ${\cal F}$ but with the symmetric state (which does not satisfy clustering in the large volume limit). They show the EE for a large region of size $R$, contains precisely a term of the form (\ref{3121}) as $R\mu \gg 1$. This explained previous lattice simulations \cite{kallin2011anomalies}. This also coincides with the expectations from our calculations since the lattice expression for the relative entropy in (\ref{3121}) is for large $R\mu$
\be 
S_{\cal F}(\omega\circ E)-S_{\cal F}(\omega)\sim \frac{{\cal G} (d-2)}{2}\log (R \mu)\,.
\ee
 The symmetric state has the new term that shows up in the calculations of \cite{Metlitski:2011pr}. Note that the mutual information difference has completely different behavior. 

\subsection{Thermal states and the thermofield double}  
\label{thermofield}

Another important context in which to apply the previous ideas is that of thermal states. For finite quantum systems we can use the usual Gibbs ensemble
\begin{equation}
\rho_{\beta}^{R}=\frac{1}{Z}e^{-\beta H_{R}}=Z^{-1}\sum\limits_{i}e^{-\beta E_{i}}\vert E_{i}^{R}\rangle\langle E_{i}^{R}\vert\;,
\end{equation}
where $R$ stands for `right' system. The thermofield double (TFD) arises by duplicating the system with a `left' side, and it is defined by the following natural purification
\begin{equation}
\vert\textrm{TFD}\rangle =Z^{-1/2}\sum\limits_{i}e^{-\beta E_{i}/2}\vert E_{i}^{R},E_{i}^{L}\rangle\,.
\end{equation}
As described in \cite{Haag:1992hx}, thermal states and TFD can be naturally described in algebraic terms. Technically, one notices that thermal states can be defined through the KMS condition, which can be seen as a periodicity of correlation functions under shifts $\tau\rightarrow \tau +\beta$ of the imaginary time axis, and that such KMS states have two natural and commuting GNS representations, which become the previous right and left systems. More importantly, given an operator with support only on the left system $V_{L}$, we can find a dual operator in the right system which acts on the same way on the TFD,
\begin{equation}
V_{L}\vert\textrm{TFD}\rangle =JV_{R}J \vert\textrm{TFD}\rangle\;,
\end{equation}
where $J$ is an antiunitary operator.

In this context, we again seek to compute the difference between mutual informations associated to complete and neutral algebras. This context is somewhat simpler, in the sense that we do not need to partition the systems to define such mutual informations. We can directly compute the mutual information between right and left systems $I^{RL}_{\mathcal{F}}$ and $I^{RL}_{\mathcal{O}}$ respectively.\footnote{These are mutual informations between type I algebras describing Hilbert spaces in a tensor product.} By the very same reasons, the diference is given by the following relative entropy
\begin{equation}
I_{\mathcal{F}}^{RL}-I_{\mathcal{O}}^{RL}=S_{\mathcal{F}}(\omega_{\textrm{TFD}},\omega_{\textrm{TFD}}\circ E_{L}\otimes E_{R})\;.
\end{equation}
Also as before, this quantity can be computed in two dual ways. Let's do first the intertwiner version. To be precise, we again assume to have the charge creating operators of the regular representation $V^{i}$, with $i=1,\cdots, \vert G\vert$. These operators allow us to construct the subalgebra discussed in~(\ref{sec:esa}), defined as
\be
(a)\equiv\sum_{ij} a_{ij} V^i (V^j)^\dagger\;,
\ee
generated by the projectors $P_{ij}=V^i (V^j)^\dagger$. A lower bound to the relative entropy appears when restricting to such subalgebra. We thus need to compute the following correlation functions
\begin{equation}\label{PrPl}
\rho^{\omega_{\textrm{TFD}}}_{jl,ik}=\langle\textrm{TFD}\vert P^R_{ij}P^L_{kl}\vert\textrm{TFD}\rangle\,.
\end{equation}
To maximize such correlation functions we choose $P^L_{ij}=JP^R_{ij} J$. We are using also the relation $e^{-\beta (H_{R}-H_{L})/2}V_{R}\vert\textrm{TFD}\rangle =J V_{R}^{\dagger} J \vert\textrm{TFD}\rangle$ to arrive to
\begin{equation}
\rho^{\omega_{\textrm{TFD}}}_{jl,ik} =Z^{-1}\textrm{Tr}(e^{-\beta H_{R}/2}P^R_{ij}e^{-\beta H_{R}/2}(P^R_{kl})^{\dagger})=Z^{-1}\textrm{Tr}(e^{-\beta H_{R}}P^R_{ij}(-\beta/2)P^R_{lk})\;,
\end{equation}
where $P^R_{ij}(-\beta/2)\equiv e^{\beta H_{R}/2}P^R_{ij}e^{-\beta H_{R}/2}$ is the operator evolved over imaginary time. This expression is very convenient to study the high and low temperature limits of the difference in entropies. At high temperatures, $\beta\rightarrow 0$ we have $P^R_{ij}(-\beta/2)\rightarrow P^R_{ij}$ so that:
  \be
\rho^{\omega_{\textrm{TFD}}}_{jl,ik}\simeq  Z^{-1}\textrm{Tr}(e^{-\beta H_{R}} P^R_{ij} P^R_{lk}) = \delta_{jl} \, Z^{-1}\,\textrm{Tr}(e^{-\beta H_{R}} P^1_{ik})\,.
\ee
Neutrality of the Gibbs ensemble implies that: 
\be 
Z^{-1}\,\textrm{Tr}(e^{-\beta H_{R}} P^1_{ik})=Z^{-1}\,\textrm{Tr}(e^{-\beta H_{R}} E( P^1_{ik}))=\frac{1}{|G|}\delta_{ik}\,,
\ee
and
\be
\rho^{\omega_{\textrm{TFD}}}_{jl,ik}=|G|^{-1} \, \delta_{ik}\delta_{jl}\,.\label{md2}
\ee
We thus arrive to the same story as in section (\ref{lower}), where we should associate high temperatures with small distance $\epsilon$ between subregions. The state~(\ref{md2}) is invariant under conjugation with any unitary transformation operator of the form 
\be
D\otimes D^*\,.
\ee
This is a pure state 
\be
S(\omega)=0\,,\label{twisted112}
\ee
and $\omega_{\textrm{TFD}}$ is maximally entangled between the $L$ and $R$ sides in charge space at sufficiently high temperatures.

On the other hand, since the state obtained~(\ref{md2}) is the same as~(\ref{md}), the computation of the state $\omega_{\textrm{TFD}\circ E_{L}\otimes E_{R}}$ and its entropy is exactly the same resulting in 
\begin{equation}
I_{\mathcal{F}}^{RL}-I_{\mathcal{O}}^{RL}=S_{\mathcal{F}}(\omega_{\textrm{TFD}},\omega_{\textrm{TFD}}\circ E_{L}\otimes E_{R})\geq \log |G|\;.
\end{equation} 
 Since the relative entropy is bounded by above by the same number we conclude that at high temperatures the difference of mutual informations saturates to $\log |G|$.

The behaviour at low temperatures is markedly different. At low temperatures $Z^{-1}\,e^{-\beta H_{R}}$ approximates a projector into the vacuum state. In particular,
\begin{equation}
e^{-\beta H_{R}/2}V^i e^{-\beta H_{R}/2}\rightarrow 0\,.
\end{equation}
Equivalently, for the projectors $P_{ij}=V^i (V^{j})^{\dagger}$ we have
\begin{equation}
e^{-\beta H_{R}/2}V^i (V^{j})^{\dagger}e^{-\beta H_{R}/2}\rightarrow \frac{1}{|G|}\delta_{ij}e^{-\beta H_{R}}\,.
\end{equation}
This implies that the correlation function~(\ref{PrPl}) factorizes and the state is just
\be
\rho^{\omega_{\textrm{TFD}}}_{jl,ik}=\frac{1}{|G|^{2}} \, \delta_{ij}\delta_{kl}\,.
\ee
This is the identity matrix in the subalgebra $(a)$, and it is of course invariant under $ E_{L}\otimes E_{R}$. We conclude that at sufficiently low temperatures the state on the intertwiners is unperturbed by the conditional expectation and the associated relative entropy vanishes.

Therefore, as $T$ goes from zero to infinity, the relative entropy goes from zero to $\log |G|$. There is a priori no critical temperature for the transition, and indeed it can be smooth. Basically, the entropy increases whenever the temperature crosses a threshold in which particles of a given representation become thermally excited.

The discussion here technically applies for finite groups, but one could extend this intertwiner version to continuous scenarios as well. But for continuous groups, it is again easier to consider the dual twist version of the story. This twist version can be used to arrive at the previous $\log |G| $ result for finite groups, but it gives a simpler result in the general case. In the TFD context, the twist algebra is easily defined. We do not need to invoke the split property. It is just the globally defined unitary representation of the symmetry group $\tau_{g}=U_{g}$ acting on the first Hilbert space. In this case, there are no choices for the twist algebra. The density matrix in the twist algebra is invariant under the group transformations and then must be an element of the center of the group algebra determined by the probabilities of different sectors $q_r=\langle P_r \rangle$. These are the expectation values associated with the Casimirs of the group at temperature $T$. The result (\ref{sed}) directly applies, but now this is not an inequality but an equation,
 \be
I_{{\cal F}}(1,2)- I_{{\cal O}}(1,2)= -\sum_r q_r \log q_r+\sum_r q_r \log(d_r^2) \,.\label{chis}
\ee

At low temperatures, the TFD is just the unentangled product of vacuums, and each vacuum has to be invariant by itself. Therefore $q_r\rightarrow \delta_{r,1}$ and  (\ref{chis}) goes to zero. As we increase the temperature the entropy increases since the TFD is a now a coherent mixture of different irreducible representations,
\begin{equation}
\vert\textrm{TFD}\rangle =\sum\limits_{ij}e^{-\beta E_{i}/2}\vert E_{i}, r_{j}\rangle\otimes \vert E_{i}, \bar{r}_{j}\rangle \;.
\end{equation}
As the temperature increases, the twist algebra gains entropy, since its expectation values get contributions from different representations. At temperatures for which all representations are excited with the probabilities of the regular representation we get the $\log |G|$. We have discussed at the end of section \ref{lower} the reasons why these probabilities will be approached rapidly once we have enough excited states. 

Let us do a couple of comments here. We first remark that the consideration of the TFD at different temperatures $\beta$ is analogous to the consideration of the mutual information for two subregions in a QFT for a variable distance $\epsilon$ between them. In both cases, when the distance parameter goes to zero, the difference saturates to $\log G$, while for high enough distances it goes to zero. Second, the present approach gives a new perspective to the problems described in \cite{Harlow:2015lma}, concerning the CFT operators describing wormhole threading gauge fields in the bulk. In the present approach, these gauge fields are dual to our gauge invariant intertwiners formed by charged operators in the right and left sides of the thermofield double. Notice that in the model $\mathcal{F}$, such operators can be formed in an additive manner from the tensor product of the two CFT's. We will discuss this issue in more detail in Part II, where the necessary tools to consider local symmetries are developed.

\subsection{Replica trick}
\label{RE}   
   
Though this paper is focused on the operatorial approach, here we briefly describe the Euclidean approach using the replica trick. The modifications on the replica trick that are appropriate to compute quantities in ${\cal O}$ have been implicitly or explicitly used in the literature before (see for example \cite{calabrese2009entanglement,calabrese2011entanglement,Balasubramanian:2016xho,Agon:2013iva}).   

First, in the model ${\cal F}$ one uses the replica trick without any particular change to compute bare entropies and mutual information as a combination of entropies. The density matrix has an expression in terms of a path integral in the full Euclidean plane with boundary conditions on the two cuts just above and below the spatial region $W$. The boundary conditions are the value of the fields where the density matrix is evaluated. Let us call $W=\cup_{i=1}^m W_i$ to the different connected components and $W_i^\pm$ to the two boundaries of the cut in each region. 

Given this density matrix, for ${\cal O}$ one should take into account that the reduced density matrix should be projected to the additive algebra of ${\cal O}$ in each disjoint region. This can be done transforming the fields on each cut $W_i^\pm$ by the group element $g$, summing over $g$, and dividing by $|G|$ (or just averaging over the invariant measure of the compact group). Then the replica trick continues by computing the powers $\rho^n$ and taking the trace $\textrm{tr}\rho^n$. This is done by taking $n$ copies of the density matrix, which amounts to $n$ copies of the cut plane, and sewing the different boundaries in a periodic order. We have the freedom to redefine the fields on each cut plane by a transformation $g$ without changing the path integral, because of the invariance of the action, but we have to equalize the fields on the boundaries that are sewn together. This implies there are some of the transformations with elements $g$ in each region $W_i$ that can be eliminated but there are some combinations that cannot. The final result is that $\textrm{tr}\rho^n$ in ${\cal O}$ consists of an average of several different partition functions in the original model ${\cal F}$. If we have $m$ connected components for $W$ there are $n\, m$ independent group elements over which we average. But $n$ of them, one for each copy, can be eliminated by redefinition. We finally get $n (m-1)$ group element averaging. Each of these partition functions can be pictured as given by the expectation value of a Renyi twist operator as usual. But this is the Renyi twist operator in ${\cal F}$ combined with group twist operators across the different $W_i$ and copies.         

\subsection{The case of $d=2$}
\label{dosd}
In two dimensions there are two differences with respect to the previous discussions. The first one is that $W_1 W_2$ and the shell have the same topology of two intervals. The second is that due to the non-trivial statistics of the charged sectors, the SS do not necessarily come from a symmetry group, and they have to be described more generally by their dimensions $d_i$ and fusion rules.   

If we still consider the case where ${\cal O}$ is the fixed point of ${\cal F}$ under a symmetry group $G$, the case of $W_1$ a single interval and $W_2$ containing two semi-infinite regions covering the rest of the space, with a separation distance $\epsilon\rightarrow 0$ to $W_1$, for a massive theory we get twice the value corresponding to higher dimensions
\be
\Delta I= 2 \log G\,. 
\ee
This can be thought as an instance of (\ref{instance}), since the boundary of a single interval has now two disjoint connected components, and there are two sets of independent intertwiners connecting $W_1$ with the two parts of $W_2$. However, if the theory is conformal $W_2$ can be thought of as a single interval. In this scenario we obtain
\be
\Delta I=\log |G|\,.
\ee
If two intervals $W_1$ and $W_2$ touch each other on one side we get $\Delta I=\log |G|$ in both cases, conformal and massive.

 A case which can be computed exactly is the algebra of the current $j(x)$ in the line, that we identify with the chiral derivative of a massless free scalar in $d=2$, that is $j(x^+)=\partial_+ \phi(x^+)$, with $x^+=t+x$. The line we are considering can be thought of as a null line in the $d=2$ model. 
By bosonization, this is the same model as the one obtained by restricting the algebra ${\cal F}$ of a free chiral Dirac fermion field to the algebra ${\cal O}$ generated by the current. The group symmetry is the global charge $U(1)$ group. This is an example of the discussion in section \ref{U1}.  
  The field $j(x)$ is free with commutator 
\be
[j(x),j(y)]=i\, \delta'(x-y)\,,\label{commut}
\ee
and Hamiltonian
\be
H=\frac{1}{2}\int dx\, j^2(x)\,.
\ee
For any conformally invariant model as this one, the mutual information in vacuum is a function of the cross ratio of the end-points of the intervals
\be
\eta=\frac{(b_1-a_1)(b_2-a_2)}{(a_2-a_1)(b_2-b_1)}\in (0,1)\,,
\ee
where we have written $I_1=(a_1,b_1)$, $I_3=(a_2,b_2)$, as intervals on the real line. The mutual informations for two intervals in both models can be computed exactly. For the chiral fermion \cite{Casini:2005rm,Casini:2009vk,Longo:2017mbg} we have 
\be
I_{{\cal F}}(\eta)=-\frac{1}{6}\log(1-\eta)\,,\label{ffermion}
\ee
while for the current 
\be
I_{\cal O}(\eta)=-\frac{1}{6}\log(1-\eta) -g(\eta)\,,  
\ee
where $g(\eta)>0$ to have $I_{\cal O}(\eta)<I_{{\cal F}}(\eta)$.\footnote{$g(\eta)$ is called $-U(\eta)$ in \cite{Arias:2018tmw}. } The difference is 
\be
I_{{\cal F}}(\eta)-I_{\cal O}(\eta)=g(\eta)
\ee
and was computed for any $\eta$ in \cite{Arias:2018tmw}. In the limit of small $\epsilon$,  $ 1-\eta\sim (\epsilon/R)^2$, with $R$ the size of the interval, and we have 
\be
I_{{\cal F}}(\eta)-I_{\cal O}(\eta)=g(\eta)\sim \frac{1}{2}\log(-\log(1-\eta))\sim \frac{1}{2} \log(\log(R/\epsilon))\,.\label{sistea}
\ee
This coincides with the general result of section \ref{U1}.

If we have the usual relations that the entropy is the same for complementary regions in a pure global state we would have 
\be
S(I_1\cup I_3)=S(I_2\cup I_4)\;, \label{ruin}
\ee
where we are thinking in a compactified real line divided in four intervals. Completing this relation with the entropies of the intervals to get mutual informations,\footnote{The single interval entropies are $S(r)=(c/6) \log(r/\epsilon)+k$, where $r$ is the size of the interval and $k$ some constant.}  for a CFT in $d=2$ this translates into the symmetry relation \cite{Casini:2004bw}
\be
I(\eta)= I(1-\eta)- \frac{c}{6} \log\left(\frac{1-\eta}{\eta}\right)\,,\label{fds}
\ee
where $c$ is the central charge (summed over both chiralities). This symmetry property does not hold when there are superselection sectors which ruin (\ref{ruin}), and in particular is badly broken for the chiral current (not for the chiral fermion). This symmetry leads to
$g(\eta)=g(1-\eta)$  
while for the current we have $g(0)=0$ because large distance mutual information vanishes, and $g(1)$ is divergent. For the case of a finite symmetry group $I_{{\cal F}}(0)-I_{\cal O}(0)=0$ while $I_{{\cal F}}(1)-I_{\cal O}(1)= \log |G|$.

Relation (\ref{fds}) can be shown from the replica trick using modular invariance for non chiral models \cite{Cardy:2017qhl}.  In connection to this, it has been shown more generally that modular invariant models are complete (duality holds for two intervals) and the symmetry property (\ref{fds}) holds \cite{Xu:2018fsv}. 

It is to be noted that in $d=2$ it does not hold any more 
$\Delta I=S_{({\cal O}_{34})'}(\omega|\omega\circ E)
$
 where $E$ maps $({\cal O}_{34})'$ to ${\cal O}_{12}$. This is because in $d=2$ the algebra  $({\cal O}_{34})'$, in addition to the intertwiners, contains the twists, which are not in ${\cal F}_{12}$. It is expected that $S_{({\cal O}_{34})'}(\omega|\omega\circ E)$ has limit $2 \log |G|$ at the point of contact, instead of $\log |G|$ as happens with $\Delta I$.  

The case of general SS not necessarily coming from a symmetry group was treated in \cite{Longo:2017mbg,Xu:2018uxc,Xu:2018fsv}. Each sector has a statistical dimension $d_r$, which can be non integer. Generalizing the result for a group  we have for the limit of contact between the complementary intervals
 in a 2d-CFT
\be
S_{({\cal O}_{34})'}(\omega|\omega\circ E)= \log {\cal D}^2=\log \sum_r d_r^2\,.\label{redh} 
\ee   
${\cal D}^2=\sum_r d_r^2$ is called the quantum dimension of the model. In terms of the quantum dimensions of the SS, this formula is the same in any spacetime dimension. It is also the index of inclusion of algebras ${\cal O}_{12}\subset ({\cal O}_{34})'$ \cite{Longo:1994xe,Longo:1989tt,Kawahigashi:1999jz}. The result (\ref{redh}) holds for finite index. The dimensions $d_r$ is greater than $1$ and can be non integer. It can only take some specific values for $d_r<2$ \cite{Longo:1989tt}.

\section{Examples of intertwiner and twist bounds}
\label{bounds}

In this section, we compute some concrete examples of intertwiner and twists expectation values. The objective is to build up intuition on how they generally behave, in order to provide the best bounds available.  We describe how intertwiners are assimilated to edge modes localized near the boundary and how they tend to minimize the modular energy for nearly complementary regions, while they spread out for more distant regions. We also describe how the charge creating operators in the same region try to minimize their mutual entanglement, and so repel each other. Finally, concerning the squeezed twists we show that in the short $\epsilon$ limit they are exponentially suppressed by the area and have Gaussian expectation values for Lie groups.     

\subsection{Free fermion. Minimizing the modular Hamiltonian and edge modes}

In this subsection we explain, for the fermion field, how the unitary
intertwiner $\mathcal{I}$ of two complementary regions can be suitable
chosen in order to maximize its vacuum expectation value, i.e $\left\langle \mathcal{I}\right\rangle \simeq1$.\footnote{In general $|\left\langle \mathcal{I}\right\rangle| \leq1$ since $\mathcal{I}$
is unitary.} We consider the theory of a free fermion field and the $\mathbb{Z}_{2}$ 
symmetry $\psi\rightarrow -\psi$ discussed in section \ref{DHR}. The intertwiner
between a region $W_{1}$ and its complement $W_{2}=W_{1}^{'}$
can be written as
\begin{equation}
\mathcal{I}=V_{1}V_{2}^{\dagger}\,,\label{int_5}
\end{equation}
where $V_{i}\in\mathcal{F}\left(W_{i}\right)$ are fermionic unitary
operators made out of the fermion fields
\begin{equation}
V_{i}=\int d^{d-1}x\,\left[\alpha_{i}^{\dagger}\left(x\right)\psi\left(x\right)+\psi^{\dagger}\left(x\right)\alpha_{i}\left(x\right)\right]\,,\label{eq: ferm_int}
\end{equation}
where $\alpha_{i}$ are spinor valued functions supported in the region
$W_{i}$. Automatically we have $V_{i}=V_{i}^{\dagger}$ and in order
to have $V_{i}^{-1}=V_{i}^{\dagger}$ we must also impose 
\begin{equation}
\int d^{d-1}x\,\alpha_{i}^{\dagger}\left(x\right)\alpha_{i}\left(x\right)=1\,.\label{norm_u}
\end{equation}
Because the local field algebra $\mathcal{F}\left(W_{1}\right)$ satisfies
Haag duality, we can choose the unitary $V_{2}\in\mathcal{F}\left(W_{2}\right)$
as the (vacuum) modular conjugated of some new unitary operator $\tilde{V}_{1}\in\mathcal{F}\left(W_{1}\right)$\footnote{For fermionic nets, the modular conjugation $J$ must be replaced
by the twisted modular conjugation $ZJ$ \cite{carpi2008structure}. Regardless this
technicality, the outcome of the argument below holds. }
\begin{equation}
V_{2}=-i \,(Z J)^\dagger\tilde{V}_{1}^{\dagger} Z J\,,
\end{equation}
where $Z$ is $i^{F}$ and $F$ the fermion number. 
Since modular conjugation respects statistics, $V_{2}$ is
a fermionic operator if $\tilde{V}_{1}$ is fermionic.\footnote{Moreover, for free fields $V_{2}$ is of the form \eqref{eq: ferm_int}
if $\tilde{V}_{1}$ is of this form too.} Using the relations for the modular operator and modular conjugation for a fermionic model \cite{carpi2008structure},
we can rewrite the vacuum expectation value of the intertwiner \eqref{int_5}
as
\begin{equation}
\left\langle V_{1}V_{2}^{\dagger}\right\rangle =\left\langle V_{1}\Delta^{\frac{1}{2}}\tilde{V}_{1}^{\dagger}\right\rangle =\left\langle V_{1}\mathrm{e}^{-\frac{1}{2}\mathcal{H}_{F}}\tilde{V}_{1}^{\dagger}\right\rangle \,,\label{exp_int_mod}
\end{equation}
where 
\begin{equation}
\mathcal{H}_{F}=\mathcal{H}_{W_{1}}-\mathcal{H}_{W'_{1}}\,,
\end{equation}
is the full modular Hamiltonian. Expression \eqref{exp_int_mod} indicates that we can search for a maximum in the expectation value within the choices $\tilde{V}_{1}=V_{1}$. 
 The modular Hamiltonian is quadratic in the fermion
field operators\footnote{From now on, we omit the subscript $W_{1}$.}
\begin{eqnarray}
\mathcal{H} & = & \int_{W_{1}\times W_{1}}d^{d-1}x\,d^{d-1}y\,\psi^{\dagger}\left(x\right)H\left(x,y\right)\psi\left(y\right)\,,\\
H\left(x,y\right) & = & \sum_{k}\int_{-\infty}^{+\infty}ds\,u_{s,k}\left(x\right)\,2\pi s\,u_{s,k}^{\dagger}\left(y\right)\,,
\end{eqnarray}
where the spinor-valued functions $u_{s,k}$ are the (simultaneous)
eigenfunctions of the correlator kernel $C\left(x,y\right)=\left\langle \psi\left(x\right)\psi^{\dagger}\left(x\right)\right\rangle $
and the modular Hamiltonian kernel (for a review see \cite{Casini:2009sr})
\begin{eqnarray}
\int_{W_{1}}d^{d-1}y\,C\left(x,y\right)u_{s,k}\left(y\right) & = & \frac{1}{1+\mathrm{e}^{-2\pi s}}\,u_{s,k}\left(y\right)\,,\label{eig_corr}\\
\int_{W_{1}}d^{d-1}y\,H\left(x,y\right)u_{s,k}\left(y\right) & = & 2\pi s\,u_{s,k}\left(x\right)\,.
\end{eqnarray}
The modular energy of each mode is $2\pi s$. This is unbounded from above and below, $s\in (-\infty,\infty)$. The index $k$ labels possible degeneracies. We may choose them satisfying the ordinary
orthogonality and completness relations
\begin{eqnarray}
\sum_{k}\int_{-\infty}^{+\infty}ds\,u_{s,k}\left(x\right)\,u_{s,k}^{\dagger}\left(y\right)=\mathbf{1}_{N}\delta\left(x-y\right)\,, &  & x,y\in W_{1}\,,\\
\int_{-\infty}^{+\infty}dx\,u_{s,k}^{\dagger}\left(x\right)\,u_{s',k'}\left(x\right)=\delta\left(s-s'\right)\delta_{k,k'}\,,\label{orto_s}
\end{eqnarray}
and  $N$ is the spinor space dimension. To compute 
\eqref{exp_int_mod} we introduce new fermion operators
\be
\tilde{\psi}\left(s,k\right) = \int_{W_{1}}d^{d-1}x\,u_{s,k}^{\dagger}\left(x\right)\psi\left(x\right)\,,\hspace{1cm}
\psi\left(x\right) = \sum_{k}\int_{-\infty}^{+\infty}ds\,u_{s,k}\left(x\right)\,\tilde{\psi}\left(s,k\right)\,,\label{new_modes}
\ee
that satisfy the usual anticommutation relations 
\begin{eqnarray}
\left\{ \tilde{\psi}\left(s,k\right),\tilde{\psi}^{\dagger}\left(s',k'\right)\right\} =\delta\left(s-s'\right)\delta_{k,k'}\,, &  & \left\{ \tilde{\psi}\left(s,k\right),\tilde{\psi}\left(s',k'\right)\right\} =0\,.
\end{eqnarray}
The modular Hamiltonian is diagonal in these new modes
\begin{equation}
\mathcal{H}=\sum_{k}\int_{-\infty}^{+\infty}ds\,\tilde{\psi}\left(s,k\right)^{\dagger}\,2\pi s\,\tilde{\psi}\left(s,k\right)\,.\label{new_H}
\end{equation}
Using  \eqref{new_modes} we can rewrite \eqref{eq: ferm_int} as\footnote{From now on, we omit the subscripts $i$ in $V_{i}$ and $\alpha_{i}$.}
\begin{equation}
V=\sum_{k}\int_{-\infty}^{+\infty}ds\left[\tilde{\psi}\left(s,k\right)\tilde{\alpha}\left(s,k\right)^{*}+\tilde{\psi}^{\dagger}\left(s,k\right)\tilde{\alpha}\left(s,k\right)\right]\,,\label{new_v}
\end{equation}
where
\be
\tilde{\alpha}\left(s,k\right)  =  \int_{\mathbb{R}^{d-1}}d^{d-1}x\,u_{s,k}^{\dagger}\left(x\right)\alpha\left(x\right)
 \,,\label{mod_ft}
\ee
and the normalization condition \eqref{norm_u} is translated into 
\begin{equation}
\sum_{k}\int_{-\infty}^{+\infty}ds\,\left|\tilde{\alpha}\left(s,k\right)\right|^{2}=1\,.
\end{equation}
The vacuum correlators for these new modes can be easily obtained
from \eqref{eig_corr} and \eqref{new_modes}
\be
\left\langle \tilde{\psi}^{\dagger}\left(s,k\right)\tilde{\psi}\left(s',k'\right)\right\rangle   =  \frac{1}{1+\mathrm{e}^{2\pi s}}\delta_{kk'}\delta\left(s-s'\right)\,.\label{new_corr1}
\ee
Replacing \eqref{new_H} and \eqref{new_v} into \eqref{exp_int_mod}
and using \eqref{new_corr1} and the fact that
\begin{eqnarray}
\left[\mathcal{H}_{W'_{1}},V\right]=0 & \textrm{and} & \mathcal{H}_{F}\left|0\right\rangle =0\,,
\end{eqnarray}
an straightforward computations gives 
\begin{equation}
\left\langle V_{1}V_{2}^{\dagger}\right\rangle =\left\langle V_{1}\Delta^{\frac{1}{2}}V_{1}^{\dagger}\right\rangle =\sum_{k}\int_{-\infty}^{+\infty}ds\frac{\left|\tilde{\alpha}\left(s,k\right)\right|^{2}}{\cosh\left(\pi s\right)}\,.
\end{equation}
It is not possible to choose a charge creating operator that commutes with the modular Hamiltonian, which would imply $\left\langle V_{1}V_{2}^{\dagger}\right\rangle=1$ precisely, because it creates charges and the modular Hamiltonian is charged neutral. But we can choose a charged mode with a very small modular energy. As this formula clearly displays,  in order to maximize the expectation value of the intertwiner we have to construct
a wave packet with small modular energy, by localizing $\tilde{\alpha}\left(s,k\right)$ sharply around $s=0$. 
There is a lot of freedom in approaching this limit. 
For example, we can choose Gaussian wave packets 
\begin{equation}
\tilde{\alpha}\left(s,k\right)=\frac{\sqrt{p_{k}}}{\left(2\pi\right)^{\frac{1}{4}}\sqrt{\sigma}}\mathrm{e}^{-\frac{s^{2}}{4\sigma^{2}}-i\lambda_{k}s}\,, \label{gausi}
\end{equation}
with $\lambda_{k}\in\mathbb{R}$, $\sigma>0$, $0\leq p_{k}\leq1$
and $\sum p_{k}=1$, and we have 
\begin{eqnarray}
\left\langle V_{1}V_{2}^{\dagger}\right\rangle  & \underset{\sigma\rightarrow0}{\longrightarrow} & 1\,.\label{int_to_one}
\end{eqnarray}
As an example, we have $\left\langle V_{1}V_{2}^{\dagger}\right\rangle >0.99$
for $\sigma=\frac{1}{22}$.  Notice we still have the freedom to choose different phases and probabilities for different degeneracy parameter $k$, and the degeneracy parameters are in correspondence with the variables describing the boundary of the region \cite{Arias:2017dda}. 

After we have shown that the expectation value of the free fermion
intertwiner can be (asymptotically) maximized, we want to see how are
these ``maximized'' wave packets localized in position space. We
expect that such wave packets are more and more supported around the
boundary $\partial W_{1}$ as long as the expectation value \eqref{int_to_one}
approximates to $1$. Certainly, the modular conjugated operator $V_{2}$
will be located around $\partial W_{2}$. Here we will show some examples.

\subsubsection{$d=2$ chiral fermion and Rindler wedge}
\label{subsec:1+1-chiral-fermion}

\begin{figure}[t]
\begin{minipage}{7cm}
\begin{center}
\includegraphics[width=6.5cm]{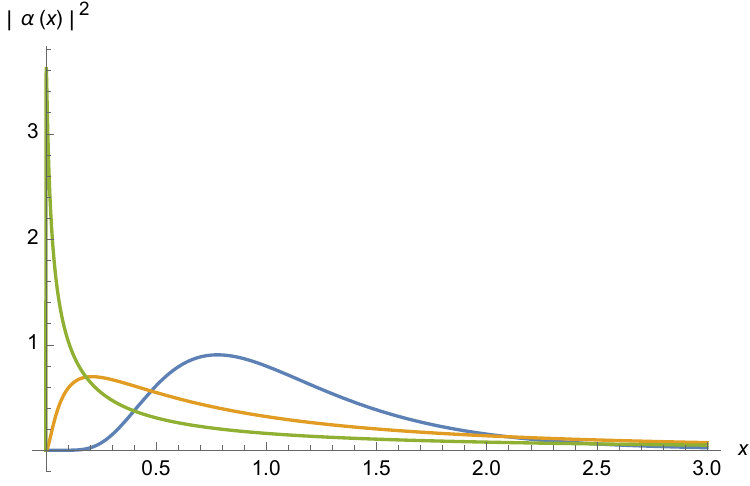}
\caption{Localization of the wave packet $\left|\alpha\left(x\right)\right|^{2}$
for $\lambda=0$ and different values of $\sigma=\frac{1}{5},\frac{2}{5},1$. }
\label{fig:2d_rw_var}
\end{center}
\end{minipage}
\ \
\hfill
\begin{minipage}{7cm}
\begin{center}
\includegraphics[width=6.5cm]{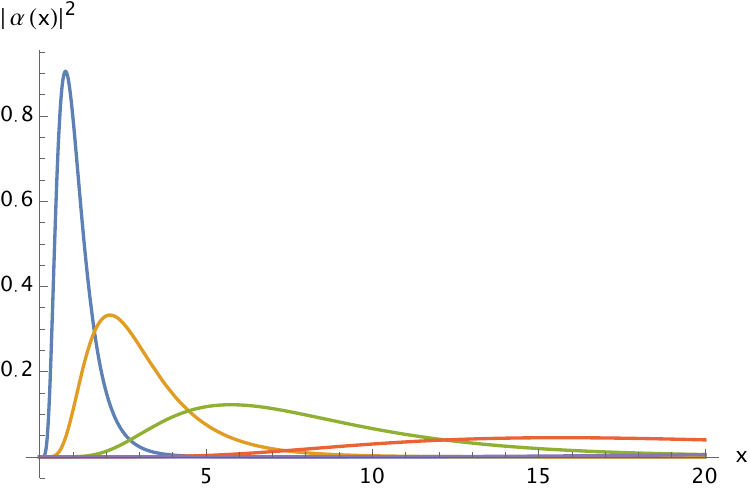}
\caption{Localization of the wave packet $\left|\alpha\left(x\right)\right|^{2}$
for $\sigma=1$ and different values of $\lambda=0,1,2,3$. }
\label{fig:2d_rw_disp}
\end{center}
\end{minipage}
\end{figure}

The normalized eigenfunctions are \cite{Arias:2018tmw}
\begin{equation}
u_{s}\left(x\right)=\frac{\mathrm{e}^{is\log\left(x\right)}}{\sqrt{2\pi x}}\,,
\end{equation}
and the ``modular'' Fourier transform \eqref{mod_ft} can be analitically
done 
\begin{equation}
\alpha\left(x\right)=\left(\frac{2}{\pi}\right)^{\frac{1}{4}}\sqrt{\frac{\sigma}{x}}\,\mathrm{e}^{-\sigma^{2}\left(\log\left(x\right)-\lambda\right)^{2}}\,.\label{wp_2d_rw}
\end{equation}
The probability densitity $\left|\alpha\left(x\right)\right|^{2}$
is an ordinary Gaussian wave packet but in the logarithmic variable $z=\log\left(x\right)$.
In figure \ref{fig:2d_rw_var} we plot the the wave packet \eqref{wp_2d_rw}
for a fixed $\lambda$ and different values of $\sigma$. Similarly,
in figure \ref{fig:2d_rw_disp} we plot the same wave packet for a
fixed $\sigma$ and different values of $\lambda$. As we can see
from figure \ref{fig:2d_rw_var}, as long as $\sigma\rightarrow0$,
the wave packet $\left|\alpha\left(x\right)\right|^{2}$ concentrates
around $x=0$. The job of the parameter $\lambda$ is to move 
the wave packet center of mass inside the region $x>0$. For a given $\lambda$,
the wave packet concentrates $\frac{1}{2}$ of the probability between
$0<x<\mathrm{e}^{\lambda}$.

\subsubsection{$d=2$ chiral fermion in one interval}

We set $W_{1}=\left(a,b\right)$ an interval. In this case, the normalized
eigenfunctions are \cite{Arias:2018tmw}
\begin{equation}
u_{s}\left(x\right)=\sqrt{\frac{z'\left(x\right)}{2\pi}}\mathrm{e}^{is z\left(x\right)}\,,
\end{equation}
where $z\left(x\right)=\log\left(\frac{x-a}{b-x}\right)$. Then,
the integral \eqref{mod_ft} can be analytically done 
\begin{equation}
\alpha\left(x\right)=\left(\frac{2}{\pi}\right)^{\frac{1}{4}}\sqrt{z'\left(x\right)}\sqrt{\sigma}\mathrm{e}{}^{-\sigma^{2}\left(\lambda-z\left(x\right)\right)^{2}}\,.\label{wp_2d_1}
\end{equation}
The probability densitity $\left|\alpha\left(x\right)\right|^{2}$
is a Gaussian wave packet in the variable $z$.
 As $\sigma\rightarrow0$ the wave packet get spread out in the variable $z$ and this means its probability is concentrated around the endpoints of the interval. The job of
the parameter $\lambda$ is to displace the center of the probability of the wave
packet in $z$ and this leads to an asymmetric distribution between the endpoints of the interval. In other
words, we can freely choose the intertwiner to be located
around $x=a$ or around $x=b$, or simultaneously around both endpoints
with some relative probability that can be chosen at will.
To be more precise, we first redefine the real parameter $\lambda$
as $\mu =\sqrt{2}\lambda\sigma$. Now, the
probability distribution
\begin{equation}
\left|\alpha\left(x\right)\right|^{2}=\left(\frac{2}{\pi}\right)^{\frac{1}{2}}z'\left(x\right)\sigma\,\mathrm{e}{}^{-\left(\mu-\sqrt{2}\sigma z\left(x\right)\right)^{2}}\,,
\end{equation}
has the limit
\begin{equation}
\lim_{\sigma\rightarrow0}\left|\alpha\left(x\right)\right|^{2}=q\cdot\delta\left(x-a\right)+\left(1-q\right)\cdot\delta\left(x-b\right)\,,
\end{equation}
where $q=q\left(\mu\right)=\frac{1-\text{erf}(\mu)}{2}\in\left[0,1\right]$
and $\text{erf}$ is the usual Gaussian distribution error function. If we combine different wave functions with different phases we see that as long as we are not interested in the precise form of the packet concentrated in the extremes of the interval, we have the freedom of a quantum mechanical wave function for a two dimensional Hilbert space seated at the two end points. 

\subsubsection{$d=2$ chiral fermion in $n$ intervals}

The multi-interval region is denoted as $W_{1}=\bigcup_{i=1}^{n}\left(a_{i},b_{i}\right)$,
where $a_{i}<b_{i}<a_{i+1}$. Following \cite{Arias:2018tmw}, we have that the
eigenfunctions 
\begin{equation}
u_{s}\left(x\right)=\frac{\left(-1\right)^{l+1}}{\sqrt{2\pi}}\frac{P\left(x\right)}{\sqrt{-\prod_{i}\left(x-a_{i}\right)\left(x-b_{i}\right)}}\mathrm{e}^{i s z\left(x\right)}\,,\quad x\in\left(a_{l},b_{l}\right)\,,
\end{equation}
where 
\begin{equation}
z\left(x\right)=\log\left(-\frac{\prod_{i}\left(x-a_{i}\right)}{\prod_{i}\left(x-b_{i}\right)}\right)\,,
\end{equation}
and $P\left(x\right)$ is a polinomial of degree $n-1$. Indeed, we
can choose $n$ linearly independent polynomials $P_{k}$ in order
to form an orthonormal basis. As happened for the one interval
case, for a wave packet in modular coordinates $\tilde{\alpha}\left(s,k\right)$
sharply localized around $s=0$, the above eigenfunctions make the
wave packet in position space 
to be highly localized around the interval endpoints $a_{i}$ and
$b_{i}$. To be more precise, lets take a single normalized eigenfunction.
Then, in analogy with the single interval case, we know that the wave
function 
\begin{equation}
\varphi\left(x\right)=\sqrt{\omega'\left(x\right)}\int_{-\infty}^{+\infty}ds\,\tilde{\alpha}\left(s\right)\mathrm{e}^{i s z \left(x\right)}
\end{equation}
is localized such as
\begin{equation}
\left|\varphi\left(x\right)\right|^{2}=\sum_{l=1}^{n}\left[q\,\delta\left(x-a_{l}\right)+\left(1-q\right)\delta\left(x-b_{l}\right)\right]\,,\label{deltas_n}
\end{equation}
in the limit when support of $\tilde{\alpha}\left(s\right)$ shrinks
to $s=0$.
The probability $q\in\left[0,1\right]$ above can be freely choosen.
Then the wave packet becomes
\begin{eqnarray}
\alpha\left(x\right) & = & \int_{-\infty}^{+\infty}ds\,\tilde{\alpha}\left(s\right)u_{s}\left(x\right)=\frac{r\left(x\right)}{\sqrt{z'\left(x\right)}}\varphi\left(x\right)\nonumber \\
 & \simeq & \sum_{l=1}^{n}\left[\sqrt{q}\frac{r\left(a_{l}\right)}{\sqrt{z'\left(a_{l}\right)}}\,\eta_{a_{l}}\left(x\right)+\sqrt{1-q}\frac{r\left(b_{l}\right)}{\sqrt{z'\left(b_{l}\right)}}\eta_{b_{l}}\left(x\right)\right]\,,\label{wp_n_1e}
\end{eqnarray}
where we have written
\begin{equation}
r\left(x\right)=\frac{\left(-1\right)^{l+1}}{\sqrt{2\pi}}\frac{P\left(x\right)}{\sqrt{-\prod_{i}\left(x-a_{i}\right)\left(x-b_{i}\right)}}\,,
\end{equation}
and $\eta_{c}\left(x\right)$ is a wave packet sharply concentrated
around $x=c$ and normalized according to $\int\left|\eta_{c}\left(x\right)\right|^{2}dx=1$.
For example, if we choose conveniently the normalized eigenfunction
given by
\begin{equation}
P_{k}^{\left(a\right)}\left(x\right)=\sqrt{\frac{\prod_{i}\left(b_{i}-a_{k}\right)}{\prod_{i\neq k}\left(a_{i}-a_{k}\right)}}\frac{\prod_{i\neq k}\left(x-a_{i}\right)}{\sqrt{-\prod_{i}\left(x-a_{i}\right)\left(x-b_{i}\right)}}\,,
\end{equation}
we have that $\frac{r_{k}\left(a_{l}\right)}{\sqrt{z'\left(a_{l}\right)}}=\delta_{kl}$.
Choosing $q_{k}=1$, eq. \eqref{wp_n_1e} simplifies to
\begin{equation}
\alpha\left(x\right)\simeq\eta_{a_{k}}\left(x\right)\,.
\end{equation}
Similarly, we can choose a  normalized eigenfunction such  that  $\frac{r_{k}\left(b_{l}\right)}{\sqrt{z'\left(b_{l}\right)}}=\delta_{kl}$. Choosing $q_{k}=0$, eq. \eqref{wp_n_1e} simplifies to
\begin{equation}
\alpha\left(x\right)\simeq\eta_{b_{k}}\left(x\right)\,.
\end{equation}
In other words, in the limit of small modular parameter $s\simeq0$,
there is always a wave packet for the fermion intertwiner localized
 around  any chosen endpoint. A general intertwiner
can be constructed as a superposition of such endpoints localized wave
packets 
\begin{equation}
\alpha\left(x\right)\simeq\sum_{l=1}^{n}\left(e^{i \phi_{a_l}}\,\sqrt{p_{a_{l}}}\,\eta_{a_{l}}\left(x\right)+e^{i \phi_{b_l}}\sqrt{p_{b_{l}}}\eta_{b_{l}}\left(x\right)\right)\,.
\end{equation}
According to the normalization relation \eqref{norm_u} for the wave
packet and the localization properties of the functions $\eta_{a_{l}}\left(x\right)$
and $\eta_{b_{l}}\left(x\right)$, the probablities $p_{a_{l}}$ and
$p_{b_{l}}$ and the phases $\phi_{a_l}$, $\phi_{b_l}$, can be freely choosen with the exception that they must
satisfy
\begin{equation}
\sum_{l=1}^{n}p_{a_{l}}+\sum_{l=1}^{n}p_{b_{l}}=1\,.
\end{equation}
Thus, we can picture these wave functions as a quantum mechanical degree of freedom in a Hilbert space with one basis vector for each end point.

\subsubsection{Massive fermion in the Rindler wedge}

We first treat the case $d=2$.
In this case, the theory does not decouples in chiralities and hence
the eigenfunction are $2$-dimensional spinor functions. We get
\begin{equation}
u_{s}\left(x\right)=\left(\begin{array}{c}
u_{s,+}\left(x\right)\\
u_{s,-}\left(x\right)
\end{array}\right)=\frac{1}{\pi}\sqrt{m\,\cosh\left(\pi s\right)}\left(\begin{array}{c}
K_{\frac{1}{2}-is}\left(mx\right)\\
-i\,K_{\frac{1}{2}+is}\left(mx\right)
\end{array}\right)\,,\label{ef_2d_m}
\end{equation}
where $K_{\nu}\left(z\right)$ is the modified Bessel function of
2nd kind. 
For $x\simeq0$ the eigenfunctions \eqref{ef_2d_m} become
\begin{eqnarray}
u_{s,+}\left(x\right) & \simeq & \frac{1}{\sqrt{2\pi x}}\mathrm{e}^{is\log\left(\frac{mx}{2}\right)}\textrm{ ,}\\
u_{s,-}\left(x\right) & \simeq & \frac{\left(-i\right)}{\sqrt{2\pi x}}\mathrm{e}^{is\log\left(\frac{mx}{2}\right)}.
\end{eqnarray}
Then the probability density $\left|\alpha\left(x\right)\right|^{2}=\left|\alpha_{+}\left(x\right)\right|^{2}+\left|\alpha_{-}\left(x\right)\right|^{2}$
behaves for wave packets localized near $x\simeq0$ as the sum of two distributions in
the variable $z=\log\left(\frac{mx}{2}\right)$. The analysis of the
localization of such wave packet in this regime follows from the massless
case (subsection \eqref{subsec:1+1-chiral-fermion}). 
 We again can form a Gaussian wave packet as in (\ref{gausi}). The result is the following.  In the limit of small dispersion in $s\sim 0$, $\sigma\sim 0$, we have a wave packet localized near $x\sim 0$, but the probability is $p_+\sim 1$ and $p_-\sim 0$, where $p_{\pm}$ are the probabilities for the two chiralities. The probabilities of the two chiralities are interchanged for $\lambda \gg 1/\sigma$. Hence we have the freedom to choose the proportion of chirality at will, but the wave packet becomes sharply localized around $x=0$ for $\sigma\rightarrow 0$.

For $d>2$, by dimensional reduction, the eigenfunctions can be decomposed into plane waves $\mathrm{e}^{i\bar{k}_{\Vert}\cdot\bar{x}_{\Vert}}$ in the parallel directions times the $d=2$ solutions with a mass $\sqrt{\bar{k_\Vert}^{2}+m^{2}}$. Then, given an arbitrary wave function $\alpha(\bar{x}_{\Vert})$ in the parallel direction, we can Fourier decompose it, and for each $\bar{k}_{\Vert}$ choose a mode in the $x^1$ direction concentrated in the origin and with high probability. Therefore the modes with high probability in $d$ dimensions will be localized near  $x^1=0$ but have arbitrary wave functions in the parallel directions.

\subsubsection{Spheres in CFT}
Suppose we have a sphere in a CFT and we have an Abelian sector with unitary charge creating operator inside the sphere. This has to be chosen such as to have almost zero modular energy. The sphere can be conformally mapped to a hyperbolic space \cite{Casini:2011kv}, with curvature scale $R$ and temperature $(2 \pi R)^{-1}$. Then the modular energy is just $2 \pi R H$, where $H$ is the ordinary Hamiltonian in the hyperboloid. To produce a $V$ with small modular energy the excitation has to be of low momentum. This requires it to be spread on regions much bigger than the curvature radius. On the other hand, it can be placed anywhere in the translational invariant hyperbolic space. However, once mapped back to the sphere it will be highly concentrated on the boundary of the sphere in Minkowski coordinates, but can be spread in angular coordinates.    

\subsection{Free examples for finite groups}
\label{freexamples}

In this section, we study a simple example of intertwiner lower bound for finite groups.  Let us think we have independent fermion fields $\psi_i$, $i=1,\cdots, N$, and consider symmetries that interchange the different fields. We can build charge generating operators using the same type of operator $V$ of eq. (\ref{siste}) used in the preceding section. To simplify calculations, and since we are not interested in the fermion character of the fields here, but on the permutation symmetries between different fields,  
we are going to use bosonic operators $B$ for each field. These we construct out of the product of two of the $V$ operators corresponding to non overlapping test functions in the same region, $B=i  V_x V_y$, 
 such that $B^2=1$, $B^\dagger =B$.\footnote{We could also think in scalar fields, where $B$ is the generator of the charge under the $\phi\rightarrow -\phi$ symmetry.} We also want these operators to have very small expectation value $\langle B \rangle\simeq 0 $, what can be done by taking modes with small correlation.
     
 We take operators $B_i$ for the different fields corresponding to the same mode, 
 where $i$ refers to the $i^{\textrm{th}}$ the field. These operators commute for different fields. As discussed above, we can choose $B_i^{1}$ and $B_i^2$ in  two complementary regions such that $\langle B_i^1 B_i^2\rangle\simeq 1 $. It is also clear that the expectation value of any $B_i B_j$ with $i\ne j$  vanishes. 

Let us consider the group $Z_3$ of ciclic permutations of the fields with $N=3$ fields, with $|G|=3$. Let us take the algebra generated  by the unitaries $B_1, B_2, B_3$ of the three fields corresponding to a single mode in $W_1$. The algebra contains $2^3=8$ operators. The algebra can also be described by orthogonal projectors. Calling
\be
P_i^\pm=\frac{1\pm B_i}{2}\,,\hspace{.5cm} (P_i^\pm)^2=(P_i^\pm)\,,\hspace{.5cm} (P_i^\pm)^\dagger=P_i^\pm\,,\hspace{.5cm} P_i^+ \,P_i^-=0\,,
\ee
we have the following set of $8$ orthogonal projectors as basis elements 
\be
P_{\pm\pm\pm}=P_1^\pm P_2^\pm P_3^\pm\,.
\ee 
We call more simply $P_\beta$ to these projectors, where $\beta$ is an index that can take $8$ values. 
There is an analogous algebra in region $W_2$. The vacuum state just gives non zero expectation value to the same projector in $W_1$ and $W_2$, and we have
\be
\omega(P_\beta^1 P^2_{\beta'})=\frac{1}{8} \delta_{\beta,\beta'}.
\ee  
This has entropy $S(\omega)=\log(8)$. 

Under the action of the group the eight projectors in $W_1$ are interchanged in the following form.  There are two regular representations (of three elements) spanned by the projectors with $\beta$ having two plus signs or with two minus signs, and two trivial representations due to the projectors with all signs equal. Each representation matches with one corresponding representation in $W_2$. Under the state $\phi=\omega\circ E$ all projectors $P^1_{\beta}P^2_{\beta'}$ of the same regular representation will have the same expectation value $(1/8)\times (1/3)$ because the conditional expectation mixes $\beta, \beta'$ on all the possible values of the representation. Then, each regular representation adds 
$-9 \times 1/8\times 1/3 \log(1/8 \cdot 1/3)$ to $S(\phi)$, while the trivial representations adds the same as for the entropy of $\omega$, that is, $-1/8 \log(1/8)$ each. Then we get the bound 
\be
\Delta I \ge S(\phi)-S(\omega) = \frac{3}{4} \log(3)\,.
\ee
This coincides with the general result (\ref{twisted}).\footnote{In section \ref{lower} we got $S(\omega)=0$ because we choose a bigger non-commutative algebra containing the projectors to the diagonal elements we are using here. Even if the two entropies change when we enlarge the algebra in this way, the relative entropy given by the difference $S(\phi)-S(\omega)$ does not change.} In the factor $3/4$ we recognize the total probability of the regular representations, which equals $6/8$. 

To improve this bound we add a new site (on each region), that is, we take two operators for each field, call then $B^\alpha_i$, where $\alpha=a,b$ represent two different modes. Let us assume that the modes are decoupled, that is $\langle B^a_i B^b_i\rangle\simeq 0$ such that there is no entanglement between the two modes. This will be automatic if the modes commute with each other, for example, if they are spatially separated since by monogamy of entanglement they cannot have correlations between them if they are maximally entangled with modes in the complementary region.  The algebra is spanned by a set of projectors 
\be
P_\beta=P_{\pm\pm,\pm\pm,\pm\pm}=(P^a_1)^\pm (P^b_1)^\pm (P^a_2)^\pm (P^b_2)^\pm (P^a_3)^\pm (P^b_3)^\pm\,.
\ee 
Now, when we apply the group transformations we will have a larger proportion of regular representations because there are four possibilities $(\pm\pm)$ for each field, and the three fields have to have equal this index in order not to have a regular representation. In general, taking $N$ independent sites we get that the probability of the regular representation is $\left(1-\frac{1}{2^{2N}}\right)$, and following the same calculation as above we arrive at
\be
\Delta I\ge \left(1-\frac{1}{2^{2N}}\right) \, \log(3)\,.
\ee
That is, our bound can approach $\log|G|$ as much as we want.

Different groups can be treated similarly. Let us take for example the non-Abelian group of permutations ${\cal S}_3$ of the three fields which has $|G|=6$. Using $N$ sites we again get for each field $2^N$ labels for the projectors. In order that, starting with one of the projectors, the permutations of the fields do not generate $3!=6$ different projectors, and hence the regular representation, it must be that at least two of the labels for the different fields are equal. Then the probability of the regular representation is the same as the probability of having the three labels different. As shown in section \ref{lower} the regular representation will always contribute $\log|G|$.  This  gives
\be
\Delta I\ge \frac{(2^N-1)(2^N-2)}{2^{2N}} \log 6\,.  
\ee
We need more an more sites for better precision, but the approach is exponentially fast.

It is evident that an example for the permutation group ${\cal S}_n$ can be constructed in the same way by using $n$ fields. Since each finite group is a subgroup of a permutation group, and the regular representation of the permutation group decomposes into regular representations of the subgroups, an example can be devised in the same lines for any finite group.

\subsection{Intertwiners at a finite distance. Repulsion of charged modes.}
We have seen the intertwiners are concentrated on the boundary for complementary regions. Here we want to show they will spread out in the coordinates orthogonal to the boundary if the two regions are separated. We cannot use now the modular reflection to obtain a good charge creating operator partner. Then we simply minimize the expectation value to obtain the optimal intertwiner. We still deal with the simple case of the symmetry $\mathbb{Z}_2$ of the free fermion. 

The vacuum expectation value of the intertwiner is
\bea
\langle V_{1}V_2\rangle & = & \int dx\, dy\, \alpha_1^i(x)\, \alpha_2^j(y)\, \left(\langle \psi_i(x)\psi_j^\dagger(y)\rangle -\langle \psi_j(y)\psi_i^\dagger(x) \rangle \right)\nonumber \\ & = & \int dx\, dy\, \alpha_1(x) (C(x-y)-C^*(x-y))\alpha_2(y)\,, \label{var_ifd}
\eea
where $C(x-y)=\langle \psi(x)\psi^\dagger(y)\rangle$ is the fermion correlator, and we have used that the support of the two functions is disjoint. For two balls with supports of size $R$ separated by a distance $L\gg R$ we have that this expectation value falls as $(R/L)^{d-1}$ in the massless case and exponentially in the massive one.  

Taking variations in \eqref{var_ifd} with respect to $\alpha_1$ and $\alpha_2$ with the constraints $\int \alpha_1^2=\int \alpha_2^2=1$ we get 
\be
\lambda_2 \,\alpha_2(y)=\int dx\, \alpha_1(x) (C(x-y)-C^*(x-y)) \,,
\ee
where $\lambda_2$ is a constant, the Lagrange multiplier. We have an analogous equation for $\alpha_1$. The solutions of these integral equations are generally not easy to obtain, but we can think for example in the easy case of very separated regions. In that case, the correlator function is almost constant when $x,y$ belong to each of the regions. Then it follows that the optimal distribution is given by constant functions $\alpha_1, \alpha_2$. Hence, the charged modes have spread as much as possible. One can easily compute the contribution to the entropy of this intertwiner and check for example that it is less than the mutual information for the fermion at large distances, while it has the same falling $L^{d-1}$ (in the massless case) with the distance $L$ between regions. 

As an example, for a chiral fermion, we can compute the relative
entropy for one intertwiner mode. In this case, we have a two dimensional
abelian algebra $\mathcal{A}_{1}=\left\{ \mathbf{1},U\right\} $ with the intertwiner\footnote{We introduce a $i$ prefactor in the definition \eqref{int_uni}
in order to $U$ be a unitary operator.}
\begin{equation}
U=i\mathcal{I}=iV_{1}V_{2}^{\dagger}=i\int dx\,dy\,\alpha_{1}\left(x\right)\alpha_{2}\left(y\right)\left[\psi\left(x\right)+\psi^{\dagger}\left(x\right)\right]\left[\psi\left(y\right)+\psi^{\dagger}\left(y\right)\right]\,.\label{int_uni}
\end{equation}
 For simplicity we have used real funtions $\alpha_{i}\left(x\right)$.
The vacuum expectation value, in the algebra $\mathcal{F}$, of such
operator $U$ is
\begin{eqnarray}
\left\langle U\right\rangle  & = & i\int dx\,dy\,\alpha_{1}\left(x\right)\alpha_{2}\left(x\right)\left[C\left(x-y\right)-C\left(y-x\right)\right]\nonumber \\
 & \simeq & -\frac{1}{\pi L}\left[\int dx\,\alpha_{1}\left(x\right)\right]\left[\int dy\,\alpha_{2}\left(y\right)\right]\,.
\end{eqnarray}
In the above expression we have used $C\left(x-y\right)=\left\langle \psi\left(x\right)\psi^{\dagger}\left(y\right)\right\rangle =\delta\left(x-y\right)+\frac{i}{2\pi}\frac{1}{x-y}\simeq\frac{i}{2 \pi L}$
for two far separated regions by a distance $L\gg R_{1},R_{2}$, where
$R_{i}$ are the sizes of the intervals $W_{i}$. For constants
functions $\alpha_{i}$ normalized according to $\int dx\,\alpha_{i}\left(x\right)^{2}=1$
we have $\int dx\,\alpha_{i}\left(x\right)=\sqrt{R_{i}}$, and hence
\begin{equation}
\left\langle U\right\rangle =-\frac{\sqrt{R_{1}R_{2}}}{\pi L}
\end{equation}
On the other hand, the vacumm expectation value of the opeartor $U$
in the algebra $\mathcal{O}$ is 
\[
\left\langle E_{12}\left(U\right)\right\rangle _{\mathcal{}}=\omega\left(E_{12}\left(U\right)\right)=0\,.
\]
The classical probability distribution of any state $\varphi$ in
the abelian algebra $\mathcal{A}_{1}$ is $\left(p_{1},p_{2}\right)$
where $p_{1}-p_{2}=\left\langle U\right\rangle $ and $p_{1}+p_{2}=\left\langle \mathbf{1}\right\rangle =1$.
Then we have the following two probabilities distributions
\begin{eqnarray}
\omega & \rightarrow & \left(\frac{1}{2}-\frac{\sqrt{R_{1}R_{2}}}{2\pi L},\frac{1}{2}+\frac{\sqrt{R_{1}R_{2}}}{2\pi L}\right)\,,\\
\omega\circ E_{12} & \rightarrow & \left(\frac{1}{2},\frac{1}{2}\right)\,.
\end{eqnarray}
Then, the relative entropy, restricited to the algebra $\mathcal{A}_{1}$,
between such states is
\begin{equation}
\left.S\left(\omega\left|\omega\circ E_{12}\right.\right)\right|_{\mathcal{A}_{1}}=\frac{R_{1}R_{2}}{2\pi^{2}L^{2}}\,,\label{srel_int_alg}
\end{equation}
which is strictly smaller than the mutual information in the field
algebra $I_{\mathcal{F}}\left(1,2\right)=\frac{R_{1}R_{2}}{6L^{2}}$
(see (\ref{ffermion})). The model ${\cal O}$ does not contain the fermion and its mutual information at large distances falls with a larger power than the one of the fermion. Hence, we can speculate on two reasons why \eqref{srel_int_alg}
does not coincide with the fermion mutual information at large distances. First, the two dimensional algebra $\mathcal{A}_{1}$
may be too small and second, our election of the intertwiner is not good
enough (for example, we can still make different elections multiplying
$U$ by any unitaries in $\mathcal{O}_{W_{1}\vee W_{2}}$).

As a commentary to the previous calculation in section \ref{freexamples}, if we choose the different charged modes $B_a$ on each region such that they are not independent to each other we clearly get a less optimal result. In the limit when these modes are maximally entangled with the complementary region, this just means we have to take non-overlapping modes for the different sites. However, if the two regions $W_1$, $W_2$ do not touch each other we cannot produce maximally entangled modes, and the charged modes will have a finite width in the direction perpendicular to the boundary. In this case, even if we have several sites on $W_1$ that are spatially separated, in general, the correlation of these modes will not vanish. To improve the result we need to diminish these correlations as much as possible since these correlations between charged modes in $W_1$  are not intertwiner correlations. This means the modes tend to repel each other in the direction parallel to the boundary in order to maximize the bound.  

\subsection{Sharp twists have Gaussian correlations with area law}
\label{twistcontinuo}
In general it is difficult to obtain an exact explicit expression for the twists that has all the desired properties. For example,  
for a $U(1)$ symmetry with a current $J^\mu$ we can write a twist operator for a shell around a ball $W_1$  and for the element $e^{i k Q}$ of the symmetry group as
\bea
\tau_k &= &e^{i\, k\, Q_1}\,,\\
Q_1 &=& \int d\Omega\, \int dr\, r^{d-2}\, \int dt\, \alpha(t)\,\gamma(r)\, e\, J^0(x)\,, \label{esp}
\eea
where $\gamma$, $\alpha$ are smooth smearing functions, $\gamma(x)=0$ for $r>R+\epsilon$, $\gamma(r)=1$ for $r<R+\epsilon/2$, and $\alpha(t)=0$ for $|t|>\epsilon/2$, while $\int dt\, \alpha(t)=1$. These elements form a group of unitaries, $\tau_{k_1} \tau_{k_2}=\tau_{k_1+k_2}$, and transform the charged elements inside the ball in the same way as the global symmetry group. However, $\tau_k$ is not periodic with period $2\pi$. To obtain this periodicity one should deform the twist inside the shell. This can be accomplished using the split property (see \cite{Doplicher:1984zz}) but the result would have a less transparent expression. For the $U(1)$ case, as was discussed in \ref{U1}, it turns out that the expectation values will fall fast with $|k|$ for small $\epsilon$ and the actual compactification radius in the variable $k$ does not affect the leading term in $\epsilon$ of the entropy. Hence, we will use the expression (\ref{esp}) in the following, and we consider the $\epsilon \rightarrow 0$ limit.

Let us consider the $U(1)$ symmetry first. Because of CPT, expectation values of odd powers of $Q_1$ vanish. For computing $\langle Q_1^2 \rangle$
 we use that, because of conservation, the correlation function of the currents writes
\be
\langle J^\mu(0) J^\nu(x)\rangle= (g_{\mu\nu} \nabla^2- \partial_ \mu \partial_\nu) \, H(|x|)\,.
\ee
For a CFT it is $H(|x|)\propto |x|^{-2 (d-2)}$.  Integrating by parts we get
\be
\langle Q_1^2 \rangle=\int d^dx\, d^dx'\,  \alpha(t)\, \alpha(t')\, \beta(r)\, \beta(r')\, H(|x-x'|)\,,
\ee
where $\beta(r)=\gamma'(r)$ has support in the shell. Keeping one point fixed and moving the other on the shell, the result is seen to be proportional to the area times the remaining integral. Because the result is dimensionless, in a CFT is universally given by 
\be
 \langle Q_1^2 \rangle= c \frac{R^{d-2}}{\epsilon^{d-2}}\,,
\ee
where the dimensionless constant $c$ depends on the precise shape of the smearing functions. 
If there are mass scales in the theory nothing changes for the leading term as far as $\epsilon$ is in the UV regime. 

To compute $\langle Q_1^4 \rangle$ exactly we should know the four-point functions of the current and these functions depend on the specific details of the theory. However, if we want to compute the leading term in $\epsilon\rightarrow 0$ we can argue as follows. Because of conservation and translation invariance, the four-point function of the charge density $J^0$ can be written as a combination of spatial derivatives of some functions $H$ of the coordinate differences. The bulk integral can then be integrated out to get integrals on the shell. One way to convince oneself of this is that each of the four $Q_1$ operators do not depend on the smearing inside the ball and the flux of the current can be written in a different Cauchy surface, while the shell part cannot be changed. Then, as above, in the thin shell the leading contribution comes from points of coincidence of the correlator functions $H$. But the behavior of $H$ at coincidence points can be read off from the points of coincidence of the correlators of $J^\mu$, and satisfy clustering properties. Then the leading term comes from two pairs of coincidence points and for each coincidence points we have the same contribution as for the two-point function. There are also three and four-point coincidences but these give subleading terms since we lose powers of the area. Since we have $3$ possible pairings between the four points the leading term should read
\be
  \langle Q_1^4 \rangle\simeq  3 \,c^2 \left(\frac{R^{d-2}}{\epsilon^{d-2}}\right)^2\,. 
\ee
With the same reasoning we see that for the purpose of computing the leading term for small $\epsilon $ in $\langle Q_1^n\rangle$ we can use Wick's theorem and think $Q_1$ is a free operator with Gaussian statistics. The same conclusion arises from thinking the charged fluctuations as a sum over a large number of independent fluctuations along the surface, and then using the central limit theorem. We then arrive to a Gaussian distribution\
\be
 \langle e^{i  \kappa Q_1} \rangle\sim e^{-k^2 \frac{\langle Q_1^2\rangle }{2}}\,.
\ee
For small enough $\epsilon$ only small $k$ hase non zero expectation value and the compactification radius does not affect the leading term in the entropy. As explained in section \ref{U1} this leads to 
\be
\Delta I\simeq \frac{1}{2}\log\langle Q_1^2\rangle\simeq \frac{d-2}{2}\log(R/\epsilon)\,.
\ee

More generally we expect that in the $\epsilon\rightarrow 0$ limit, twists for any finite group symmetry that affects the UV fix point should also have an area law
\be
\langle \tau \rangle \sim e^{-c \frac{R^{d-2}}{\epsilon^{d-2}}}\,.\label{leat}
\ee  
We can argue this has to be the case in the following way. For a sharp twist  
 the charges that the twist measures are formed by the tensor product of a large number $\sim R^{d-2}/\epsilon^{d-2}$ of independent charge fluctuations (representations) along the surface. These form a large representation of the group, and because of the arguments in section \ref{lower} this representation is mainly formed by copies of the regular representation except for a fraction of the Hilbert space that is exponentially small in the number of fused representations. The expectation value of the twist, for any element of the group except the identity, is zero for the regular representation. Then we get the leading behavior (\ref{leat}).

A well-known example is to take a QFT  and replicate it $N$ times. We can then take the orbifold by the symmetry under cyclic permutations of operators between copies. The twist operator for this symmetry is the Renyi twist operator \cite{Calabrese:2009qy}, with expectation value for small $\epsilon$
\be
\langle \tau_n\rangle= \textrm{tr} \rho_1^n=e^{-(n-1)S_n(W_1)}\simeq e^{-c \frac{R^{d-2}}{\epsilon^{d-2}}}\,, 
\ee 
where $S_n$ is the Renyi entropy of the region in the original model. Thus, the expectation value of the twist is exponentially small with an area law for the exponent. This coincides with the area law for EE.\footnote{The results of this paper for this particular scenario give
 $
 N I_{QFT}-I_{\textrm{Renyi orbifold}}=\log(N) 
$.
}

\section{Holographic EE and superselection sectors}
\label{holography}

In this section we want to describe how the previous approach to entanglement entropy in quantum field theory with SS gives a new perspective in the context of holographic entanglement entropy \cite{Ryu:2006bv}. The proposal made by Ryu and Takayanagi, later covariantly generalized in \cite{Hubeny:2007xt}, concerns the computation of entanglement entropy for holographic theories. In these scenarios, to compute the entanglement of a certain subregion, we just need to extremize the area over all bulk surfaces anchored on the boundary of such subregion. Once such surface is found, the entanglement entropy is just given by the usual Bekenstein-Hawking expression
\be\label{RT}
S=\frac{A}{4G}\;,
\ee
where $G$ is Newton's constant. 
 The holographic EE is an important generalization to black hole entropy formula and further, it gives an interpretation of the black hole entropy in terms of entanglement in the boundary theory. This is concretely realized in the thermofield double, as had been previously explained in \cite{Maldacena:2001kr}.

 The challenge is to understand this expression. There has been real progress in this direction. First, in \cite{Casini:2011kv}, it was shown that for spherical regions and the CFT vacuum, the RT proposal just reduces to the usual Bekenstein-Hawking formula of an unconventional black hole, so-called hyperbolic black holes \cite{Emparan:1999gf}. In this sense, for such spherical scenarios, the proposal reduces to one of the entries of the AdS/CFT dictionary, namely the one that relates thermal entropy with black hole entropy. This was extended for small perturbations of the state around the vacuum \cite{Blanco:2013joa}, where the matching of bulk and gravity calculations depend upon the validity of the Einstein equations in the bulk \cite{Lashkari:2013koa,Faulkner:2013ica}. 
 
 The holographic entropy for generic boundary regions and states was computed by Lewkowycz and Maldacena in \cite{Lewkowycz:2013nqa}, becoming the first generic proof of the previous formula. 
 The calculation in \cite{Lewkowycz:2013nqa} computes a Euclidean quantum gravity path integral that allows the computation of the entropy in a thermodynamic-like way and clarifies that the EE is universally given by a minimal area.  However, it rests on a bulk replica trick which does not provide a more detailed statistical origin. In particular, the reason why the non-local EE in the boundary should be given by a local expression on the minimal surfaces is still unclear. In other words, we lack a more transparent connection with the idea that the dominant part of entanglement should appear locally around the horizon \cite{Bombelli:1986rw}. 

New advances to understand the physics of such relation have arrived from two directions. In \cite{Freedman:2016zud,Harper:2018sdd,Cui:2018dyq}, such formula has been shown to be equivalent to a maximization procedure. One maximizes over vector fields with a fixed maximal density per unit area. The vector field is thought as representing a flux of ``bit threads'' crossing the bulk connecting the two entangled boundary regions. Such maximization procedure boils down to finding the minimal surface area traversed by the vector field, the `bottleneck', where the vector field is maximally packed. Although the result is finally the same, one interesting aspect of this approach is that it suggests there is nothing special at the bulk entangling surface, this surface being a property of all vector fields with fixed maximal density. The problem with this approach is that it is not understood what are these vector fields or bit threads, why their packing needs to be bounded, and what is their relation to actual field theory entanglement.

Another approach was developed in \cite{Almheiri:2014lwa,Pastawski:2015qua,Harlow:2016vwg}, where the EE formula, including the quantum corrections\cite{Faulkner:2013ana}, has been shown to be a generic feature of error correcting codes, connecting aspects of bulk reconstruction with holographic entanglement. The problem with this approach lies in its generality and, therefore, in its inability to understand the geometric nature of entanglement entropy in holographic scenarios and the physical origin of the main area contribution to entropy, as opposed to the previous approaches. Nevertheless, this approach fits very nicely with toy models of AdS/CFT based on tensor networks \cite{Pastawski:2015qua}.

In this section, we draw a parallelism between the ideas described in this paper on EE based on the superselection structure of QFT and holographic theories. Our proposal is that the holographic theories should be thought as complete theories but where there is a sub-theory describing the semiclassical physics with a very large number of superselection sectors. We do not know if this picture can be made exact for some models or has to be understood in an approximate sense in the large $N$ expansion. As we will see, the present perspective captures the advantages of the previous approaches at once, while it could potentially make the physical picture in the QFT side more transparent and concrete.  
After we describe the main idea we endeavour to make some precise constructions and compare with some 2d CFT's and large-N vectors models.

\subsection{A picture of minimal areas as measures of duality violation}
\label{area1}

Our first observation is that
holographic EE, with the RT prescription and its quantum correction, does not show problems of duality for any region. This is because we have equality for the entropies for arbitrary complementary regions in a pure global state.  Then we expect the boundary QFT (and its dual quantum gravity) to be a complete theory at the microscopic level, in the sense discussed through the present article. This excludes superselection sectors since any such structure would entail an entropy difference between complementary regions. As an example, this is the case of SYM theories in $d=4$. 

\begin{figure}[t]
\begin{center}  
\includegraphics[width=0.6\textwidth]{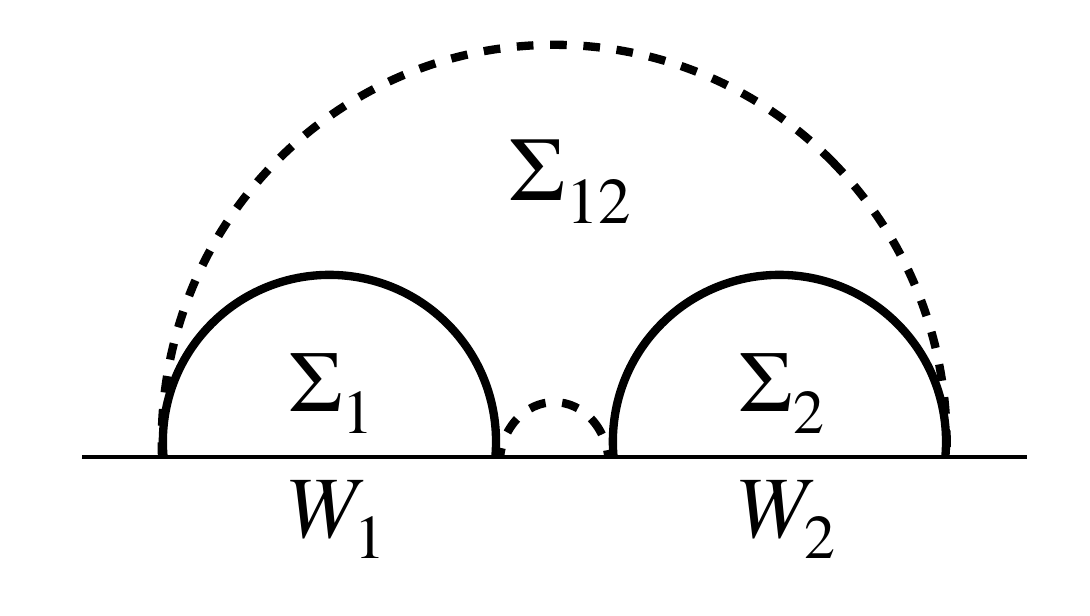}
\captionsetup{width=0.9\textwidth}
\caption{Two regions $W_1$ and $W_2$ in the boundary of AdS with its entangleent wedges $\Sigma_1$, $\Sigma_2$. The entanglement wedge $\Sigma_{12}$ of the union $W_1\cup W_2$ is bigger than $\Sigma_1\cup \Sigma_2$.}
\label{holo1}
\end{center}  
\end{figure}

While a Holographic theory is a complete theory, its gravity representation display in fact what appears to be a set of severe problems concerning the relations between algebras and regions for the semiclassical degrees of freedom.  To observe this aspect, we can resort to subregion-subregion duality, shown in fig~\ref{holo1}. We have three algebras-regions to consider. We call $W_1$, $W_2$, and $W_{12}=W_1\cup W_2$ to the boundary regions, $\Sigma_1$, $\Sigma_2$, and $\Sigma_{12}$, to the respective entanglement wedges, and  $A_1$, $A_2$, and $A_{12}$, to the areas of their respective boundaries.  
Let us also call ${\cal O}_{\Sigma}$ to the algebra of semiclassical bulk fields in the bulk surface $\Sigma$. If ${\cal F}_W$ is the full algebra of the boundary QFT fields in $W$, by subregion/subregion duality \cite{Czech:2012bh,Dong:2016eik,Jafferis:2015del} we have ${\cal O}_{\Sigma_W}\subset {\cal F}_W$. 
If we consider the algebra generated by the bulk semiclassical fields in $\Sigma_1$ and $\Sigma_2$, as usual we call it $\mathcal{O}_{\Sigma_1}\vee \mathcal{O}_{\Sigma_2}={\cal O}_{\Sigma_1\cup \Sigma_2}$. In the bulk this is represented as the union of the entanglement wedge associated to $W_{1}$ and the one associated to $W_{2}$, see fig \ref{holo1}. This gives rise to a disconnected bulk. Notice this algebra is not the algebra ${\cal O}_{\Sigma_W}$ for any boundary region $W$. Another algebra is the bulk field algebra ${\cal O}_{\Sigma_{12}}$ corresponding to $W_{12}$. This is the algebra of a connected surface on the bulk, as depicted in fig~\ref{holo1}. Then we have
\be
 \mathcal{O}_{\Sigma_{1}}\vee \mathcal{O}_{\Sigma_{2}}\subset {\cal O}_{\Sigma_{12}}\,,
\ee
but we do not have equality.
On the other hand $\mathcal{O}_{\Sigma_{12}}\subset {\cal F}_{W_{12}}= {\cal F}_{W_{12}'}' \subset (\mathcal{O}_{\Sigma_{W_{12}'}})'$
Therefore 
\begin{equation}
\mathcal{O}_{\Sigma_{W_1}}\vee \mathcal{O}_{\Sigma_{W_2}}\subset (\mathcal{O}_{\Sigma_{W_{12}'}})'\;,
\end{equation}
but not equality between these algebras.
This is one of the main observations of this section. The RT prescription, together with subregion/subregion duality, predicts a violation of duality for the subnet of bulk fields associated with the boundary regions. In turn, this indicates a non-trivial inclusion of algebras with an associated structure of superselection sectors. Hence, our proposal is that there is a sub-theory ${\cal O}$ of the full theory ${\cal F}$ which contains, at least, the bulk fields.

\begin{figure}[t]
\begin{center}  
\includegraphics[width=0.75\textwidth]{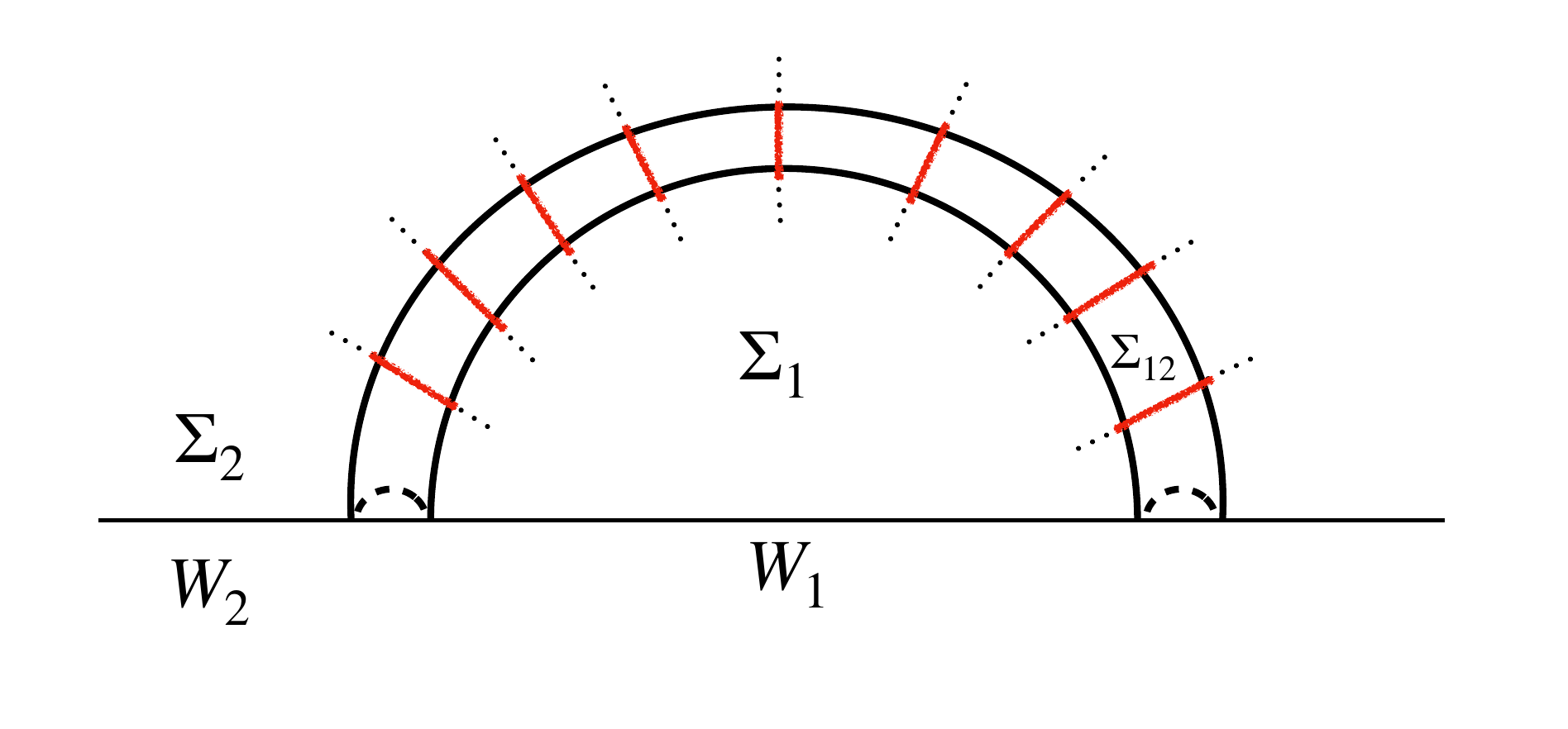}
\captionsetup{width=0.9\textwidth}
\caption{The region of $\Sigma_{12}$ that is not included in $\Sigma_1\cup \Sigma_2$ is a measure of duality failure. The intertwines crossing this region should carry an entropy bounded by the holographic entropy bound.}
\label{holo2}
\end{center}  
\end{figure}

As usual, one is interested in the size of this duality violation.  This size roughly measures the dimension of the space of intertwiners that belong to $ {\cal O}_{\Sigma_{12}}$ and not to $\mathcal{O}_{\Sigma_{1}}\vee \mathcal{O}_{\Sigma_{2}}$. This question has a natural holographic answer. Stripping out the bulk dual of $\mathcal{O}_{\Sigma_{1}}\vee \mathcal{O}_{\Sigma_{2}}$ from that of ${\cal O}_{\Sigma_{12}}$, we are left with the spacetime domain $\Sigma_{12}-\Sigma_1-\Sigma_2$ depicted in fig~\ref{holo2}. All bulk local fields with support in such domain should represent holographic intertwiners between regions $W_1$ and $W_2$. They do not belong to the algebra of bulk fields ${\cal O}$ formed additively in $W_{1}$ and $W_2$ but still commute with algebra ${\cal F}$ in the complementary region $W_{12}'$. The size of this domain is a clear measure of the amount of intertwiners we can find, and when $W_1$ and $W_2$ are nearly complementary to each other, this is deeply accounted by the area of the minimal surface (see figure \ref{holo2}).

To make this measure more precise we look at the entropies. Note the RT prescription states that EE is computed by the area $A_{\Sigma_W}$ of the boundary of the minimal surface $\Sigma_W$ anchored at the appropriate region $W$ on the boundary. Such an area is of $\mathcal{O}(c)$, where $c$ is the central charge. Quantum corrections to such formula first arise by considering the EE $S^0_{\Sigma}$ of bulk fields across the boundary of the surface $\Sigma$. Given a boundary region $W$ we have to compute $S^0(W)\equiv S^0_{\Sigma_W}$, the entanglement entropy of bulk fields across the minimal surface \cite{Faulkner:2013ana}. These corrections are of $\mathcal{O}(1)$. The final formula reads
\begin{equation}
S_{W}=\frac{A_{\Sigma_W}}{4G}+S^0_{\Sigma_W}\,.
\end{equation}
As usual with EE, this formula is technically ill-defined, since it is just infinity, the typical UV divergence in QFT seen holographically as an infinite volume in AdS space. The second term, the EE of bulk fields,  also provides an independent divergence local on the boundary of $\Sigma_W$, which is expected to be regularized by quantum gravity effects. There are counterterms in $S^0$ eliminating divergences which are fixed by the precise value of the Newton constant in the area term. 
To eliminate divergences we can consider mutual information instead of entropy. This must be well defined in the boundary and must therefore have a well defined bulk expression. This is
\be
I(W_1,W_2)=S_{W_1}+S_{W_2}-S_{W_1W_2}=(4 G)^{-1} \left(A_{\Sigma_{1}}+A_{\Sigma_2}-A_{\Sigma_{12}}\right) + \left( S^0_{\Sigma_1}+S^0_{\Sigma_2}-S^0_{\Sigma_{12}}\right)\,. \label{clesiano}
\ee
In the situation of fig \ref{holo2} the boundaries of the entangling surfaces in the bulk do not match and the term formed by $S^0$ still needs to be renormalized to compute the mutual information with (\ref{clesiano}).
  With a view in our proposal we rewrite this formula as
\bea
 I(W_1,W_2) &=& (4 G)^{-1} \left(A_{\Sigma_{1}}+A_{\Sigma_2}-A_{\Sigma_{12}}\right) +\left(S^0_{\Sigma_{1}\cup \Sigma_{2}}-S^0_{\Sigma_{12}}\right) \nonumber\\
 & + & \left(  S^0_{\Sigma_1}+S^0_{\Sigma_2}-S^0_{\Sigma_{1}\cup \Sigma_{2}}\right)\,.\label{dmn}
\eea
 Notice that $S^0_{\Sigma_{1}\cup \Sigma_{2}}$ is not the quantum correction for the EE of a minimal surface corresponding to any region in the boundary.     
This formula fits nicely with the idea that in the semiclassical holographic description of the full holographic theory ${\cal F}$, a certain subalgebra ${\cal O}$ is singled out which is related to bulk locality.  This subalgebra has superselection sectors. Hence, as was described in the paper, the formula for the mutual information in such a situation is\footnote{As we will discuss in the companion article, the relative entropy term, in appropriate scenarios, is responsible of the topological entanglement entropy.  Therefore, our proposal seems to be related to the computation of black hole entropy as a topological contribution \cite{McGough:2013gka}. }
\be
I_{\cal F}(W_1,W_2)= S_{{\cal F}_{12}}(\omega|\omega\circ E_1\otimes E_2)+I_{\cal O}(W_1,W_2)\,. \label{piry}
\ee
In this formula we are thinking there is a conditional expectation $E_{W}$ that maps the algebra ${\cal F}_W$ to the one ${\cal O}_{\Sigma_W}$ for any region.
The second term $I_{\cal O}(W_1,W_2)$ is the mutual information in the ``low energy'' sector containing the bulk quantum fields. This is
\be
I_{\cal O}(W_1,W_2)=S^0_{\Sigma_1}+S^0_{\Sigma_2}-S^0_{\Sigma_{1}\cup \Sigma_{2}}=\tilde{S}^0_{\Sigma_1}+\tilde{S}^0_{\Sigma_2}-\tilde{S}^0_{\Sigma_{1}\cup \Sigma_{2}}\,, \label{qqq}
\ee
where now we can use bare entropies $\tilde{S}^0$ instead of the renormalized ones $S^0$. 
 To compute this mutual information we have to use the entropy of the algebra ${\cal O}_{W_1}\otimes {\cal O}_{W_1}$ for the two regions, which is represented by $\tilde{S}^0(\Sigma_{W_1}\cup \Sigma_{W_2})$ in (\ref{qqq}), instead of the entropy of ${\cal O}_{\Sigma_{12}}$. This mutual information does not suffer phase transitions as we move the regions apart. Looking at (\ref{dmn})
we interpret the first term in (\ref{piry}) as
\be
S_{{\cal F}_{12}}(\omega|\omega\circ E_1\otimes E_2)= (4 G)^{-1} \left(A_{\Sigma_{1}}+A_{\Sigma_2}-A_{\Sigma_{12}}\right) +\left(S^0_{\Sigma_{1}\cup \Sigma_{2}}-S^0_{\Sigma_{12}}\right)\,.
\ee
 The entropies $S^0$ must be renormalized in this term.  This is dominated by the contribution of a ``high energy'' sector containing the intertwiners. It is essentially given by the area term at leading order. The subleading term must be there since for an excitation of the quantum fields in $\Sigma_{12}-\Sigma_1-\Sigma_2$, well separated from boundaries and thus not affected by renormalization, the entropy is interpreted as intertwiner entropy and has to contribute to the relative entropy.   

This interpretation means the area term is then precisely the order parameter for the failure of duality in the model ${\cal O}$. One special feature of holography is that the entropy would be essentially dominated by the intertwiner term that gives the leading contribution proportional to the central charge. This is of course not the case in ordinary theories with a small number of degrees of freedom, nor the case of theories with a large number of degrees of freedom but where the subalgebra is also large. We need a large number of superselection sectors, and that the charged fields with different representations should be most of the fields.   

Another simple consequence of this interpretation is that 
 the relative entropy between two states that are produced one from the other by acting with bulk quantum fields (elements of ${\cal O}$) is equal in the field algebra and the subalgebra, 
 \be
S_{\cal F}(\rho|\omega)=S_{\cal O}(\rho| \omega) \;.
\ee
 In the context of holography, this is the statement that boundary relative entropy equals bulk relative entropy \cite{Jafferis:2015del}.

A puzzling feature of the gravitational entropy is that it is generally believed to be microscopically determined by the high energy sector of the theory, and at the same time has an expression in terms of areas given by the low energy gravity field. In the present picture, this could be naturally explained in that the gravitational field (dual to the stress tensor) is part of ${\cal O}$, and the correlation functions in ${\cal O}$ in fact determine the full set of correlations functions in the full model. We give a more concrete picture of how can this happen in section \ref{cft2} below.  
According to the general theory of superselection sectors, the model ${\cal F}$ could in principle be reconstructed from ${\cal O}$.
 The would-be gravitational entropy  $S_{{\cal F}_{12}}(\omega|\omega\circ E_{12})$ would  be equal to $S_{{\cal O}_{(12)'}'}(\omega|\omega\circ E_{12})$ computable in ${\cal O}$. This identification, as well as formula (\ref{piry}), holds for any state invariant under $E$. However, it does not hold for states that are not invariant under the global conditional expectation $E$. For a state that is not invariant under $E$ we have to use eq. (\ref{ecui}), and the geometrical meaning of the different terms would be generally lost. However, we can be interested only in the effective state for low energy observables, and hence use $\omega\circ E$, where (\ref{piry}) holds again.   

There are however some differences with respect to the DHR case studied in the rest of the paper. One of this is that in principle the algebra ${\cal O}$ is a subalgebra of the full theory only in an approximate sense. Related to this approximation we also have that the net ${\cal O}$ is in general not additive for overlapping single component regions (see \cite{Duetsch:2002hc}). This departs from the simple DHR picture. Thus some generalization is needed for the holographic case. We will say a bit more on this on section \ref{gff} below. However, we note there is a simple scenario of CFT's in $d=2$ where a description in terms of SS as the one developed in this paper might apply exactly. This is discussed in section \ref{cft2}. Before these attempts to put the present ideas in more concrete grounds, we want to display the surprising qualitative connections between the intertwiners and the bit thread picture.

\subsection{Intertwiners, bit threads, and edge modes}

An interesting consequence of the proposal is that the main part of the EE comes from the intertwiners. Let us recall what we know qualitatively about the contribution of the intertwiners in a general case to the mutual information in the formula (\ref{piry}). This nicely fits with, and further may potentially clarify,  other discussions in the literature, in particular the bit thread picture to holographic EE of Freedman and Headrick \cite{Freedman:2016zud} and the idea of edge modes \cite{Lin:2017uzr,Donnelly:2016auv,Camps:2018wjf} that was inspired by some ideas about regularizing the entropy for gauge fields (see for example \cite{Casini:2013rba,Ghosh:2015iwa,Donnelly:2011hn}).

First, there is the fact that we are able to change the intertwiner algebra in many ways, and the best approximation to the correct result follows from maximizing their expectation values or maximizing the relative entropy with respect to the trivial state where they have zero expectation value.  Hence, as with bit threads, we have to maximize the contribution to entropy, and we are able to relocalize intertwiners if, for example, we change the position or the shape of a region.

Another feature of intertwiners is that the best ones are such that the modular energy of the created particles should be as small as possible (for the case of nearly complementary regions $W_1$ and $W_2$). While this means they tend to be packed near the boundary of the region in the perpendicular direction to the boundary, they can be very delocalized in the parallel direction to the boundary. In the dual holographic picture, this means their contribution can sense deep regions of the bulk. Which region of the bulk is precisely determined by the condition of having as low modular energy as possible. Then we should pack them near the minimal surface. The operators localized near the minimal surface in the bulk are barely moved by the modular flow, and represent large scale operators with low modular energy in the QFT (see \cite{Faulkner:2017vdd}).     

  As we have seen in section \ref{bounds}, intertwiners tend to be assimilated to edge modes in the boundary QFT near the boundary of the region that can be non-local in the parallel direction. These are physically macroscopic operators that connect in the bulk the two boundary regions $W_1$ and $W_2$. This seems to be very different to degrees of freedom in a local center in the bulk minimal surface due to the splitting of gauge degrees of freedom. These are cutoff ambiguities that should not modify the physics and cannot trade physical information between $W_1$ and $W_2$. In fact, these type of ambiguities are present for any cut of the bulk, and the information that the surface has to be minimal is lost in this idea.  

Another feature is that the maximization of entanglement entropy between the charge creating operators located at different regions generally implies that the entanglement between the different charge creating operators in the same region should be minimized. As discussed above, intertwiners repel each other. In the holographic picture, it is naturally suspected that this repulsion will take a local form and intertwiners will organize themselves when crossing the minimal surface to be Planck scale separated. This would lead to the bit thread picture. However, an attempt to show this expected feature cannot escape a more detailed understanding of the holographic QFT. 

A feature that one could expect to probe is that the density of intertwiner entropies should satisfy the holographic principle and entropy bound \cite{tHooft:1993dmi,Bousso:2002ju,Susskind:1994vu} through any surface in the bulk. This leads to a covariant version of the bit thread picture. In this way, one should arrive at a microscopic picture in which the non-local entanglement entropy in the boundary QFT turns out to be a local area in the bulk, but not any area but just the minimal one.   
From our perspective, the intertwiners are physical objects, operators that cross from region $W_{1}$ to $W_{2}$. Being physical objects, they carry a physical entropy which cannot violate the holographic principle in the bulk. This connects the majorization over intertwiner configurations with the minimization over bulk surfaces anchored in the boundary entangling surface, suggesting a max-min construction similar to \cite{Freedman:2016zud}.

\subsection{CFT families as superselection sectors}
\label{cft2}

Conformal field theories are surely the most important ground to test the previous ideas. In particular, they are the starting point to advance in the understanding of the entanglement structure in quantum gravity. As we describe now, there is an interesting approach to model the holographic algebras ${\cal F}$ and $\cal O$ in conformal field theories based on the theory developed so far. We will later argue that this approach sets the ground for further discussions around their holographic counterpart.

Notice that, given the specific models we have been considering, one might be tempted to conclude that all this framework is only applicable to systems with certain symmetries. This is indeed not always the case. We could have situations in which there is a certain conditional expectation $E$ not related to any symmetry group, but such that its net effect is to partition the system into an ensemble of superselection sectors. The question of whether a certain structure of superselection sectors can be understood as arising from an associated symmetry group has been extensively studied in the mathematical literature, see \cite{Haag:1992hx}. The answer lies in the so-called reconstruction theorems \cite{Doplicher:1990pn,Longo:1994xe}, and states that this is true for DHR sectors when the dimension is greater than $2$, and for BF sectors when the dimension is greater than $3$. Indeed, the discussion we present here can only be made precise for CFT's in $d=2$ (for specific reasons we comment in a moment). But, interestingly, we will be able to enlarge the approach to higher dimensions in an approximate sense by applying it to generalized free fields, to be described below.

Let's start with 2d CFT's.  The operator algebra of such theories is given by a set of primary fields $\mathcal{V}_{\Delta}$ and their descendants. Interestingly, for a given primary, all descendants can be obtained by linear combinations of the following basis of operators
\begin{equation}
\mathcal{V}_{\Delta}^{f}=U_{f}\mathcal{V}_{\Delta} U_{f}^{-1}\;,
\end{equation}
where $f$ is some conformal transformation (a diffeomorphism of the circle). In other words, if we are to generate the whole conformal family (the Verma module), we just need the primary and the generators of conformal transformations. In 2d this is exactly the algebra of the smeared energy-momentum tensor. In this way, every operator of the CFT can be written as
\begin{equation}\label{expansionCFT}
\mathcal{V}=\sum\limits_{\Delta}\mathcal{T}_{\Delta}\mathcal{V}_{\Delta}\;,
\end{equation}
where the sum runs over smeared primary fields $\mathcal{V}_{\Delta}$ and we have denoted by $\mathcal{T}_{\Delta}$ a generic operator constructed solely in terms of the energy-momentum tensor, i.e with the Virasoro algebra of the CFT. Such expression is clearly reminiscent of eq. (\ref{esis}), with  $\mathcal{T}_{\Delta}$ playing the role of observable algebra and $\mathcal{V}_{\Delta}$ the charged operators. Indeed, each primary defines a sector of the theory $\vert \Delta \rangle = \mathcal{V}_{\Delta}\,\vert 0\rangle$, and since each CFT family is an irreducible representation of the Virasoro algebra, applying $\mathcal{T}_{\Delta}$ to such sector will not take us away from it. Besides, since each primary is a local operator, we are in the DHR case, as could have been expected from the discussion in section \ref{DHR} since we are in $d=2$.

This perspective can now be taken one step further by defining a conditional expectation from the CFT algebra to the algebra of the energy-momentum tensor\footnote{In turn, such conditional expectation defines an inclusion of algebras $\mathcal{T}\subset \mathcal{F}_{\textrm{CFT}}$. In general, for theories in which the number of primary fields is infinite, the index associated to such inclusion will be infinite (see \cite{Longo:1989tt} for the definition and uses of the index), but for minimal models it might be an interesting quantity to study on its own.}:
\begin{equation}
E(\mathcal{V})=\mathcal{T}_{1}\;,
\end{equation}
where $\mathcal{T}_{1}$ denotes the (operator) coefficient of the identity in the expansion of the operator~(\ref{expansionCFT}). Notice that this is a true conditional expectation
\begin{eqnarray}
E(1)&=&1\nonumber\\
E(\mathcal{T}\mathcal{V}\mathcal{T'})&=&\mathcal{T}E(\mathcal{V})\mathcal{T'}\;,
\end{eqnarray}
albeit in general it does not arise as an average over any symmetry group.

These observations motivate the search for the subspace of states which is invariant under the conditional expectation. When considering symmetry groups, this was the space of group invariant states. Here we are going to argue that it is the subspace generated by all conformal transformation acting on the vaccum,\footnote{Such conformal transformations can be parametrized by two diffeomorphisms of the circle, corresponding to each light cone direction. Above we just labeled them by a generic $f$. Such continuous set of states $\vert f\rangle =U_{f}\vert 0\rangle$ can be seen as the generalized coherent states associated to the energy-momentum sector (see \cite{Caputa:2018kdj} for an application of this states to define quantum complexity in 2d CFT's).}
\begin{equation}
\vert f\rangle =U_{f}\vert 0\rangle\;.
\end{equation}
To show such claim we compute the one-point function of a generic operator $\mathcal{V}$ in such states:
\begin{equation}
\langle 0\vert U_{f}^{-1} \mathcal{V}  U_{f}\vert  0\rangle = \sum\limits_{\Delta} \langle 0\vert U_{f}^{-1} \mathcal{T}_{\Delta}\mathcal{V}_{\Delta}  U_{f}\vert  0\rangle = \langle 0\vert U_{f}^{-1} \mathcal{T}_{1}  U_{f}\vert  0\rangle = \langle 0\vert U_{f}^{-1} E(\mathcal{V})  U_{f}\vert  0\rangle\;.
\end{equation}
In the second equality we have used the fact that the inner product between states lying in different superselection sectors is zero. Therefore, for all sectors different from the vacuum, we know that $ U_{f}^{-1} \mathcal{T}_{\Delta}\mathcal{V}_{\Delta}  U_{f}\vert  0\rangle \in \mathcal{H}_{\Delta}$ and has vanishing inner product with the vacuum.

So if we focus on the vacuum sector, we have that, as discussed previously, the relative entropy between different states in such subspace is
\begin{equation}
S_{\cal F}(\rho\vert\omega)=S_{\cal T}(\rho_{\mathcal{T}}\vert\omega_{\mathcal{T}})\;,
\end{equation}
where $\rho_{\mathcal{T}}$ and $\omega_{\mathcal{T}}$ are the original states restricted to the algebra of the stress tensor. As discussed previously, we again see how this relation has little to do with gravity itself, and it is of much more general scope. In particular, notice that here it is valid for any CFT with any central charge.

Also, for the set of states $\omega^{f}\equiv\vert f\rangle \langle f\vert$, we can compute the EE by means of the formula
\begin{equation}
I_{{\textrm{CFT}}}(1,2)= S(\omega^{f}_{12}|\omega^{f}_{12}\circ E_{12}) + I_{\mathcal {T}}(1,2)\:.
\end{equation}
In the previous relation, the left-hand side is the mutual information between two intervals in the full CFT in the state $\omega^{f}$, while the second term in the right-hand side is the contribution associated to the algebra of the energy-momentum tensor. We remind that the previous observation does not mean that primary fields other than the identity do not contribute to the entanglement entropy. It just means that they contribute only through the term belonging to $\mathcal {T}$ that appears in the OPE of two primaries located one at each interval. In principle, in this $d=2$ setting, one could have a situation in which several complete consistent models ${\cal F}$ could be obtained from the subalgebra,\footnote{To make the relation with ${\cal F}$ unique may imply to take extended Virasoro algebras as subalgebras.} but the important point is that the non vanishing correlation functions are still expectation values in ${\cal T}$.  

We remark that the previous expression is exact. We leave its evaluation for future work since it requires further techniques than the ones presented so far. But the importance for us is first to notice that it has the same structure as the proposed quantum corrected version of holographic mutual information (\ref{piry}), for the case in which we have only the metric as bulk fields. In such a scenario, the relative entropy $S(\omega^{f}_{12}|\omega^{f}_{12}\circ E_{12})$, the ever-present actor of this article, is expected to compute minimal areas in the bulk. Indeed, notice that the contribution associated to the energy-momentum tensor is going to be $\mathcal{O}(1)$ in the large central charge limit, so by construction $S(\omega^{f}_{12}|\omega^{f}_{12}\circ E_{12})\propto \mathcal{O}(c)$. It can also be mentioned that the same idea also holds for larger subalgebras including the stress tensor where the fusion of the primaries closes in itself. For this case, where no approximation is made in the definition of the subalgebra, we do not have problems of additivity for overlapping single intervals in the bulk. This is related to the fact that there are no gravitons living in the bulk in $d=3$.      
\subsection{Generalized Free Fields and Holographic Entanglement Entropy}
\label{gff}

As described above, the reconstruction theorem informs us that \emph{any} type of DHR superselection sector structure in QFT in $d\geq 3$ arises as due to the existence of a field algebra and a group of symmetries acting over it. This suggests we cannot extend the previous discussion in the context of 2d CFT's to higher dimensions. Indeed, physically, the reason is that in higher dimensions the algebra of energy-momentum tensor does not close.\footnote{Though we cannot discard it will close in a subalgebra with other operators forming the neutral part under some large symmetry group. This is the case of large-N vector models, which we will treat below.} 

But although we cannot exactly apply it, it turns out that we can apply it approximately in the following way, that parallel the description in section \ref{area1}. The trick to evade the reconstruction theorem is to consider theories whose relevant degrees of freedom are generalized free fields (GFF). GFF are defined as fields which satisfy Wick's theorem so that their correlation functions factorize into products of two-point functions, but they do not obey any linear wave equation. In the context of CFT's, this implies that the fields can have any scaling dimension we wish.\footnote{Notice that true scalar free fields in CFT's must have dimension $\Delta =\frac{d-2}{2}$ due to the wave equation.}

One's inner desire is that such GFF close an algebra. In such case, we could rigorously import the developed techniques to this important scenario. But as it is well known this is not the case. From the GFF point of view, the reason was nicely explained in \cite{ElShowk:2011ag}, and it is due to the following fact. For the correlation functions of a primary field $\mathcal{V}_{\Delta}$ to obey Wick's theorem, the spectrum of its four-point function must contain a tower of fields $\mathcal{V}_{nl}$ with dimensions $\Delta_{nl}=2\Delta +2n+l$. For free fields obeying a wave equation, such tower contains the stress-tensor, but for generic GFF this is not the case. This is a problem, since every primary field couples to the energy-momentum tensor with an OPE coefficient $C_{T\mathcal{V}\mathcal{V}}\propto \frac{\Delta}{\sqrt{c}}$, and this coupling destroys Wick's factorization. The only left-out possibility for this GFF to exist is that they emerge approximately in CFT's with a large central charge. In such a scenario, it is clear that multiplying enough GFF will contain fields with scaling dimension of $\mathcal{O}(c)$. These fields couple strongly to the energy-momentum tensor and do not satisfy Wick's theorem.

In this scenario, we can follow two almost equivalent ideologies. The first is to approximately divide the field spectrum into those fields with $\Delta\sim\mathcal{O}(1)$ and those with $\Delta\sim\mathcal{O}(c)$. The problem is that the low dimension set is not a proper algebra. To convert it into a proper algebra we can further define a subspace of the Hilbert space in which we are going to consider the evaluation of EE. This is  the `code subspace' in \cite{Dong:2016eik,Pastawski:2015qua,Harlow:2016vwg}. Then we can project the set $\Delta\sim\mathcal{O}(1)$ to such subspace and this would produce a proper algebra. At any rate, at the level at which we will carry the discussion, what matters is that we take the following physically motivated assumption, that any operator of the theory can be written as
\begin{equation}\label{GFF}
\mathcal{V}=\mathcal{V}_{L}+\sum\limits_{\Delta\sim\mathcal{O}(c)}\mathcal{V}_{L}^{\Delta}\mathcal{V}_{\Delta}\;.
\end{equation}
In the previous expression, $\mathcal{V}_{L}^{\Delta}$ represents the low dimension operator coefficient accompanying the high dimension operator $\mathcal{V}_{\Delta}$. In turn, $\mathcal{V}_{L}$ is the low dimension operator coefficient accompanying the identity. One should compare~(\ref{GFF}) with~(\ref{expansionCFT}). Indeed, as in~(\ref{expansionCFT}), we define the conditional expectation as the projector onto such identity coefficient
\begin{equation}
E(\mathcal{V})=\mathcal{V}_{L}\;.
\end{equation}
To show this is a true conditional expectation for \emph{all} observables we need to define a code subspace and project the low dimension set there. But for us, it will be enough that indeed
\begin{eqnarray}
E(1)&=&1\,,\nonumber\\
E(\mathcal{V}_{L}\mathcal{V}\mathcal{V}_{L}')&=&\mathcal{V}_{L}E(\mathcal{V})\mathcal{V}_{L}'\;,
\end{eqnarray}
whenever the dimension of $\mathcal{V}_{L}E(\mathcal{V})\mathcal{V}_{L}'$ is of $\mathcal{O}(1)$ in the large central charge limit.

As for the stress tensor in 2d CFT's, the important thing now is to identify the set of states invariant under the conditional expectation. The answer here is simpler, this is the set of semiclassical states, plus small deviations generated by the GFF. This can be seen in two parallel ways. First, notice that correlation functions are exponentially suppressed by the dimensions of the associated operators, so they will vanish for heavy operators in the large central charge limit. We can again use the intuition coming from superselection sectors. Applying some heavy operator to certain semiclassical state takes us to a different semiclassical state, and so orthogonal to the one we started with. Again, this cannot be precisely correct, since semiclassical states have non-vanishing inner products. But these are usually exponentially suppressed in the central charge and we can disregard them.

So again, if we consider the relative entropy between two states separated by the action of GFF we obtain
\be
S(\rho|\omega)=S(\rho_{\textrm{GFF}}| \omega_{\textrm{GFF}}) \;,
\ee
where the subscripts stand for the states restricted to the GFF algebra. In the context of holography, we recognize this as the statement that boundary relative entropy equals bulk relative entropy \cite{Jafferis:2015del}. But as we have seen through the article, this is not really a special feature of gravity, but a more generic structure that appears whenever there is a conditional expectation and the states considered are invariant under it. Such conditional expectations arise naturally in systems with symmetries, but also should be the case of holographic CFT's with large central charges. 

Also, for the set of weakly perturbed semiclassical states $\omega\equiv\vert \textrm{GFF},s\rangle \langle \textrm{GFF},s\vert \equiv U_{\textrm{GFF}}\vert s\rangle \langle s\vert U_{\textrm{GFF}}^{-1}$, where $s$ stands for the semiclassical state and $U_{\textrm{GFF}}$ is a weak perturbation constructed by means of GFF, we can compute the EE by means of the formula
\begin{equation}
I_{{\textrm{CFT}}}(1,2)= S(\omega_{12}|\omega_{12}\circ E_{12}) + I_{\textrm{GFF}}(1,2)\:.
\end{equation}
As before, it is important to remember that this does not imply that heavy fields do not contribute to EE, just that they do so through the term belonging to $\textrm{GFF}$ that appears in their OPE.

In relation to holographic EE, our claim is now obvious. The second term in such expression is the mutual information of bulk fields. This is true by construction since we define entanglement in the bulk by the associated relative entropy. This is of $\mathcal{O}(1)$ in the large central charge limit. The first term should then be the area term in (\ref{dmn}). 

Before moving into a more holographic description of the physics, let us do several observations. First, notice that this first term is the `topological' term in the DHR discussions. 
Second, this perspective deepens the connection between entanglement and geometry, as proposed in \cite{Maldacena:2001kr,VanRaamsdonk:2009ar,Maldacena:2013xja}. In this case, the area appears as a measure of the macroscopic difference between the actual state $\omega_{12}$ and the state $\omega_{12}\circ E_{12}$, which arises when we disentangle all high energy operators. It seems that this entanglement between high dimension CFT operators is the glue of spacetime and it is measured by such relative entropy. 

Finally, to compute the entropy, we would need to find an algebra of intertwinners and configurations that maximize this relative entropy. This algebra of intertwinners is basically the algebra of high dimension operators. Consider a product of charge creating operators $U_{\Delta}^{1}U_{\Delta'}^{2}$, each operator located on the different domains. Since this is an operator in the CFT it can be expanded as
\begin{equation}
U_{\Delta}^{1}U_{\Delta'}^{2}=\mathcal{V}_{L}+\sum\limits_{\Delta\sim\mathcal{O}(c)}\mathcal{V}_{L}^{\Delta}\mathcal{V}_{\Delta}
\end{equation}
Using the invariance under the conditional expectation we have
\begin{equation}
\omega_{12}(U_{\Delta}^{1}U_{\Delta'}^{2})=\omega_{12}(E(U_{\Delta}^{1}U_{\Delta'}^{2}))=\omega_{12}(\mathcal{V}_{L})\;,
\end{equation}
while 
\begin{equation}
\omega_{12}\circ E_{12}(U_{\Delta}^{1}U_{\Delta'}^{2})=\omega_{1}\circ E_1(U_{\Delta}^{1})\,\omega_{2}\circ E_2(U_{\Delta'}^{2})=0\;.
\end{equation}
Therefore, the only difference from the two states arises due to the inprint from the space of intertwiners $U_{\Delta}^{1}U_{\Delta'}^{2}$ on the GFF fields left by the conditional expectation.  Therefore, such relative entropy, although arising from states in the full CFT, is \emph{fully determined} by the GFF algebra. Again, this seems to answer the question as to why the low energy Einstein-Hilbert gravity action knows so well about the entropy of its high energy states.

Let's describe the properties of the intertwiner imprint on the GFF algebra. This imprint is the operator $\mathcal{V}_{L}=E(U_{\Delta}^{1}U_{\Delta'}^{2})$. This operator is not generated additively in regions $W_{1}$ and $W_{2}$, but it commutes with the complement domain $W_{12}'$. This is proven as follows. First, the CFT intertwiner $U_{\Delta}^{1}U_{\Delta'}^{2}$ is additively generated in regions $W_{1}$ and $W_{2}$. This implies
\begin{equation}
[U_{\Delta}^{1}U_{\Delta'}^{2},\mathcal{V}_{L}^{(12)'}]=0\;,
\end{equation}
where $\mathcal{V}_{L}^{(12)'}\subset W_{12}'$. But since $E(0)=0$ we have:
\begin{equation}
E([U_{\Delta}^{1}U_{\Delta'}^{2},\mathcal{V}_{L}^{(12)'}])=[E(U_{\Delta}^{1}U_{\Delta'}^{2}),\mathcal{V}_{L}^{(12)'}]=[\mathcal{V}_{L},\mathcal{V}_{L}^{(12)'}]=0\;.
\end{equation}
This shows that GFF fields have the problems of algebras and regions of theories with SS, their intertwiners and associated relative entropies being controlled by the projection of the CFT algebra into the GFF algebra. In holography, these GFF imprints have well-known duals in the bulk. We thus need to maximize correlations over configurations of GFF in the appropriate region.  Since the entropy that such GFF fields can carry is bounded by the holographic principle  \emph{throughout the whole bulk}, the relative entropy is bounded by the area of the minimal surface crossed by the GFF. What it would lack to be proven is that such a bound can indeed be saturated. But it is more interesting to see the boundedness of $S(\omega_{12}|\omega_{12}\circ E_{12})$, together with the majorization of intertwiner contributions, as the microscopic origin of the holographic entropy bound.

\subsection{Large N vector models}
A simpler example than large $N$ gauge theories is given by large $N$ vector models. These models also come with a natural subalgebra which is the one of invariant operators. For example, we can take a $SO(N)$ symmetry group with $\frac{N(N-1)}{2}$ generators. We take the full theory as ${\cal F}$ and the invariant operators as ${\cal O}$.

There are two interesting regimes. The first one is when we take $\epsilon\rightarrow 0$ between complementary regions first, for a given fixed  $N$ that can be large.  
This should give for complementary regions, according to section \ref{U1} 
\be
\Delta I=  \frac{N(N-1)(d-2)}{4} \log \frac{R}{\epsilon}+\textrm{subleading}\,.\label{susy}
\ee
This increases with the number of generators of the symmetry $\sim N^2$ while the central charge increases with $N$. Besides, the area term of the mutual information is the same in both models. The entropy in intertwiners that gives the difference between the complete model and the orbifold is not enough to affect the area term. These features are very different from the holographic case.

However, there is another regime that appears when we take the $N\rightarrow \infty$ first, and then allow $\epsilon\rightarrow 0$. 
The above formula cannot apply anymore in this case. For $N$ large enough $\Delta I$ in (\ref{susy}) will overcome the mutual information of the full model that increases only with the central charge $\sim N$.  
For example, for $N$ identical independent free fields the logarithmic term of the full theory grows like $N$ in even dimensions and does not exist in odd dimensions. Hence, the leading contribution in $N^2$ for the logarithmic term of the orbifold and the logarithmic contribution from the intertwiners exactly cancel. This gives a contribution $\sim - N^2 \log \frac{R}{\epsilon}$ for the orbifold. This negative term cannot overcome the area term $\sim N (R/\epsilon)^{d-2}$ by the positivity of mutual information. Therefore we expect a change of regime before
\be
N\sim \frac{1}{\log \frac{R}{\epsilon}}\left(\frac{R}{\epsilon}\right)^{d-2}\,.
\ee
For such a large $N$ there are too many sectors and the probability of the fluctuations is relatively small such that different fluctuations will typically not add to the same sector. In other words, for the leading area term the full mutual information will coincide with the ``holographic term''  $S(\omega_{12}|\omega_{12}\circ E_{12})$, and the orbifold will have a subleading contribution, as is expected in holography. However, in contrast, in holography we can have for fixed large $N$ entropies in the area term as large as we want taking smaller $\epsilon$ without changing regime. This should be related to the very different density of superselection sectors as we move to larger energies.    

If for some large $N$ vector model we would have a holographic dual with the same interpretation as above, the results of section \ref{thermofield} would suggest the BH entropy is half the expression (\ref{chis}) depending on the sector probabilities on the Abelian algebra of the Casimirs of the group.   

\subsection{Monogamy}

The holographic entanglement entropy is monogamous \cite{Hayden:2011ag}. This is the property that the tripartite information is negative,
\bea
I(A,B,C)&=&S(A)+S(B)+S(C)-S(AB)-S(BC)-S(AC)+S(ABC)\label{last}\\
&&\hspace{-.7cm}=I(A,B)+I(B,C)+I(A,C)-S(\rho_{ABC}|\rho_A\otimes \rho_B\otimes \rho_C)\le 0\,.\nonumber 
\eea

It is interesting to see what can be said about the difference of this quantity for the models ${\cal F}$ and ${\cal O}$ for a general DHR case. Using the last expression in (\ref{last}) and the same tools as in section \ref{entropyDHR} we get
\bea
  \Delta I(A,B,C)&=&S(\omega_{AB}|\omega_{AB}\circ E_{AB})+S(\omega_{AC}|\omega_{AC}\circ E_{AC})+S(\omega_{BC}|\omega_{BC}\circ E_{BC})\nonumber \\  
 &&\hspace{4cm} -S(\omega_{ABC}|\omega_{ABC}\circ E_{ABC})\,.
\eea

For a finite group the first three terms can be at most $\log|G|$, but this bound can be simultaneously saturated, for example, for three regions that touch each other. The last term with the minus sign is bounded above by $2\log |G|$ because of the same convexity reasons used in section \ref{upper}. For regions that touch this again will be saturated.  Therefore, $\Delta I(A,B,C)=\log |G|$ in this case. This is positive, but is consistent with the negative sign in topological entanglement entropy for topological models \cite{Kitaev:2005dm,Levin:2006zz} since this term should be attributed to the ``gauged'' model ${\cal O}$ as a negative term.  

If the holographic entropy is dominated by this difference we see the negativity of $I(A,B,C)$ is quite different from the topological case. An independent argument for monogamy is necessary since $\Delta I(A,C,B)$ can be positive. In particular, in the holographic case, we are always far from saturation and the freedom of rearrangement of the intertwiners can play an important role for this property.      

As a final commentary, we note that monogamy is not ensured by large $N$ limits in vector models. We can take $N$ identical free scalar fields with very large $N$ such that the $I_3$ of the full model is the same as $N$ times the mutual information of a single scalar. The free scalar is not monogamous \cite{Casini:2008wt}.  

\section{Summary and conclusions}
\label{conclusion}

In the context of QFT, the definition and computation of meaningful information theoretic quantities can become extremely complicated. The reason is simple.  The most basic building block, the entanglement entropy, is infinite and therefore ill-defined. To overcome this obstacle, two natural avenues have been pursued in the past. The first and most natural thing to do is to regularize the QFT with a lattice, which makes entanglement entropy finite. The problem is that we should only trust aspects of such entanglement entropy that do not depend on the regularization scheme. Unfortunately, in several examples, it turns out that to obtain the expected universal results one needs to fine-tune the UV lattice definition, for example by ad hoc choices of boundary operators/algebras. The second and most rigorous avenue is to consider mutual information or related quantities, which can be considered either directly in the continuum QFT or as limits of lattice quantities \cite{Casini:2015woa}.  The advantage of this approach is that it is in principle free from ambiguities, but the surprise is that in some case it apparently turns out not to provide the expected universal results. The questions are thus clear: How do we extract the universal terms in the expansion of the entanglement entropy correctly and unambiguously? What are the new physical features involved?

The main objective of this article has been to study these problems for the case of theories with global symmetries. These symmetries have the property that charged operators can be constructed locally. In the context of algebraic QFT, these charged superselection sectors are called DHR (because of Haag, Doplicher, Roberts \cite{Doplicher:1969tk,Doplicher:1969kp,Doplicher:1973at}).

The solution to the problem stated above starts with the key observation that theories with DHR sectors have certain ambiguities in the assignation of algebras to regions. These ambiguities have been known for a long time, see \cite{Haag:1992hx}, and we have described them in section~\ref{algebra-regions} in fair generality. The main important message in this regard is that in theories with DHR sectors it is not possible to assign algebras to regions in a satisfactory way, where this means a way satisfying the properties of isotonia~(\ref{isotonia}), duality~(\ref{duality}), additivity~(\ref{additivity}) and intersection~(\ref{intersection}).\footnote{Duality for two intervals in CFT in $d=2$ is related to modular invariance. Then duality in higher dimensions and different regions can also be thought as requirements generalizing the ones of modular invariance for $d=2$ to other QFT and dimensions.} 
More concretely, for global symmetries, there is a clash between duality and additivity for certain topologically non-trivial regions. In particular, for two disconnected regions, such as the ones used to define entanglement entropy through mutual information, the additive algebra of regions $1$ and $2$, defined as usual as $\mathcal{O}_{W_{1}}\vee \mathcal{O}_{W_{2}}$, is not equal to the commutant algebra of the complementary region. Calling such complementary region $(12)'$, we have a violation of duality
\be
\mathcal{O}_{W_{1}}\vee \mathcal{O}_{W_{2}}\subset (\mathcal{O}_{W_{(12)'}})'\;.
\ee
The reason for such proper inclusion is that one can find neutral operators $I_{r}$, which are called  intertwiners for group theoretic reasons, which do not belong to the additive algebra $\mathcal{O}_{W_{1}}\vee \mathcal{O}_{W_{2}}$ but commute with the algebra of the complementary region $\mathcal{O}_{W_{(12)'}}$. Basically, for localized charge creating operators $V^{i}$, transforming in certain representation $r$ with dimension $d_{r}$ of the symmetry group, one can form the neutral operator
\be
I_{r}=\sum\limits_{i=1}^{d_{r}}V_{1}^{i}(V_{2}^{i})^{\dagger}\;,
\ee
where the subscript indicates the localization properties of the operator. From this expression it is transparent that $I_{r}\in (\mathcal{O}_{W_{(12)'}})'$ but $I_{r}\notin \mathcal{O}_{W_{1}}\vee \mathcal{O}_{W_{2}}$.

Crucially, there is a loss of duality for the complementary region as well. More concretely we have
\be
\mathcal{O}_{W_{(12)'}}\subset  (\mathcal{O}_{W_{1}}\vee \mathcal{O}_{W_{2}})'\;.
\ee
In turn, this is due to the existence of twist operators $\tau_{[g]}$, labeled by the conjugacy classes of the global group, which basically implement the symmetry transformation just in one of the connected components, but they belong to the neutral algebra as well, even in the non-abelian case. These twists do not belong to the additive algebra of the complementary region $(12)'$, but since it is a symmetry transformation, it commutes with all $\mathcal{O}_{W_{1}}\vee \mathcal{O}_{W_{2}}$, which is composed of products of neutral operators. Most importantly, as it has been described in several places in the article, these twists operators do not commute with the intertwiners.

We want to remark that these observations, the appearance of these intertwiners and twists when considering topologically non-trivial regions, do not depend on the regularization scheme. In particular, it does not depend on algebra choices in a lattice regularization. It is a true physical feature of the continuum QFT, a macroscopic manifestation of the underlying global symmetry group. It is also important to notice that these observations are purely made within the vacuum sector of the theory, no charge creating operator is needed, since both $I_{r}$ and $\tau_{[g]}$ are neutral operators that indeed belong to the additive algebra of a sufficiently big ball in spacetime.

The solution to the problem stated above is rooted in the implications of the existence of such operators for mutual information. From both a technical and physical perspective, the whole article has been devoted to analyzing the modifications to the mutual information due to this enlarged operator algebras. The main tool that has been used is the following wonderful formula
\be
S(\omega|\phi\circ E)-S(\omega_{\cal O}|\phi)=S(\omega|\omega_{\cal O}\circ E)\,, \label{dfh1}
\ee
which is proved and described in depth in \cite{ohya2004quantum}. In this formula $E:\cal F\rightarrow \cal O$ is a conditional expectation between algebras satisfying an inclusion relation $\cal O\subset \cal F$, and $\omega_{\cal O}$ is just the restriction to $\cal O$ of the state $\omega$. When applied to QFT, $\cal F$ is the field algebra, which includes charge creating operators in all irreducible representations, the symmetry group operations and the neutral algebra, which is $\cal O$ in our case. By choosing the conditional expectation appropriately, when computing the mutual information between regions $1$ and $2$ the previous formula becomes
\be
 I_{\cal F}- I_{\cal O}=S_{\cal F}(\omega|\omega_{\cal O}\circ E)=S_{(\mathcal{O}_{(12)'})'}(\omega|\omega_{\cal O}\circ E)\;,
\ee
implying that such relative entropy difference can be computed solely from the neutral algebra in the vacuum sector of the theory. Even $ I_{\cal F}$ has a natural and direct definition in $\mathcal{O}$, see~(\ref{IO}).

The fact that the difference of mutual informations is itself a relative entropy greatly simplifies the analysis of such an object since one can resort to monotonicity and convexity to constraint it in several ways. More concretely, we have found two dual ways to attack the problem. In the first approach, we compute a lower bound to such relative entropy by restricting to a certain finite algebra of intertwiners, constructed basically from~(\ref{esa1}). The challenge is to find the best finite intertwiner subalgebra, i.e a finite subalgebra providing the best lower bound to the relative entropy. Interestingly, this maximization procedure requires two concrete physical ingredients. First, from a group theory point of view, we need to choose the intertwiner subalgebra associated with the regular representation of the group. Second, from the point of view of QFT, once such regular representation is chosen we have to make sure we maximize the correlation functions in the vacuum state. This forces us to choose the intertwiners so as to commute as much as possible with the modular Hamiltonian. Explicit examples of this maximization of correlation functions, and of how the regular representation is inherently present in the vacuum have been described in section~(\ref{bounds}). The identification of these two physical features, the regular representation and choosing intertwiners that commute with the modular Hamiltonian, are two of the most important physical messages of the article.

The second line of attack uses the equality of entanglement entropies for complementary algebras to relate the previous relative entropy to another relative entropy in the complementary algebra, which includes the additive algebra and the twists operators $\tau_{g}$. From this perspective the problem is similar, we need to find the best subalgebra that provides the best upper bound. The connection with the intertwiner version is rooted in the fact that the group algebra has the same dimension as the regular representation. While the intertwiners are labeled by irreducible representations, the invariant twists can be labeled by conjugacy classes, and both labels run over the same number of elements.

Moreover, such twist/intertwiner duality is best described by both the entropic certainty and uncertainty relations derived in section~(\ref{certainty}), which nicely codify the non-commuting character of the twist/intertwiner algebra in an information theoretic manner. These uncertainty relations are also in between the most important physical messages of the article.

Using these features, we have been able to compute the modifications to universal contributions to the mutual information associated with finite and continuous (Lie) groups, including large-N vector models for example. We have also computed the universal contributions when considering different topologies (more subregions), excitations, scenarios with spontaneous symmetry breaking, thermofield double states and analyze the particularities of two-dimensional theories. All these results have been described in section~(\ref{entropyDHR}). Some of the results were found previously in the literature and some of them are new. But we want to stress that all of them arise from the same basic physical principles discussed above. So in this sense, the present approach provides a unification of all these seemingly disconnected results.

Finally, the last important message of the article is that the same set of ideas seem to apply in the context of holographic entanglement entropy. In this context, it is simple to observe how the RT prescription can be pictured as a statement about Haag-duality violation in the semiclassical limit. In particular, minimal areas measure the relative entropy associated with a set of generalized intertwiners. In the bulk, this is a precise statement, while in the boundary it requires a somewhat imprecise definition of what is a low dimension operator and what is not. This arbitrariness is codified in the definition of the conditional expectation, which basically tells how Newton's constant gets renormalized and how one separates `quantum' from `classical' contributions. Interestingly, there are two cases where everything is well defined. The first is the case of two dimensions with the subalgebra being that of the energy-momentum tensor. The second is the case of large-N vector models, which is really in the set of cases associated with continuous global symmetries. It would be a natural path to follow to try to develop approximate generalizations of the exact picture of the SS structure where the conditional expectation is the object that take the place of the symmetry.  

Our analysis shows that some of the featured properties of holographic entanglement entropy, such as the JLMS relation \cite{Jafferis:2015del} and the modified RT prescription that includes quantum corrections \cite{Faulkner:2013ana} arise naturally in our approach. Indeed, it is transparent that such properties are not particular of holographic entanglement and have a much more general scope. They basically apply to any situation in which we have a natural inclusion of algebras and the state considered is invariant under the conditional expectation that effects such inclusion. On the other hand, we have shown how monogamy of mutual information is not a generic property of theories displaying such classical versus quantum structure, and not even of large-N theories, since it is violated in large-N vector models. Monogamy is thus a true dynamical feature that needs to be studied on a case by case basis. Another interesting output of our analysis is that in usual holographic scenarios the set of generalized intertwiners provide the \emph{physical hardware} of the bit-threads proposal done in \cite{Freedman:2016zud}. This conclusion has an important consequence. Since intertwiners are here physical entities, they carry real entropy which is then bounded by the holographic entropy bound. Such bound is obviously tightest on the minimal area surface. This suggests both that the bound is at the core of holographic entanglement entropy, and that the completeness of the boundary theory is at the core of the holographic entropy bound.

We want to end with some important remarks. It is sometimes said that the problems we have been considering in the present work arise in theories with gauge symmetries, and are due to a certain arbitrariness in the choice of algebras in lattice regularizations. Our first important remark is that this is wrong. The problems only appear when the operator algebra considered is incomplete and the theory has a structure of superselection sectors. To sense the difference, we could have a ``gauge'' theory with charges in all representations. This theory has no problems of assignations of algebras to regions in any meaningful sense, where meaning is always related to properties of the continuum QFT. An important example in this regard is holographic theories, which are expected to be complete theories. Such gauge theories and completeness topics will be further discussed in the companion article. Indeed, the converse is also true, we can have theories with no gauge symmetry which actually show macroscopic ambiguities in the definition of the mutual information. All the cases considered in this paper are examples of such a scenario. The second important remark is that whenever we have a structure of superselection sectors, their contribution to the mutual information can be obtained only by focusing on the vacuum sector. This is pretty impressive and indeed it can be related to the fact that the neutral algebra is an example of a sufficient algebra, whenever the state considered is invariant under the symmetry, see \cite{petz2007quantum,ohya2004quantum} for the definition of a sufficient algebra. In the context of holography, this observation would explain how the low energy theory contains information about the full entropy in quantum gravity, i.e the information about the entropy of the complete theory. 

\section*{Acknowledgements}
We thank discussions with Pablo Bueno, Gonzalo Torroba, and an inspiring communication by Roberto Longo. 
This work was partially supported by CONICET, CNEA
and Universidad Nacional de Cuyo, Argentina. The work of H. C. and J. M. is partially supported by an It From Qubit grant by the Simons foundation. 
\appendix

 \section{Constructing the regular representation}
\label{regular}

In this appendix we show how to construct the endomorphisms corresponding to the regular representation of the group. We use constructions developed in \cite{Longo:1994xe}. The regular represenation is defined as the direct sum of all irreducible sectors, each appearing a number of times equal to their dimension:
\begin{equation}\label{regend}
\rho_{\textrm{R}}\simeq \oplus_{s} \,d_{s}\, \rho_{s}\;.
\end{equation}
By `constructing' such reducible representation we mean to provide a set of charged intertwiners $V^{si}_{\textrm{R}}$ satisfying ($i$ labels potential internal multiplicities, see below),
\bea
V_{\textrm{R}}^{si}: \iota &\rightarrow & \rho_{\textrm{R}}\\
(V_{\textrm{R}}^{si})^\dagger V_{\textrm{R}}^{rj} &=& \delta_{ij}\delta_{sr}\,,\\
\sum_{si} V_\textrm{R}^{s,i} (V_\textrm{R}^{si})^{\dagger} &=& 1\,.
\label{regv}
\eea
The first relation\footnote{Intertwiners from one representation $\rho$ to another $\sigma$, denoted by $T:\rho\rightarrow\sigma$, are operators $T_{i}$ satisfying $
T_{i}\,\rho =\sigma\, T_{i}$.} says that all $V^{si}_{\textrm{R}}$ intertwine the vacuum representation $\iota$ to the regular one $\rho_{\textrm{R}}$. Equivalently, $V^{si}_{\textrm{R}}\vert 0 \rangle$ is a state that transforms under the regular representation of the group. The second and third relations ensure that the regular endomorphism can be explicitly written as
\be 
\rho_{\textrm{R}}(b) =  \sum_{s,i}  V_{\textrm{R}}^{si} b (V_{\textrm{R}}^{si})^{\dagger}\;.
\ee
More importantly for us, such relations allow to construct the closed algebra
\be 
(a)=\sum_{s,r,i,j} a_{srij} V_{\textrm{R}}^{si} (V_{\textrm{R}}^{rj})^\dagger\;,
\label{closedVr}
\ee
used in the main text to find lower bounds for the relative entropy.

There are two possible avenues to construct such a space of intertwiners. The first is to use the method described in the text for constructing the irreducible sectors. This approach requires to have some operator $\mathcal{O}_{\textrm{R}}$ that takes us from the vacuum to the given charged sector, in this case the regular one. This approach is quite sensible and physical when such operators are found easily. For example, in gauge theories, it is simple to consider Wilson lines in any given representation. 

There is also a complementary approach, that mostly requires knowledge of the charged intertwiners associated to the irreducible sectors $\rho_{s}$. These are the $V_{s}^{i}$, with $i=1,\cdots , d_{s}$, that were described in the main text. They satisfy
\bea
V_{s}^{i}: \iota &\rightarrow & \rho_{s}\,,\\
(V_{s}^i)^\dagger V_{s}^j &=& \delta_{ij}\,,\\
\sum_i V_s^i (V_s^i)^\dagger &=& 1\,,\\
\rho_{s}(b) &=& \sum_{i} \, V_{s}^{i} \,b \,(V_{s}^{i})^{\dagger}\,.
\label{vrel}
\eea
Now, the regular repesentation, as defined above~(\ref{regend}), implies the existence of partial isommetries $\omega_{s}^{i}$, with $i=1,\cdots , d_{s}$, with the following properties
\bea
\omega_{s}^{i}:\rho_{s}&\rightarrow&\rho_{\textrm{R}}\,,\\
(\omega_{s}^{i})^{^\dagger} \omega_{r}^{j} &=& \delta_{ij}\delta_{sr}\,,\\
\sum_i \omega_{s}^{i}(\omega_{s}^{i})^\dagger &=& 1\,,\\
\sum_{s i} \, \omega_{s}^{i}\rho_{s}(\omega_{s}^{i})^{\dagger} &=& \rho_{\textrm{R}}\;.
\label{omegrel}
\eea
Indeed, these operators were explicitly constructed in ref.\cite{Longo:1994xe}, with a particular charged intertwiner to the regular endomorphism $v: \iota \rightarrow  \rho_{\textrm{R}}$.  Its explicit construction might be cumbersome, but its existence is guaranteed for finite groups, see \cite{Longo:1994xe}. Assuming we have such an operator, ref.\cite{Longo:1994xe} shows that there is an anti-isomorphism between the $V_{s}^{i}$ and the $\omega_{s}^{i}$
\be
\omega_{s}^{i}=\vert G \vert \, E( v (V_{s}^{i})^{\dagger} )\;,
\ee
where $E$ is the conditional expectation and $\vert G\vert$ is the order of the group.

So given $V_{s}^{i}$ and $v$, we can construct all the $\omega_{s}^{i}$. It is simple now to find the charged operators of the regular representation $V_{\textrm{R}}^{si}$. Since $V_{s}^{i}: \iota \rightarrow  \rho_{s}$ and $\omega_{s}^{i}:\rho_{s}\rightarrow\rho_{\textrm{R}}$, it is clear that
\be 
\omega_{s}^{i}V_{s}^{j} \,\iota =\rho_{\textrm{R}}\omega_{s}^{i}V_{s}^{j} \;,
\ee
or equivalently:
\be
\omega_{s}^{i}V_{s}^{j}:\iota \rightarrow  \rho_{\textrm{R}}\,.
\ee
Moreover, given~(\ref{vrel}) and~(\ref{omegrel}), it is simple to verify that relations~(\ref{regv}) hold with $V_{\textrm{R}}^{si}=\omega_{s}^{i}V_{s}^{j}$, so that~(\ref{closedVr}) is indeed a closed algebra.

\bibliography{EE}{}
\bibliographystyle{utphys}

\end{document}